%% file: charged_matter_amplitudes_ult.tex
\documentclass[12pt,a4paper]{article}
\usepackage{jheplikemod,ifpdf,tocloft,rotating,amsmath,textcomp,mathrsfs,mathtools,braket,rotating,graphicx,multirow,bbm,shuffle}
\usepackage[bb=libus]{mathalpha}
\usepackage[T1]{fontenc}
\usepackage{lmodern}
\usepackage{appendix}
\usepackage{tikz}\usetikzlibrary{calc}
\hypersetup{pdftitle={Tree Amplitudes with Charged Matter in Pure Gauge Theory},pdfcreator={},linkcolor=[rgb]{0.15,0.35,0.75},colorlinks=true,citecolor=[rgb]{0.675,0,0.2},urlcolor=[rgb]{0.15,0.35,0.65}}
\usepackage{makeidx}
\usepackage[totoc,itemlayout=singlepar,column sep=50pt]{idxlayout}
\usetikzlibrary{calc,decorations,decorations.markings,decorations.pathmorphing,decorations.pathreplacing,shapes.geometric,patterns,shapes}

\graphicspath{{figures/}}
\input{header_data.tex}

\makeindex
\title{\texorpdfstring{{\huge \mbox{Tree Amplitudes with Charged}\\[-5pt]\mbox{Matter in Pure Gauge Theory}}}{Tree Amplitudes with Charged Fermions in Pure Gauge Theory}}
\author{\vspace{-24pt} Jacob~L.~Bourjaily,}\emailAdd{bourjaily@psu.edu}
\author{\vspace{-2pt} Michael~Plesser,}\emailAdd{plesser@psu.edu}
\author{\vspace{-2pt} Philip~Velie}\emailAdd{velie@psu.edu}
 
\affiliation{Institute for Gravitation and the Cosmos, Department of Physics,\\Pennsylvania State University, University Park, PA 16802, USA}
\abstract{%
We describe the implementation and usage of \fpackage, a \textsc{Mathematica} package for the computation of tree amplitudes involving arbitrary numbers of gauge bosons and arbitrarily-charged massless fermions of (possibly) distinct flavours in pure (non-supersymmetric) gauge theory. These are given in terms of a basis of partial amplitudes involving distinct-flavoured fermions dressed by specific colour tensors. Distinct-flavour partial amplitudes are expressed as linear combinations of those involving only a single flavour, which may be evaluated as component amplitudes of (maximally) supersymmetric Yang-Mills theory. All relevant colour tensors can be realized as explicit, numeric arrays given any choice of charge generators (for any gauge theory---including $\mathfrak{u}_1$); from these, all colour contractions relevant to cross sections may be readily computed.\\

The complete package and a notebook demonstrating its primary usage and functionality are included in this work's submission's ancillary files on the \texttt{arXiv}.
}

\preprint{}

\begin{document}

\maketitle\thispagestyle{empty}
\pagenumbering{roman}

\setcounter{section}{0}

\pagenumbering{arabic}
\vspace{0pt}%

\section{Introduction}

Recent decades have been witness to enormous progress in our understanding of and our ability to compute perturbative scattering amplitudes in a variety of quantum field theories. This is especially the case for theories involving massless particles in four dimensions. Throughout these developments, concrete and difficult calculations, often well beyond (in order or multiplicity) or outside what would be required for real-world experiments, have led to discoveries of shocking simplicity and novel mathematical structures which, once understood and implemented, give way to more powerful computational tools that drive these developments even farther. This progress has been fueled by (theoretical) \emph{data}, and publicly available built upon reliable computational tools for computing amplitudes and the ingredients required in their representation.

Scattering amplitudes of arbitrary multiplicity in pure (or supersymmetric) gauge theory \emph{without} charged matter can today be effectively determined using a variety of modern techniques at tree-level (and often beyond (see \emph{e.g.}~\mbox{\cite{ArkaniHamed:2008gz,ArkaniHamed:2010kv,ArkaniHamed:2012nw,Arkani-Hamed:2013jha,Bourjaily:2013mma,Bourjaily:2015jna,Bourjaily:2017wjl,Bourjaily:2023apy}})); and these representations have been built into widely-available computational tools including \package~\cite{Bourjaily:2023uln} (see also \mbox{\cite{Dixon:2010ik,Bourjaily:2010wh}}). Surprisingly, however, this has not been the case for amplitudes involving charged or massive matter \emph{even at tree-level}---other than through the Feynman expansion itself, of course (see \emph{e.g.}\ \mbox{\cite{Mertig:1990an,Shtabovenko:2020gxv,Gleisberg:2003xi,Alwall:2014hca}}).\footnote{To be fair, this gap in our collective computational toolbox does not obviously limit our predictive reach for amplitudes most relevant to real-world experiments---those involving relatively few numbers of particles. In many cases, the inefficiencies of the Feynman expansion are more than compensated by the regular advance in raw computer power.}\\

In this work, we aim to fill this hole in our public toolbox by implementing the seminal ideas of Melia \cite{Melia:2013bta,Melia:2013epa}, Johansson and Ochirov \cite{Johansson:2015oia} (among others) into the package \fpackage---which is included with the submission files for this work on the \texttt{arXiv}.\\[-6pt]

\subsection{\emph{Spiritus Movens}}

Considering that tree-amplitudes involving (even massive) fermions and gauge bosons can be obtained via on-shell recursion (see \emph{e.g.}\ \cite{ArkaniHamed:2008yf,Cohen:2010mi}), it is natural to wonder why tools to compute them in this way have not yet been made available---especially considering the work of \cite{Dixon:2010ik} which showed how the kinematic dependence of amplitudes involving relatively few \emph{distinguishable} fermions could be faithfully represented in terms of the component amplitudes of maximally supersymmetric ($\mathcal{N}\!=\!4$) Yang-Mills theory (`sYM'), which are themselves well known and widely available. To appreciate the challenges involved in implementing a general solution to this problem, it is worthwhile to step back and first address the relevance of supersymmetry to this problem.

Amplitudes in \emph{pure} Yang-Mills theory are unaffected by the existence of supersymmetry at tree-level for the simple reason that any states other than gauge bosons can arise only via loops. But why is supersymmetry valuable in the representation of non-supersymmetric amplitudes? The answer is in part: because supersymmetry relates particles of different helicities, \emph{super}amplitudes encode a large number of particular helicity amplitudes as `components' (see e.g.~\cite{ArkaniHamed:2008gz,Bourjaily:2025tpr}); but almost as important is the fact that on-shell recursion is easier to implement in the case of sYM relative to pure YM, as there are no `bad-shifts' which must be avoided recursively and depend sensitively on the helicities of particles involved.

Of course, sYM includes fermions (and scalars) in its spectrum. Even if these particles are \emph{necessarily} charged under the adjoint representation of the gauge group (as is the case for any massless spin-1 particles \cite{Benincasa:2007xk}), surely the `colour-stripped' \emph{partial}-amplitudes (or `primitives') of sYM should encode the kinematic dependence of amplitudes involving fermions and gauge bosons. And indeed, they \emph{sometimes} do. However, while it is reasonable (if not preferable) to consider amplitudes involving an arbitrary number of distinguishable fermions in gauge theory, the number of distinguishable gauginos of sYM is strictly bounded by $\mathcal{N}\!\leq\!4$. But there are reasons to be optimistic about the situation. 

Consider, for example, any amplitude involving a single fermion line and any number of gauge bosons. At tree-level, all Feynman diagrams contributing to such an amplitude are obviously the same irrespective of supersymmetry---as there is only a single `flavour' of fermion, and no scalars of sYM can arise as intermediate particles.

For amplitudes involving two or more \emph{distinguishable} fermions, however, the partial amplitudes of sYM often include the unwanted contributions from scalar particles. In \cite{Dixon:2010ik}, the authors showed that such unwanted contributions could be systematically removed for amplitudes involving a small number of distinguishable fermions, and they suggested optimism that this would continue to be possible for arbitrarily many fermions.

The work of \cite{Dixon:2010ik} undoubtedly inspired the investigations of Melia \cite{Melia:2013bta,Melia:2013epa}, who in \cite{Melia:2013epa} discovered the surprising fact that partial amplitudes involving $n_f$ \emph{distinctly}-flavoured fermions in pure gauge theory are always expressible as linear combinations of those involving at most a \emph{single} flavour of fermion. (The converse is trivial: partial amplitudes involving fewer flavours of fermion can always be expressed as a sum over distinct flavour-pairings.) And because all single-flavoured partial amplitudes are identical to those of $\mathcal{N}{=}1$ supersymmetric gauge theory (which are in turn identical to those of sYM for any fixed choice of gaugino `flavour'), this automatically ensures that all partial amplitudes for charged matter in pure gauge theory can be computed using supersymmetric bookkeeping.\\

\newpage
The seminal work of Melia on partial amplitudes involving matter left open the question of what precise colour tensors should `dress' each partial amplitude. This was answered (for same-charged fermions) by Johansson and Ochirov in \cite{Johansson:2015oia} (see also \cite{Melia:2015ika}) using consistency across factorization channels. Importantly, their work made no reference to the peculiarities of any particular representation of any particular gauge theory; as such, their colour tensors may be constructed directly from the generators of any representation of any gauge group.

Despite (or perhaps because of) the generality of their result, the colour tensors that appear are several steps removed from those most familiar to physicists, requiring substantial efforts to make use of them in the determination of cross sections, for example.\\[-6pt]

One final comment is in order regarding the {colour}-dependence of amplitudes and the relationship between amplitudes involving distinguishable and indistinguishable fermions. While the colour-ordered partial amplitudes involving indistinguishable fermions are trivially the same as those of $\mathcal{N}{=}1$ sYM, the colour-tensors required to decorate such `colour-stripped' partial amplitudes are not generally known (beyond some select cases of fermion charge and gauge-group (see \emph{e.g.}\ \cite{Maltoni:2002mq})). Moreover, while same-flavoured, colour-dressed amplitudes are obtainable from multi-flavoured ones, the converse is not so clear.\\[-6pt]

The colour tensors described by Johansson and Ochirov \cite{Johansson:2015oia} are those for the case of partial amplitudes involving \emph{distinguishable} fermions, leading to a form of colour-dressed amplitudes in gauge theory coupled to arbitrarily charged (but distinguishable) fermions of the form
\eq{\mathcal{A}(\{{\color{flavour1}\psi_{1}},{\color{flavour2}\psi_{2}},\ldots,{\color{flavour1}\bar{\psi}_{n}}\})=\sum_{\vec{\sigma}}\mathcal{C}[{\color{flavour1}\underline{1}}\,\vec{\sigma}\,\,{\color{flavour1}\overline{n}}]\,\mathcal{A}({\color{flavour1}\psi_{1}}\,\vec{\sigma}\,\,{\color{flavour1}\bar{\psi}_{n}})\,,\label{colour_decomposition_of_fermionic_amplitudes}}
where the sum ranges over the $(n\text{-}2)!/(n_f!)$ \emph{ordered} partial amplitudes of Melia's (`all-plus') basis for distinctly-flavoured fermions, which we define below in (\ref{all_plus_basis_of_partial_amps}). From this, we can easily obtain the representation of an amplitude involving indistinguishable fermions by simply summing over all $n_f!$ flavour-pairings (or appropriate subsets thereof):
\eq{{A}(\{{\color{flavour0}\psi_1},{\color{flavour0}\psi_2},\ldots\})=\sum_{\vec{\sigma}\in\mathfrak{S}([n_f])}\mathcal{A}(\{{\color{flavour1}\psi_1},{\color{flavour2}\psi_2},{\color{flavour3}\psi_3},\ldots,{\color{flavour3}\bar{\psi}_{\text{-}\sigma_3}},{\color{flavour2}\bar{\psi}_{\text{-}\sigma_2}},{\color{flavour1}\bar{\psi}_{\text{-}\sigma_1}}\})\,.\label{same_flavour_amp_expansion}}

Notice that even for the case of indistinguishable fermions (\ref{same_flavour_amp_expansion}), only the partial amplitudes of distinctly-flavoured fermions appear. Today, it is not known how to express (\ref{same_flavour_amp_expansion}) \emph{directly} in terms of colour tensors decorating \emph{single-flavoured} partial amplitudes. (It would be worthwhile to seek such a form; but we must leave this to future work.)\\[-6pt]

Even knowing the colour-tensors required by (\ref{colour_decomposition_of_fermionic_amplitudes}), the multi-flavoured partial amplitudes required are themselves not easy to compute and not widely available from any public set of computational tools. (This is true even for the cases studied in \cite{Dixon:2010ik}.) We overcome this final obstacle by implementing Melia's flavour-reduction algorithm, expressing all multi-flavoured partial amplitudes---needed for \emph{both} multi-flavoured (\ref{colour_decomposition_of_fermionic_amplitudes}) and single-flavour (\ref{same_flavour_amp_expansion}) amplitudes---in terms of single-flavoured partial amplitudes, which in turn are identical to components of sYM.\\[-4pt]

Our goal in the present work is to combine the representations of amplitudes in (\ref{colour_decomposition_of_fermionic_amplitudes}) and (\ref{same_flavour_amp_expansion}) with all the tools required to convert these formulae into explicit analytic expressions for amplitudes for any set of charged, massless fermions. Specifically, we have implemented Melia's flavour-reduction algorithm \cite{Melia:2013bta} to allow all relevant partial amplitudes to be extracted from those of sYM, and we have built the tools required to both represent and \emph{concretely \textbf{build}} (and contract) the requisite colour-tensors---for \emph{any} choice of gauge group and fermion charge representation (including for $\mathfrak{u}_1$ gauge theory!).

\subsection{Organization and Outline}\vspace{-4pt}
This work is organized as follows. In \mbox{section~\ref{section_review}} we review the ingredients required to parse (\ref{colour_decomposition_of_fermionic_amplitudes}), starting with a general discussion of partial amplitudes (`primitives') and their relations in \mbox{section~\ref{relations_among_partial_amplitudes_general}}. We discuss partial amplitudes involving \emph{indistinguishable} fermions in \mbox{section~\ref{review_indistinguishable_partial_amps}}, those involving \emph{distinguishable} fermions in \mbox{section~\ref{subsection_review_of_multiflavoured_partials}}, the labeling conventions we use in \mbox{section~\ref{state_labeling_conventions}}, and then outline Melia's flavour-reduction algorithm \cite{Melia:2013epa} in \mbox{section~\ref{section_flavour_reduction_algorithm}}. 

In \mbox{section~\ref{colour_tensors}} we review the colour-decomposition of (\ref{colour_decomposition_of_fermionic_amplitudes}), and review the precise colour tensors described by Johansson and Ochirov \cite{Johansson:2015oia} in \mbox{section~\ref{jo_tensors_review}}; we clarify the notation we use in \mbox{section~\ref{tensor_labeling_conventions}}.  

In \mbox{section~\ref{section:fermionic_amplitudes_package}} we describe our implementation of these ideas into the \textsc{Mathematica} package \fpackage. We discuss how to download, initialize, and install the package in \mbox{section~\ref{subsec:obtaining_and_installing}}, and describe the general syntax and conventions used by the package in \mbox{section~\ref{subsec:syntax_and_conventions_for_the_package}}. The main functionality of the \fpackage\ package is summarized in \mbox{section~\ref{subsec:primary_functionality_of_package}}. 

We conclude in \mbox{section~\ref{sec:conclusion}} with a discussion of other gaps in our collective toolbox for computing scattering amplitudes, with an eye toward some which are being developed today.\\[-6pt]

The primary functions provided by the \fpackage\ package are documented in some detail in the \hyperlink{context_organization_of_appendix}{appendices}. These document each function's syntax and usage (often with illustrative examples). A complete list of these functions (organized alphabetically) can be found in the \mbox{the \hyperlink{index}{index}}.

In addition to the source code for the package \fpackage, we have included a \textsc{Mathematica} \built{Notebook} \texttt{walkthrough\_and\_demonstrations.nb} which guides the user through many examples of the package's use. In particular, this notebook includes a number of randomized consistency checks on our implementation.

\newpage
\section{Review: Gauge Theory Amplitudes with Charged Matter}\label{section_review}

In this section, we review the work of Melia \cite{,Melia:2013bta,Melia:2013epa}, Johansson and Ochirov \cite{Johansson:2015oia}, and others (see \emph{e.g.}~\cite{Melia:2015ika}) leading to the colour-decomposition of amplitudes involving $n_f$ \emph{distinguishable} fermions given above---namely
\eq{\mathcal{A}(\{{\color{flavour1}\psi_{1}},{\color{flavour2}\psi_{2}},\ldots,{\color{flavour1}\bar{\psi}_{n}}\})=\sum_{\vec{\sigma}}\mathcal{C}[{\color{flavour1}\underline{1}}\,\vec{\sigma}\,\,{\color{flavour1}\overline{n}}]\,\mathcal{A}({\color{flavour1}\psi_{1}}\,\vec{\sigma}\,\,{\color{flavour1}\bar{\psi}_{n}})\,.\label{colour_decomposition_of_fermionic_amplitudes_v2}}
We start with a discussion of the kinematic-dependent partial amplitudes involving indistinguishable and distinguishable fermions, before discussing the colour-dependent tensors that appear in (\ref{colour_decomposition_of_fermionic_amplitudes_v2}).\\[-6pt]

\subsection{\emph{Partial} Amplitudes and their (Linear) Relations}\label{relations_among_partial_amplitudes_general}\vspace{-4pt}

\emph{Partial amplitudes} (or, more precisely, partial amplitude \emph{primitives}) encode the \emph{kinematic dependence} (or `colour-stripped' part) of a gauge-invariant subset of Feynman diagrams consistent with some plane \emph{ordering} of the particles involved---that is, those diagrams which can be drawn without crossing on the disc with external states ordered along the boundary. All Feynman diagrams at tree-level (and at one-loop) can be drawn without crossing for at least one choice of ordering for the external particles (often, they contribute to many partial amplitudes). 

For an amplitude involving $n$ particles, each particle can be labelled by some index $a\!\in\![n]$, and we may use these labels to describe this ordering. The various partial amplitudes involving $n$ external states can therefore be labeled by an ordering $\{\sigma_1,\ldots,\sigma_n\}\equivL\vec{\sigma}\!\in\!\mathfrak{S}([n])$ (a permutation of the indices $[n]$). For a given ordering $\vec{\sigma}$, we may write the partial amplitude 
\eq{A(\sigma_1,\ldots,\sigma_n)\equivL A(\vec{\sigma})\,.\label{labeling_of_partial_amps}}

From their definition, it is clear that partial amplitudes are invariant under cyclic rotations of their indices; that is,
\eq{A(\sigma_1,\sigma_2,\ldots,\sigma_n)=A(\sigma_2,\ldots,\sigma_n,\sigma_1)=\cdots\,=A(\sigma_n,\sigma_1,\sigma_2,\ldots)\label{cyclicity_of_partials}}
for any choice of $\vec{\sigma}\!\in\!\mathfrak{S}([n])$. Dihedral reflections also leave partial amplitudes invariant (possibly up to a sign):
\eq{A(\sigma_1,\sigma_2,\ldots,\sigma_n)=(\text{-}1)^{n}A(\sigma_n,\ldots,\sigma_2,\sigma_1)\,.\label{dihedral_of_partials}}
Depending on whether or not these equivalences are taken into account, it could be said that there are $n!$, $(n{-}1)!$, or $(n{-}1)!/2$ `distinct' partial amplitudes for $n$ external particles.

Less trivially, partial amplitudes satisfy the `KK relations' described by Kleiss-Kuijf in \cite{Kleiss:1988ne}:
\vspace{5pt}\eq{{A}(\g{A\rule[-1.05pt]{14.5pt}{0.65pt}},\r{\alpha},\t{B\!\rule[-1.05pt]{14.5pt}{0.65pt}},\b{\beta},\g{C\!\rule[-1.05pt]{14.5pt}{0.65pt}})=(\text{-}1)^{1+|\g{CA}|}\hspace{-12pt}\sum_{\fwboxL{35pt}{{\vec{\sigma}}\!\in\!(\t{B})\!\shuffle\!(\g{\bar{{CA}}})}}\hspace{-10pt}{A}(\r{\alpha},{\vec{\sigma}},\b{\beta})\,;
\label{kk_relation_among_partials}}
here, `$\g{\bar{{CA}}}$' denotes the \emph{reversed} concatenation of the ordered sequence of indices $(\g{C}|\g{A})$, and `$A\shuffle B$' denotes the \emph{shuffle}-product: the set of all permutations of $A\cup B$ consistent with the ordering of each set separately. (Notice that both cyclicity (\ref{cyclicity_of_partials}) and dihedral reflection (\ref{dihedral_of_partials}) of partial amplitudes follow from the KK relations.)

From the KK relations, it is easy to see that the $n!$ partial amplitudes can be expressed as sums over many choices of $(n{-}2)!$ of them. Given any choice of of the pair $\{\r{\alpha},\b{\beta}\}$ in (\ref{kk_relation_among_partials}), the $(n{-}2)!$ partial amplitudes
\eq{\text{(KK-)basis of partial amplitudes: } \big\{A(\r{\alpha},\vec{\sigma},\b{\beta})\big\}_{\vec{\sigma}\in\mathfrak{S}([n]\backslash\{\r{\alpha},\b{\beta}\})}\,\label{kk_bases_of_partials}}
can be taken as a linearly-independent \emph{basis} of partial amplitudes for any choice of the `anchors' $\{\r{\alpha},\b{\beta}\}\!\subset\![n]$.

(In addition to the KK relations, there are also the so-called `BCJ relations' \cite{Bern:2008qj} (see also \mbox{\cite{Bern:2010ue,Bern:2019prr,Adamo:2022dcm,Bern:2023zkg}}) which linearly relate partial amplitudes but with non-constant, \emph{kinematic-dependent} coefficients.)\\[-6pt]

To be clear, depending on the quantum numbers labelling the states, it may be that some partial amplitudes vanish; this happens whenever no Feynman diagram can be drawn on the disc with respect to the cyclic ordering. For example, an amplitude involving two \emph{distinguishable} fermions---an electron ${\color{flavour1}e}$ and muon ${\color{flavour2}\mu}$, say---there are no planar diagrams at tree-level consistent with the ordering $A({\color{flavour1}e},{\color{flavour2}\mu},{\color{flavour1}\bar{e}},{\color{flavour2}\bar{\mu}})$ because there is no way to draw a line connecting ${\color{flavour2}\mu}$ to ${\color{flavour2}\bar{\mu}}$ without crossing the line connecting ${\color{flavour1}e}$ to ${\color{flavour1}\bar{e}}$. Because of this, we would say that 
\eq{A({\color{flavour1}e},{\color{flavour2}\mu},{\color{flavour1}\bar{e}},{\color{flavour2}\bar{\mu}})=0.\label{vanishing_multi_flavour_partial_amp}}
(Notice that we conventionally consider all particles as \emph{incoming}; as such, every fermion must be paired with an anti-fermion.)

We will have more to say about the partial amplitudes involving distinguishable fermions; but let us first turn our attention to the case of amplitudes involving some number of \emph{indistinguishable} fermions.\\[-6pt]

\subsubsection{Partial Amplitudes Involving \emph{Indistinguishable} Fermions}\label{review_indistinguishable_partial_amps}

As we have seen in (\ref{vanishing_multi_flavour_partial_amp}), partial amplitudes involving distinguishable fermions can sometimes vanish. This does not happen for amplitudes involving \emph{indistinguishable} fermions. For partial amplitudes involving fermions of the same flavour, it is always possible to connect every fermion to some anti-fermion in a way consistent with any ordering along the disc. 

Consider, for example, the partial amplitude involving three indistinguishable fermion lines and a single gauge boson (of `${+}$' helicity, say):
\eq{A({\color{flavour0}\psi},{\color{black}g^{+}},{\color{flavour0}\bar{\psi}},{\color{flavour0}{\psi}},{\color{flavour0}\bar{\psi}},{\color{flavour0}\psi},{\color{flavour0}\bar{\psi}})\,.\label{eg_single_flavour_partial}}
The fact that fermion lines are locally continuous in Feynman diagrams places restrictions on which sets of fermions may be paired with which anti-fermions. For this ordering of external states, there are five ways in which the fermions may connect to anti-fermions without crossing:
\eq{\begin{array}{@{}c@{}}\includegraphics[scale=0.6]{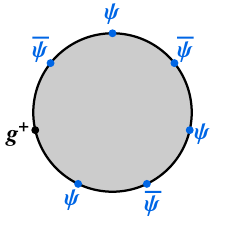}\end{array}{=}\begin{array}{@{}c@{}}\includegraphics[scale=0.6]{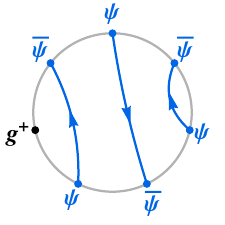}\end{array}{+}\begin{array}{@{}c@{}}\includegraphics[scale=0.6]{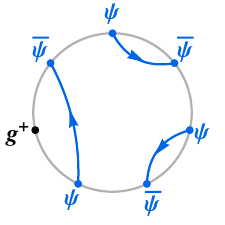}\end{array}{+}\begin{array}{@{}c@{}}\includegraphics[scale=0.6]{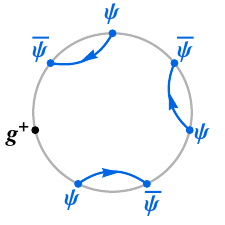}\end{array}{+}\begin{array}{@{}c@{}}\includegraphics[scale=0.6]{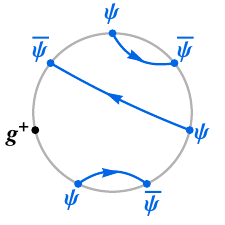}\end{array}{+}\begin{array}{@{}c@{}}\includegraphics[scale=0.6]{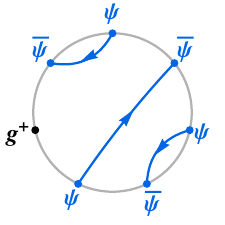}\end{array}\label{example_chords_of_single_flavour_partial_amp}}
We call pictures such as these \emph{chord diagrams}; they indicate only the plane-embedding of possible fermion lines (denoting a collection of Feynman diagrams consistent with this arrangement of external states). Any partial amplitude involving $n_f$ \emph{indistinguishable} fermions admits at least one chord diagram. Interestingly (and usefully, as we will see), there are some partial amplitudes for which the possible pairings between fermions and anti-fermions (consistent with plane embedding) is unique---\emph{e.g.},\\[-7pt]
\eq{\hspace{-80pt}A({\color{flavour0}\psi},{\color{black}g^{+}},{\color{flavour0}\bar{\psi}},{\color{flavour0}{\bar{\psi}}},{\color{flavour0}\bar{\psi}},{\color{flavour0}\psi},{\color{flavour0}\psi})\,\bigger{\;\Leftrightarrow\;}\begin{array}{@{}c@{}}\includegraphics[scale=0.6]{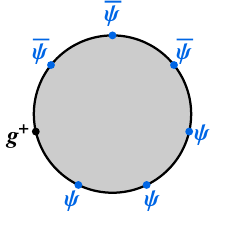}\end{array}{=}\begin{array}{@{}c@{}}\includegraphics[scale=0.6]{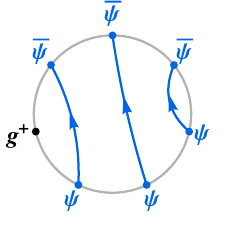}\end{array}\,.\label{highest_maturity_example}}

Partial amplitudes involving only indistinguishable fermions are said to be of `\emph{single-flavour}'. At tree-level, they are guaranteed to be identical to component amplitudes of $\mathcal{N}\!=\!1$ supersymmetric Yang-Mills (which are themselves identical to components of maximally supersymmetric Yang-Mills theory). To be clear, although the gauginos of supersymmetric theories are necessarily charged under the adjoint representation of some gauge group $\mathfrak{g}$, \emph{partial} amplitudes (or `primitives') represent only the kinematic-dependence of gauge-invariant subsets of Feynman diagrams; this kinematic dependence is unaffected by supersymmetry---only their colour-dependence would be different.\\[-6pt] 

\subsubsection{Partial Amplitudes Involving \emph{Distinguishable} Fermions}\label{subsection_review_of_multiflavoured_partials}
Let us now consider the case of partial amplitudes for which all fermions are distinguishable by their `flavour', say. If we label different flavours by distinct colours, then it is easy to see that only some flavour-pairings are consistent with plane-embedding/cyclic ordering, and some necessarily vanish. We already saw one example of this in (\ref{vanishing_multi_flavour_partial_amp}); another example would be
\eq{\hspace{-139pt}\mathcal{A}({\color{flavour1}\psi},{\color{black}g^{+}},{\color{flavour3}\bar{\psi}},{\color{flavour2}{{\psi}}},{\color{flavour1}\bar{\psi}},{\color{flavour3}\psi},{\color{flavour2}\bar{\psi}})\,\bigger{\;\Leftrightarrow\;}\begin{array}{@{}c@{}}\includegraphics[scale=0.6]{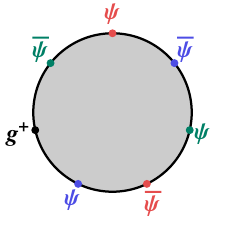}\end{array}{=}0\,}
as there is no way to connect fermions to the fermions to anti-fermions \emph{with their specified flavours} without crossing fermion lines. On the other hand, any flavour-pairing that admits \emph{some} set of non-overlapping fermion lines will be consistent with at most a single chord diagram. We give a number of examples in \mbox{Table~\ref{multiflavour_chord_diagrams_table}}.

\begin{table}[t]\caption{Illustrations of chord diagrams for multi-flavour partial amplitudes.}\label{multiflavour_chord_diagrams_table}\vspace{-20pt}$$\begin{array}{@{}c@{}}\begin{array}{@{}ccc@{}}\begin{array}{@{}c@{}}\includegraphics[scale=0.6]{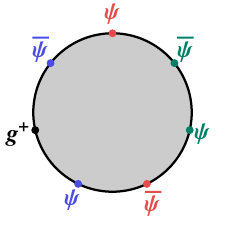}\end{array}{=}\begin{array}{@{}c@{}}\includegraphics[scale=0.6]{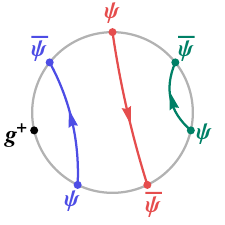}\end{array}&\begin{array}{@{}c@{}}\includegraphics[scale=0.6]{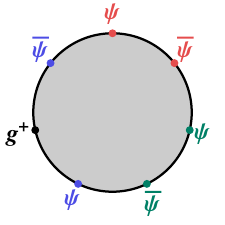}\end{array}{=}\begin{array}{@{}c@{}}\includegraphics[scale=0.6]{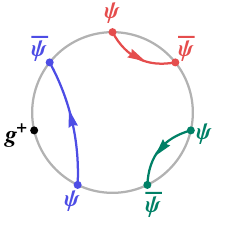}\end{array}&\begin{array}{@{}c@{}}\includegraphics[scale=0.6]{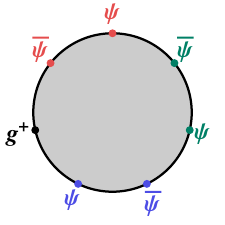}\end{array}{=}\begin{array}{@{}c@{}}\includegraphics[scale=0.6]{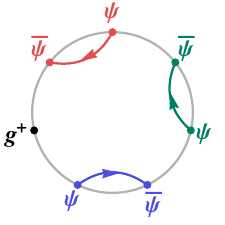}\end{array}\\[-7pt]
\mathcal{A}({\color{flavour1}\psi},{\color{black}g^{+}},{\color{flavour1}\bar{\psi}},{\color{flavour2}{{\psi}}},{\color{flavour3}\bar{\psi}},{\color{flavour3}\psi},{\color{flavour2}\bar{\psi}})&\mathcal{A}({\color{flavour1}\psi},{\color{black}g^{+}},{\color{flavour1}\bar{\psi}},{\color{flavour2}{{\psi}}},{\color{flavour2}\bar{\psi}},{\color{flavour3}\psi},{\color{flavour3}\bar{\psi}})&\mathcal{A}({\color{flavour1}\psi},{\color{black}g^{+}},{\color{flavour2}\bar{\psi}},{\color{flavour2}{{\psi}}},{\color{flavour3}\bar{\psi}},{\color{flavour3}\psi},{\color{flavour1}\bar{\psi}})
\end{array}\\[5pt]
\begin{array}{@{}c@{$\;\;\;\;\;\;$}c@{}}\begin{array}{@{}c@{}}\includegraphics[scale=0.6]{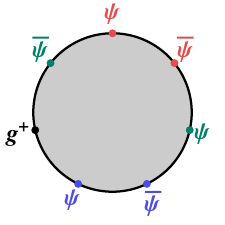}\end{array}{=}\begin{array}{@{}c@{}}\includegraphics[scale=0.6]{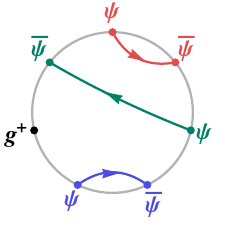}\end{array}&\begin{array}{@{}c@{}}\includegraphics[scale=0.6]{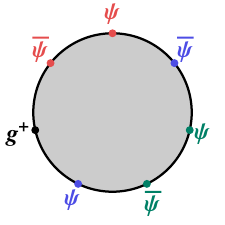}\end{array}{=}\begin{array}{@{}c@{}}\includegraphics[scale=0.6]{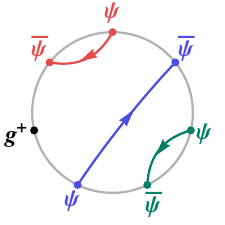}\end{array}\\[-7pt]
\mathcal{A}({\color{flavour1}\psi},{\color{black}g^{+}},{\color{flavour3}\bar{\psi}},{\color{flavour2}{{\psi}}},{\color{flavour2}\bar{\psi}},{\color{flavour3}\psi},{\color{flavour1}\bar{\psi}})&\mathcal{A}({\color{flavour1}\psi},{\color{black}g^{+}},{\color{flavour2}\bar{\psi}},{\color{flavour2}{{\psi}}},{\color{flavour1}\bar{\psi}},{\color{flavour3}\psi},{\color{flavour3}\bar{\psi}})
\end{array}\\[-0pt]~\end{array}\vspace{-40pt}$$
\end{table}

Because partial amplitudes represent merely the \emph{kinematic} dependence of a summed subset of Feynman diagrams, it is clear that any partial amplitude involving identically-flavoured fermions can be expressed as a sum over those involving distinct flavours. Thus, the partial amplitude of (\ref{eg_single_flavour_partial}) can be written as a sum over distinctly-flavoured partial amplitudes according to each of the possible chord diagrams in (\ref{example_chords_of_single_flavour_partial_amp}). These are precisely the multi-flavoured partial amplitudes given in \mbox{Table~\ref{multiflavour_chord_diagrams_table}}; as such,\\[-22pt]
\begin{align}
A({\color{flavour0}\psi},{\color{black}g^{+}},{\color{flavour0}\bar{\psi}},{\color{flavour0}{\psi}},{\color{flavour0}\bar{\psi}},{\color{flavour0}\psi},{\color{flavour0}\bar{\psi}})=&\phantom{{+}}\mathcal{A}({\color{flavour1}\psi},{\color{black}g^{+}},{\color{flavour1}\bar{\psi}},{\color{flavour2}{{\psi}}},{\color{flavour3}\bar{\psi}},{\color{flavour3}\psi},{\color{flavour2}\bar{\psi}}){+}\mathcal{A}({\color{flavour1}\psi},{\color{black}g^{+}},{\color{flavour1}\bar{\psi}},{\color{flavour2}{{\psi}}},{\color{flavour2}\bar{\psi}},{\color{flavour3}\psi},{\color{flavour3}\bar{\psi}})\nonumber\\
&{+}\mathcal{A}({\color{flavour1}\psi},{\color{black}g^{+}},{\color{flavour2}\bar{\psi}},{\color{flavour2}{{\psi}}},{\color{flavour3}\bar{\psi}},{\color{flavour3}\psi},{\color{flavour1}\bar{\psi}}){+}\mathcal{A}({\color{flavour1}\psi},{\color{black}g^{+}},{\color{flavour3}\bar{\psi}},{\color{flavour2}{{\psi}}},{\color{flavour2}\bar{\psi}},{\color{flavour3}\psi},{\color{flavour1}\bar{\psi}})\nonumber\\
&{+}\mathcal{A}({\color{flavour1}\psi},{\color{black}g^{+}},{\color{flavour2}\bar{\psi}},{\color{flavour2}{{\psi}}},{\color{flavour1}\bar{\psi}},{\color{flavour3}\psi},{\color{flavour3}\bar{\psi}})\,.\label{example_to_distinct_identity_v0}\\[-24pt]\nonumber
\end{align}
Notice that we have used `$\mathcal{A}$' to denote partial amplitudes involving $n_f$ \emph{distinctly}-flavoured fermion lines, and `$A$' to denote partial amplitudes involving indistinguishable fermions (of a `single flavour').\\[-6pt]

\paragraph{Counting Partial Amplitudes Involving \emph{Distinguishable} Fermions}~\\[-34pt]

It is worthwhile to pause and consider \emph{how many} non-vanishing partial amplitudes exist for amplitudes involving $n_f$ distinctly-flavoured fermion lines. The number of non-crossing sets of $n_f$ lines on a disc is given by the Catalan numbers $C(n_f)$:
\eq{C(n_f)\equivR\frac{1}{n_f{+}1}\binom{2n_f}{n_f}=\frac{(2n_f)!}{n_f!\,(n_f{+}1)!}\,.\vspace{5pt}\label{catalan_nf}}
(This is equal to the number of Dyck\footnote{Named for German mathematician Walther von Dyck; pronounced ``Deek''.} words of length $2\,n_f$.) As these denote fermion lines, each line should be assigned an orientation, resulting in $C(n_f)\,2^{n_f}$ \emph{oriented} pairings of fermions with anti-fermions of distinct flavours. This is the number of possible oriented chord diagrams. Allowing for gauge bosons to be placed variously around the boundary, there will be %
\eq{\text{\# gauge boson insertions: } (2n_f{+}1)(2n_f{+}2)\cdots(2n_f{+}n_g)=\frac{(2n_f{+}n_g)!}{(2n_f)!}\,}
partial amplitudes for each oriented chord diagram. Putting all this together, we find the number of (non-vanishing) partial amplitudes involving $n_f$ fermion lines and $n_g$ gauge bosons to be given by 
\eq{\fwboxR{70pt}{\begin{array}{@{}r@{}}\\[-40pt]\text{\# partial amps with}\\[-4pt]\text{$n_f$ \emph{distinct} fermions}\\[-4pt]
\text{and $n_g$ gauge bosons}\end{array}\,=\,}\overbrace{\phantom{\fwbox{0pt}{\frac{(2n_f{+}1)!}{(2n_f)!}}}\hspace{-2pt}\underbrace{C(n_f)}_{\substack{\text{chord}\\\text{diagrams}}}\,\underbrace{2^{n_f}\phantom{\fwbox{0pt}{C(n_f)}}}_{\text{orientations}}\hspace{-10pt}}^{\substack{\text{oriented}\\\text{chord diagrams}}}\,\overbrace{\phantom{\fwbox{0pt}{\frac{(2n_f{+}1)!}{(2n_f)!}}}(n_f!)}^{\substack{\text{flavour}\\\text{pairings}}}\,\overbrace{\frac{(2n_f{+}n_g)!}{(2n_f)!}}^{\substack{\text{gauge boson}\\\text{insertions}}}=\,2^{n_f}\frac{n!}{(n_f{+}1)!}\,,\label{number_of_distinct_flavoured_partials}}
where we have defined $n\equivR(2n_f{+}n_g)$ to denote the total number of particles involved. (The number of partial amplitudes would be larger by a factor of $n_f!$ if we considered all permutations of flavour indices.\footnote{On the other hand distinctly-flavoured partial amplitudes---being functions only of kinematics---do not depend on which fermion lines are assigned which flavours. Thus, while one may reasonably account for partial amplitudes differing only by their flavour assignments, these distinctions have no effect on the particular functions.})\\[-6pt]

\paragraph[Linear Relations Among Distinct-Flavoured Partial Amplitudes]{Linear Relations Among Partial Amplitudes with $n_f$ Distinct Fermions}~\\[-34pt]

Partial amplitudes involving distinctly-flavoured fermions satisfy exactly the same relations as described above---namely, they are cyclic (\ref{cyclicity_of_partials}), dihedral (up to a sign) (\ref{dihedral_of_partials}), and obey the KK relations (\ref{kk_relation_among_partials}). The primary \emph{novelty} for distinctly-flavoured partial amplitudes is that any partial amplitude for which the flavour assignments of fermion/anti-fermion pairs is \emph{inconsistent} with plane-embedding will vanish. 

This renders the KK relations considerably \emph{stronger} than in the same-flavoured case. To see this, consider first projecting all distinct-flavoured partial amplitudes into a KK basis according to: 
\eq{\big\{\mathcal{A}({\color{flavour1}\psi},\vec{\sigma},{\color{flavour1}\bar{\psi}})\big\}_{\substack{\text{(flavour-labelled) fermion lines do not cross}}}\label{kk_projected_set_of_distinct_flavour_partials}}
for any fixed (distinct-flavoured) fermion line `$({\color{flavour1}\psi},{\color{flavour1}\bar{\psi}})$'. It is not hard to see that the number of (non-vanishing) partial amplitudes included in the set (\ref{kk_projected_set_of_distinct_flavour_partials}) would be 
\eq{C(n_f{-}1)\,2^{n_f{-}1}\,(n_f{-}1)!\,\frac{(n{-}2)!}{(2n_f{-}2)!}\,{=}\,2^{n_f{-}1}\frac{(n{-}2)!}{n_f!}\,.\label{oriented_partial_amps}}
\vspace{6pt}

However, this number is too large: in \cite{Melia:2013epa,Melia:2013bta}, Melia showed that the set of partial amplitudes in (\ref{kk_projected_set_of_distinct_flavour_partials}) are \emph{not} linearly independent. Specifically, he showed that the KK-relations can in fact be used to reverse the orientation of any fermion line (recursively)---expanding any element of (\ref{kk_projected_set_of_distinct_flavour_partials}) into a true basis of partial amplitudes for which these orientations have been chosen arbitrarily for all chord diagrams. 

To show this, Melia recast the KK relations (\ref{kk_relation_among_partials}) into the following, somewhat peculiar form:
\eq{{A}(1,\g{A\rule[-1.05pt]{14.5pt}{0.65pt}},\r{\beta},\t{B\!\rule[-1.05pt]{14.5pt}{0.65pt}},\b{\alpha},C\!\rule[-1.05pt]{14.5pt}{0.65pt})=(\text{-}1)^{1+|\t{B}|}\hspace{-12pt}\sum_{\fwboxL{40pt}{\g{A}\!=\!(\!\g{a_L}\!,\!\g{a_R}\!)}}\sum_{\fwboxL{10pt}{\vec{\sigma}\!\in\![(\g{a_R}\!\shuffle\!\smash{\t{\bar{B}}})\r{\beta}]\!\shuffle\!C}}\hspace{5pt}{A}(1\,\g{a_L}\,\b{\alpha}\,\,\vec{\sigma})\,.
 \label{eqn:kkSwap}}
To be clear, this identity is valid for \emph{all} partial amplitudes---including single-flavoured ones; but it is most powerful when applied to distinctly-flavoured partial amplitudes for which many terms appearing on the RHS of (\ref{eqn:kkSwap}) vanish (as they would involve crossing fermion lines).

Taking $\{\b{\alpha},\r{\beta}\}$ in (\ref{eqn:kkSwap}) to be the positions of any choice of fermion/anti-fermion flavoured pair $\{{\color{flavour3}\psi},{\color{flavour3}\bar{\psi}}\}$, the identity can be used to reverse the orientation of the line connecting particle ${\color{flavour3}\psi}$ to ${\color{flavour3}\bar{\psi}}$. (It will also reverse the orientation of any fermion lines within the set $\{\t{B\!\rule[-1.05pt]{14.5pt}{0.65pt}}\}$). Importantly, every term in this expansion  \emph{preserves} the orientation of fermion lines outside the sequence $(\r{\beta}\,\t{B}\,\b{\alpha})$. As fermion lines are naturally organized hierarchically according to the sections of disc they partition, this identity can be iteratively applied to express any partial amplitude in (\ref{kk_projected_set_of_distinct_flavour_partials}) as a linear combination of those with \emph{fixed} fermion-line orientations. 

Thus, any fermion line may be oriented at will, reducing the number of independent partial amplitudes (\ref{oriented_partial_amps}) by a factor of $2^{n_f{-}1}$---to a basis merely of size
\eq{\frac{(n{-}2)!}{n_f!}\,.\label{melia_basis_size_for_distinct_flavours}}
It turns out that this exhausts the linear relations (with constant coefficients) satisfied by partial amplitudes involving distinctly-flavoured fermions. As such, by fixing the orientations of fermion lines in any conventional way, one will find a \emph{basis} of independent multi-flavoured partial amplitudes.\\[-6pt]

\paragraph[Melia's All-Plus Basis of Partial Amplitudes]{Melia's \emph{All-Plus} Basis of Partial Amplitudes for $n_f$ Distinguishable Fermions}~\\[-34pt]

An independent set of partial amplitudes involving $n_f$ distinguishable fermions, therefore, can be defined as the set of partial amplitudes \cite{Melia:2013bta}: 
\vspace{4pt}\eq{\text{Melia's \emph{all-plus} basis: }\big\{\mathcal{A}({\color{flavour1}\psi},\vec{\sigma},{\color{flavour1}\bar{\psi}})\big\}_{\substack{\hspace{0pt}\text{(flavour-labelled) fermion lines do not cross, \&}\\\text{for each flavour $f\!\in\![n_f]$, }\psi_f\text{ appears \emph{before} }\bar{\psi}_f\,.}}\;\;\label{all_plus_basis_of_partial_amps}\vspace{4pt}}
For partial amplitudes only involving fermions, the $\vec{\sigma}$ that appear in (\ref{all_plus_basis_of_partial_amps}) are in one-to-one correspondence with Dyck words of length $2(n_f{-}1)$ (whose orientations are all (arbitrarily) oriented in the same direction). Including also gauge bosons appearing with arbitrary positions results in a basis of size given above in (\ref{melia_basis_size_for_distinct_flavours}).\\[-6pt]

Before moving on, it is worth mentioning that (\ref{all_plus_basis_of_partial_amps}) represents a basis of partial amplitudes with a \emph{fixed} flavour-pairing between fermions and anti-fermions. For each of the $n_f!$ flavour pairings one may consider, we could construct a distinct (and linearly-independent) set of partial amplitudes according to (\ref{all_plus_basis_of_partial_amps}); the span of all such bases---over all possible flavour pairings---results in a set of $(n{-}2)!$ distinct-flavoured partial amplitudes, which is in line with the counting for the (un-flavoured) KK relations (\ref{kk_relation_among_partials}).\\[-6pt]

\subsubsection{Bookkeeping: Conventions for Labelling Partial Amplitudes}\label{state_labeling_conventions}

It is worth pausing briefly to sharpen our conventions for labeling partial amplitudes involving (possibly) distinguishable fermions and gauge bosons. We may, without loss of generality, choose to label states according to their momentum indices $a\!\in\![n]$---which we may assign to the states in whatever order we find convenient. Taking all particles to be incoming (so that the number of fermions must equal the number of anti-fermions), the state associated with the $a$th momentum $p_a$ will be denoted:
\eq{\begin{array}{l@{$\;\;\text{$\Leftrightarrow$}\;\;$}l}
\text{`$a$'}&\text{a gauge boson with momentum $p_a$}\\
\text{`${\color{flavour0}\underline{a}}$'}&\text{a \emph{fermion} (helicity $\text{+}\frac{1}{2}$) with momentum $p_a$}\\
\text{`${\color{flavour0}\overline{a}}$'}&\text{an \emph{anti-fermion} (helicity $\text{-}\frac{1}{2}$) with momentum $p_a$}\end{array}\label{tex_conventions_for_labeling_states}}
(When discussing gauge bosons with momentum $p_a$ of definite helicity, we use `$a^{\pm}$'.)\\[-6pt]

For an amplitude involving $n_f$ fermion/anti-fermion pairs and $n_g$ gauge bosons, we choose to associate the first $[n_f]\!\subset\![n]$ momenta to label fermions and the \emph{last} $[\text{-}n_f]\!\subset\![n]$ momenta to label anti-fermions. As such, the list of states will be associated with momentum indices according to
\eq{\big\{\overbrace{{\color{flavour0}\underline{1}},{\color{flavour0}\underline{\ldots}}\,,{\color{flavour0} \underline{\smash{n_f}}}}^{n_f\text{ fermions}},\overbrace{(n_f{+}1),\ldots,(n_f\text{+}n_g)}^{n_g\text{ gauge bosons}},\overbrace{{\color{flavour0}\overline{\phantom{\fwbox{0pt}{1}}{\text{-}n_f}}},{\color{flavour0}\overline{\phantom{\fwbox{0pt}{1}}\ldots\,}},{\color{flavour0}\overline{\phantom{\fwbox{0pt}{1}}\text{-}1}}}^{\hspace{-5pt}n_f\text{ anti-fermions}\hspace{-5pt}}\big\}.\label{momentum_ordering_of_states}}
To be clear, this differs from convention typically used in the literature---for which the first $[2n_f]$ momenta are assigned to an alternating list of fermions and anti-fermions, respectively (so that the first $n_f$ odd/even labels are associated with fermions or anti-fermions). Our convention is motivated \textsc{Mathematica} indexing, where the \built{Last} \built{Slot} of a \built{List} is associated with its `{-}$1$'st \built{Part}.\\[-6pt]

Using these conventions, the partial amplitude in (\ref{eg_single_flavour_partial}) `$A({\color{flavour0}\psi_1},{\color{black}g^{+}_4},{\color{flavour0}\bar{\psi}_{\text{-}3}},{\color{flavour0}{\psi}_2},{\color{flavour0}\bar{\psi}_{\text{-}2}},{\color{flavour0}\psi_3},{\color{flavour0}\bar{\psi}_{\text{-}1}})$' could be more compactly written as
\eq{A({\color{flavour0}\psi_1},{\color{black}g^{+}_4},{\color{flavour0}\bar{\psi}_{\text{-}3}},{\color{flavour0}{\psi}_2},{\color{flavour0}\bar{\psi}_{\text{-}2}},{\color{flavour0}\psi_3},{\color{flavour0}\bar{\psi}_{\text{-}1}})\;\equivL{A}({\color{flavour0}\underline{1}}\hspace{1.02pt}4^+\hspace{1.02pt}{\color{flavour0}\overline{5}}\hspace{1.02pt}{\color{flavour0}\underline{2}}\hspace{1.02pt}{\color{flavour0}\overline{6}}\hspace{1.02pt}{\color{flavour0}\underline{3}}\hspace{1.02pt}{\color{flavour0}\overline{7}})\,.}
(Notice that the `$\{\text{-}3,\text{-}2,\text{-}1\}$' momentum labels mean those of $\{5,6,7\}$ for $n{=}7$ particles.) In the \fpackage~package, we would use the expression
\eq{A({\color{flavour0}\psi_1},{\color{black}g^{+}_4},{\color{flavour0}\bar{\psi}_{\text{-}3}},{\color{flavour0}{\psi}_2},{\color{flavour0}\bar{\psi}_{\text{-}2}},{\color{flavour0}\psi_3},{\color{flavour0}\bar{\psi}_{\text{-}1}})\;\;\;\bigger{\Leftrightarrow}\;\;\;\funL{amp}\brace{\{\!\fun{f}\brace{1},\fun{p}\brace{4},\fun{fb}\brace{3},\fun{f}\brace{2},\fun{fb}\brace{2},\fun{f}\brace{3},\fun{fb}\brace{1}\,}\,.\label{example_encoding_of_sf_amp}}
where a fermion $\psi_{\var{a}}$ is encoded by `\fun{f}\brace{\var{$a$}}\,', $\bar{\psi}_{\var{\text{-}{a}}}$ is encoded by `\fun{fb}\brace{\var{$a$}}\,', and a plus-helicity gauge boson $g_{\var{a}}$ is encoded by `\fun{p}\brace{\var{$a$}}\,'. We clarify the conventions of these expressions in more detail below in \mbox{section~\ref{subsec:syntax_and_conventions_for_the_package}}.

To be clear, we are using `$\fun{fb}\brace{1}$\,' here to signify \emph{merely} the anti-fermion with momentum indexed by `$p_{\text{-}1}\equivR p_{n}$'---making \emph{no assumptions} regarding the \emph{flavour} of this anti-fermion. (Indeed: in the example (\ref{example_encoding_of_sf_amp}), all fermions are indistinguishable!)\\[-6pt]

For amplitudes involving \emph{distinguishable} fermions, we use colours to indicate fermion/anti-fermion pairs of the same flavour. Thus, the example identity (\ref{example_to_distinct_identity_v0}) would be written
\vspace{-2pt}\eq{\begin{split}
{A}({\color{flavour0}\underline{1}}\hspace{1.02pt}4^+\hspace{1.02pt}{\color{flavour0}\overline{5}}\hspace{1.02pt}{\color{flavour0}\underline{2}}\hspace{1.02pt}{\color{flavour0}\overline{6}}\hspace{1.02pt}{\color{flavour0}\underline{3}}\hspace{1.02pt}{\color{flavour0}\overline{7}})=&\phantom{{+}}\mathcal{A}({\color{flavour1}\underline{1}}\hspace{1.02pt}4^{{+}}\hspace{1.02pt}{\color{flavour1}\overline{5}}\hspace{1.02pt}{\color{flavour2}\underline{2}}\hspace{1.02pt}{\color{flavour3}\overline{6}}\hspace{1.02pt}{\color{flavour3}\underline{3}}\hspace{1.02pt}{\color{flavour2}\overline{7}}){+}
\mathcal{A}({\color{flavour1}\underline{1}}\hspace{1.02pt}4^{{+}}\hspace{1.02pt}{\color{flavour1}\overline{5}}\hspace{1.02pt}{\color{flavour2}\underline{2}}\hspace{1.02pt}{\color{flavour2}\overline{6}}\hspace{1.02pt}{\color{flavour3}\underline{3}}\hspace{1.02pt}{\color{flavour3}\overline{7}})\\
&{+}\mathcal{A}({\color{flavour1}\underline{1}}\hspace{1.02pt}4^{{+}}\hspace{1.02pt}{\color{flavour2}\overline{5}}\hspace{1.02pt}{\color{flavour2}\underline{2}}\hspace{1.02pt}{\color{flavour3}\overline{6}}\hspace{1.02pt}{\color{flavour3}\underline{3}}\hspace{1.02pt}{\color{flavour1}\overline{7}}){+}\mathcal{A}({\color{flavour1}\underline{1}}\hspace{1.02pt}4^{{+}}\hspace{1.02pt}{\color{flavour3}\overline{5}}\hspace{1.02pt}{\color{flavour2}\underline{2}}\hspace{1.02pt}{\color{flavour2}\overline{6}}\hspace{1.02pt}{\color{flavour3}\underline{3}}\hspace{1.02pt}{\color{flavour1}\overline{7}}){+}\mathcal{A}({\color{flavour1}\underline{1}}\hspace{1.02pt}4^{{+}}\hspace{1.02pt}{\color{flavour2}\overline{5}}\hspace{1.02pt}{\color{flavour2}\underline{2}}\hspace{1.02pt}{\color{flavour1}\overline{6}}\hspace{1.02pt}{\color{flavour3}\underline{3}}\hspace{1.02pt}{\color{flavour3}\overline{7}})\,.\\[-7pt]
\end{split}\label{more_compact_form_of_sf_to_mf_example}}
Notice that every term in this identity involves the same ordering of external states---namely those ordered by $\vec{\sigma}\!=\!\{1,4,5,2,6,3,7\}$; the only distinction between these terms is their flavour pairing. 

For the sake of making this information machine-readable, we encode the flavour pairs by the \built{Slot} positions of same-flavoured fermions and anti-fermions. Thus, 
\eq{\begin{split}&\mathcal{A}({\color{flavour1}\underline{1}}\hspace{1.02pt}4^{{+}}\hspace{1.02pt}{\color{flavour1}\overline{5}}\hspace{1.02pt}{\color{flavour2}\underline{2}}\hspace{1.02pt}{\color{flavour3}\overline{6}}\hspace{1.02pt}{\color{flavour3}\underline{3}}\hspace{1.02pt}{\color{flavour2}\overline{7}})\;\;\;\bigger{\Leftrightarrow}\;\;\;\text{flavour pairing: }\{\!\{1,3\},\!\{4,7\}\!,\{6,5\}\!\}\\
&\mathcal{A}({\color{flavour1}\underline{1}}\hspace{1.02pt}4^{{+}}\hspace{1.02pt}{\color{flavour1}\overline{5}}\hspace{1.02pt}{\color{flavour2}\underline{2}}\hspace{1.02pt}{\color{flavour2}\overline{6}}\hspace{1.02pt}{\color{flavour3}\underline{3}}\hspace{1.02pt}{\color{flavour3}\overline{7}})\;\;\;\bigger{\Leftrightarrow}\;\;\;\text{flavour pairing: }\{\!\{1,3\},\!\{4,5\}\!,\{6,7\}\!\}\\
&\mathcal{A}({\color{flavour1}\underline{1}}\hspace{1.02pt}4^{{+}}\hspace{1.02pt}{\color{flavour2}\overline{5}}\hspace{1.02pt}{\color{flavour2}\underline{2}}\hspace{1.02pt}{\color{flavour3}\overline{6}}\hspace{1.02pt}{\color{flavour3}\underline{3}}\hspace{1.02pt}{\color{flavour1}\overline{7}})\;\;\;\bigger{\Leftrightarrow}\;\;\;\text{flavour pairing: }\{\!\{1,7\},\!\{4,3\}\!,\{6,5\}\!\}\\
&\mathcal{A}({\color{flavour1}\underline{1}}\hspace{1.02pt}4^{{+}}\hspace{1.02pt}{\color{flavour3}\overline{5}}\hspace{1.02pt}{\color{flavour2}\underline{2}}\hspace{1.02pt}{\color{flavour2}\overline{6}}\hspace{1.02pt}{\color{flavour3}\underline{3}}\hspace{1.02pt}{\color{flavour1}\overline{7}})\;\;\;\bigger{\Leftrightarrow}\;\;\;\text{flavour pairing: }\{\!\{1,7\},\!\{4,5\}\!,\{6,3\}\!\}\\
&\mathcal{A}({\color{flavour1}\underline{1}}\hspace{1.02pt}4^{{+}}\hspace{1.02pt}{\color{flavour2}\overline{5}}\hspace{1.02pt}{\color{flavour2}\underline{2}}\hspace{1.02pt}{\color{flavour1}\overline{6}}\hspace{1.02pt}{\color{flavour3}\underline{3}}\hspace{1.02pt}{\color{flavour3}\overline{7}})\;\;\;\bigger{\Leftrightarrow}\;\;\;\text{flavour pairing: }\{\!\{1,5\},\!\{4,3\}\!,\{6,7\}\!\}\\[-6pt]\end{split}\label{flavour_pairing_convention}}
where the \emph{fermion} is always listed \built{First} and the anti-fermion \built{Last}. These pairs are conventionally ordered by flavour index where \fun{f}\brace{\var{$a$}} is assigned the $\var{a}$th index. (Of course, as discussed above, permutations of such indices leave partial amplitudes invariant.)\\[-6pt]

Notice that in the example (\ref{more_compact_form_of_sf_to_mf_example}) above, none of the distinctly-flavoured partial amplitudes are in the \emph{all-plus} basis of partial amplitudes described by Melia and defined in (\ref{all_plus_basis_of_partial_amps}): although all involve ${\color{flavour1}\psi_1}$ and $\bar{\psi}_{\text{-}1}$---denoted $\{{\color{flavour1}\underline{1}},\overline{7}\}$---in the first and last slots, either $\{{\color{flavour1}\underline{1}},\overline{7}\}$ do not share the same flavour, or some fermion lines are not oriented in the `all-plus' direction

To correct for this, we can use (\ref{kk_relation_among_partials}) to project into the (over-complete) set of partial amplitudes for which the flavour-partner of ${\color{flavour1}\psi_1}$ is put into the final \built{Slot}, and then recursively use  (\ref{eqn:kkSwap}) to orient all fermion lines into the all-plus direction. Doing this would result in the replacements,
\eq{\begin{split}\mathcal{A}({\color{flavour1}\underline{1}}\hspace{1.02pt}4^{{+}}\hspace{1.02pt}{\color{flavour3}\overline{5}}\hspace{1.02pt}{\color{flavour2}\underline{2}}\hspace{1.02pt}{\color{flavour2}\overline{6}}\hspace{1.02pt}{\color{flavour3}\underline{3}}\hspace{1.02pt}{\color{flavour1}\overline{7}})\mapsto&\phantom{{+}}\mathcal{A}({\color{flavour1}\underline{1}}\hspace{1.02pt}{\color{flavour3}\underline{3}}\hspace{1.02pt}{\color{flavour2}\underline{2}}\hspace{1.02pt}{\color{flavour2}\overline{6}}\hspace{1.02pt}4^{{+}}\hspace{1.02pt}{\color{flavour3}\overline{5}}\hspace{1.02pt}{\color{flavour1}\overline{7}}){+}\mathcal{A}({\color{flavour1}\underline{1}}\hspace{1.02pt}{\color{flavour3}\underline{3}}\hspace{1.02pt}{\color{flavour2}\underline{2}}\hspace{1.02pt}4^{{+}}\hspace{1.02pt}{\color{flavour2}\overline{6}}\hspace{1.02pt}{\color{flavour3}\overline{5}}\hspace{1.02pt}{\color{flavour1}\overline{7}})\\
&{+}\mathcal{A}({\color{flavour1}\underline{1}}\hspace{1.02pt}{\color{flavour3}\underline{3}}\hspace{1.02pt}4^{{+}}\hspace{1.02pt}{\color{flavour2}\underline{2}}\hspace{1.02pt}{\color{flavour2}\overline{6}}\hspace{1.02pt}{\color{flavour3}\overline{5}}\hspace{1.02pt}{\color{flavour1}\overline{7}}){+}\mathcal{A}({\color{flavour1}\underline{1}}\hspace{1.02pt}4^{{+}}\hspace{1.02pt}{\color{flavour3}\underline{3}}\hspace{1.02pt}{\color{flavour2}\underline{2}}\hspace{1.02pt}{\color{flavour2}\overline{6}}\hspace{1.02pt}{\color{flavour3}\overline{5}}\hspace{1.02pt}{\color{flavour1}\overline{7}})\\
\mathcal{A}({\color{flavour1}\underline{1}}\hspace{1.02pt}4^{{+}}\hspace{1.02pt}{\color{flavour1}\overline{5}}\hspace{1.02pt}{\color{flavour2}\underline{2}}\hspace{1.02pt}{\color{flavour3}\overline{6}}\hspace{1.02pt}{\color{flavour3}\underline{3}}\hspace{1.02pt}{\color{flavour2}\overline{7}})\mapsto&\phantom{{+}}\mathcal{A}({\color{flavour1}\underline{1}}\hspace{1.02pt}{\color{flavour2}\underline{2}}\hspace{1.02pt}{\color{flavour3}\underline{3}}\hspace{1.02pt}{\color{flavour3}\overline{6}}\hspace{1.02pt}{\color{flavour2}\overline{7}}\hspace{1.02pt}4^{{+}}\hspace{1.02pt}{\color{flavour1}\overline{5}}){+}\mathcal{A}({\color{flavour1}\underline{1}}\hspace{1.02pt}{\color{flavour2}\underline{2}}\hspace{1.02pt}{\color{flavour3}\underline{3}}\hspace{1.02pt}{\color{flavour3}\overline{6}}\hspace{1.02pt}4^{{+}}\hspace{1.02pt}{\color{flavour2}\overline{7}}\hspace{1.02pt}{\color{flavour1}\overline{5}}){+}\mathcal{A}({\color{flavour1}\underline{1}}\hspace{1.02pt}{\color{flavour2}\underline{2}}\hspace{1.02pt}{\color{flavour3}\underline{3}}\hspace{1.02pt}4^{{+}}\hspace{1.02pt}{\color{flavour3}\overline{6}}\hspace{1.02pt}{\color{flavour2}\overline{7}}\hspace{1.02pt}{\color{flavour1}\overline{5}})\\
&{+}\mathcal{A}({\color{flavour1}\underline{1}}\hspace{1.02pt}{\color{flavour2}\underline{2}}\hspace{1.02pt}4^{{+}}\hspace{1.02pt}{\color{flavour3}\underline{3}}\hspace{1.02pt}{\color{flavour3}\overline{6}}\hspace{1.02pt}{\color{flavour2}\overline{7}}\hspace{1.02pt}{\color{flavour1}\overline{5}}){+}\mathcal{A}({\color{flavour1}\underline{1}}\hspace{1.02pt}4^{{+}}\hspace{1.02pt}{\color{flavour2}\underline{2}}\hspace{1.02pt}{\color{flavour3}\underline{3}}\hspace{1.02pt}{\color{flavour3}\overline{6}}\hspace{1.02pt}{\color{flavour2}\overline{7}}\hspace{1.02pt}{\color{flavour1}\overline{5}})\\[-7pt]
\end{split}\label{to_all_plus_basis_examples}} 
and so-on. It is easy to see that all terms on the right hand sides above involve ${\color{flavour1}\psi_1}$ flavour-paired with whatever ${\color{flavour1}\bar{\psi}_{a}}$ is in the \built{Last} slot, and that all fermion lines are positively oriented (with $\psi$'s appearing before their flavour-paired $\bar{\psi}$'s). Even though the anti-fermion of the first flavour is different in the two examples of (\ref{to_all_plus_basis_examples}), this is fine: these terms represent different flavour-pairings of the \emph{single-flavour} partial amplitude `${A}({\color{flavour0}\underline{1}}\hspace{1.02pt}4^+\hspace{1.02pt}{\color{flavour0}\overline{5}}\hspace{1.02pt}{\color{flavour0}\underline{2}}\hspace{1.02pt}{\color{flavour0}\overline{6}}\hspace{1.02pt}{\color{flavour0}\underline{3}}\hspace{1.02pt}{\color{flavour0}\overline{7}})$'; in general, single-flavoured partial amplitudes will be represented by multi-flavoured partial amplitudes involving various, distinct flavour-pairings; for each of the $n_f!$ choices of flavour-pairing, we have a basis of multi-flavoured partial amplitudes as given by (\ref{all_plus_basis_of_partial_amps}).\\[-6pt]

\subsubsection{Melia's Flavour-Reduction Algorithm}\label{section_flavour_reduction_algorithm}
While it is clear from the examples above that any single-flavour partial amplitude can be viewed (without subtlety) as a direct sum over the \emph{distinctly}-flavoured partial amplitudes corresponding to its possible chord diagrams (see, \emph{e.g.}~(\ref{more_compact_form_of_sf_to_mf_example})), it turns out to be possible to represent all \emph{distinctly-flavoured} partial amplitudes as linear combinations of those with only a \emph{single} flavour. This surprising fact was proven by Melia in \cite{Melia:2013epa}, where he described a constructive `flavour-reduction' algorithm. Let us briefly review the structure of this algorithm here---pointing the reader to ref.~\cite{Melia:2013epa} for a more detailed discussion.

We start from the fact that any single-flavoured partial amplitude is equal to a sum over distinctly-flavoured partial amplitudes associated with its possible chord diagrams. Inverting this statement leads to an identity of the form
\eq{\underbrace{\mathcal{A}({\color{flavour1}\underline{1}},\vec{\sigma},{\color{flavour1}\overline{n}})}_{\substack{\hspace{-30pt}\text{multi-flavour partial amp}\hspace{-0pt}\\\hspace{-30pt}\text{with \emph{some} flavour-pairing}\hspace{-0pt}}}=\underbrace{{A}({\color{flavour0}\bar{1}},\vec{\sigma},{\color{flavour0}\overline{n}})}_{\substack{\text{single-flavour}\\\text{partial amp}}}{-}\sum_{\substack{\text{multi-flavour partial amps}\hspace{-70pt}\\\text{with \emph{other} flavour-pairings}\hspace{-70pt}}}\mathcal{A}({\color{flavour1}\underline{1}},\vec{\sigma},\bar{n})\,.\label{flavour_reduction_identity}}
Melia showed that once distinct-flavoured terms on the RHS above are expanded into \emph{an} all-plus basis of partial amplitudes\footnote{These partial amplitudes generally involve distinct flavour pairings, and therefore do not exist within any \emph{particular} all-plus basis.}, all the terms that appear will be of strictly `higher-maturity'\footnote{The maturity of a chord diagram is related to the branching of its Poincar\'{e} dual.} than the LHS; and thus, this identity can be recursively applied until only those terms of the \emph{highest} maturity survive. We need not concern ourselves with a precise meaning of maturity here, merely note that chord diagrams of highest maturity are those for which single-flavour partial amplitudes admit unique flavour pairings, and thus are equal to multi-flavoured partial amplitudes. 

An example of a partial amplitude of highest maturity was given in (\ref{highest_maturity_example}) above, for which we would find this algorithm to terminate with
\eq{\mathcal{A}({\color{flavour1}\underline{1}}\hspace{1.02pt}4^{{+}}\hspace{1.02pt}{\color{flavour1}\overline{5}}\hspace{1.02pt}{\color{flavour3}\overline{6}}\hspace{1.02pt}{\color{flavour2}\overline{7}}\hspace{1.02pt}{\color{flavour2}\underline{2}}\hspace{1.02pt}{\color{flavour3}\underline{3}})={A}({\color{flavour0}\underline{1}}\hspace{1.02pt}4^{{+}}\hspace{1.02pt}{\color{flavour0}\overline{5}}\hspace{1.02pt}{\color{flavour0}\overline{6}}\hspace{1.02pt}{\color{flavour0}\overline{7}}\hspace{1.02pt}{\color{flavour0}\underline{2}}\hspace{1.02pt}{\color{flavour0}\underline{3}})\,.}

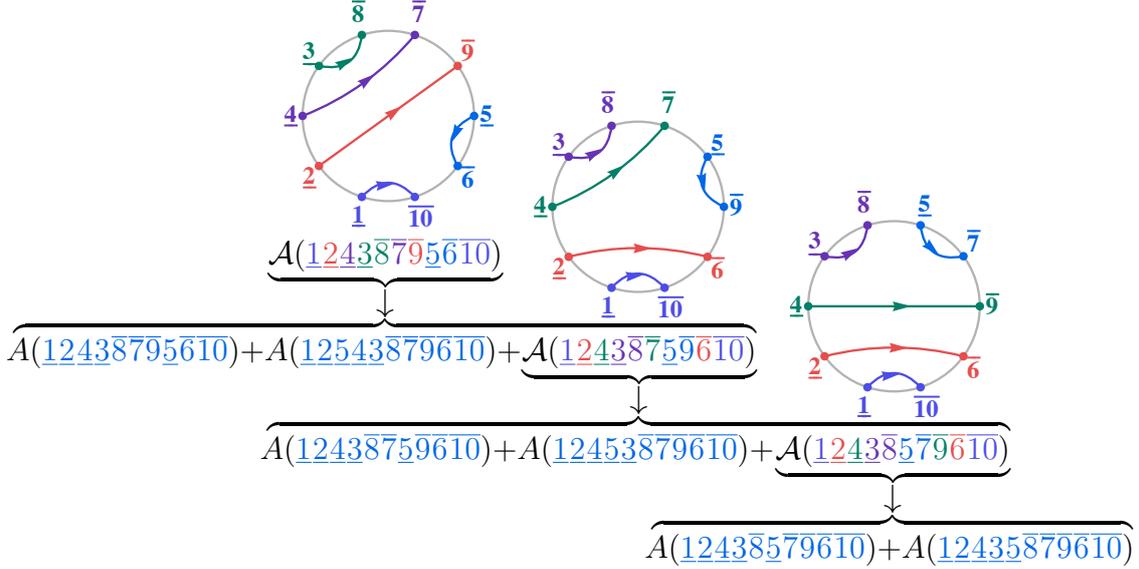
\begin{figure}[t]\caption{Illustrating the reduction of multi-flavour to single-flavour partial amplitudes.}\vspace{-14pt}\label{flavour_reduction_example_figure}
$$\hspace{-50pt}\begin{tikzpicture}[baseline=-1.5pt]\node[anchor=-90](a1)at(0,0){$\fwboxL{90pt}{\underbrace{\rule[-2pt]{0pt}{15pt}\smash{\mathcal{A}({\color{flavour1}\underline{1}}\hspace{0.5pt}{\color{flavour2}\underline{2}}\hspace{0.5pt}{\color{flavour4}\underline{4}}\hspace{0.5pt}{\color{flavour3}\underline{3}}\hspace{0.5pt}{\color{flavour3}\overline{8}}\hspace{0.5pt}{\color{flavour4}\overline{7}}\hspace{0.5pt}{\color{flavour2}\overline{9}}\hspace{0.5pt}{\color{flavour5}\underline{5}}\hspace{0.5pt}{\color{flavour5}\overline{6}}\hspace{0.5pt}{\color{flavour1}\overline{10}})}}}$};
\node[rotate=-90](m1)at($(a1.270)-(2pt,1.5pt)$){$\to$};
\node[]at($(a1.90)+(0,38.5pt)$){\includegraphics[scale=0.8]{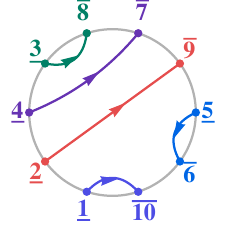}};
\node[anchor=-90](a2)at($(a1)-(1.5pt,52pt)$) {$\fwboxL{0pt}{\,\,\raisebox{9.pt}{$\overbrace{\hspace{278pt}}$}}\fwboxL{193pt}{{A}({\color{flavour0}\underline{1}}\hspace{0.5pt}{\color{flavour0}\underline{2}}\hspace{0.95pt}{\color{flavour0}\underline{4}}\hspace{0.5pt}{\color{flavour0}\underline{3}}\hspace{0.5pt}{\color{flavour0}\overline{8}}\hspace{0.5pt}{\color{flavour0}\overline{7}}\hspace{0.5pt}{\color{flavour0}\overline{9}}\hspace{0.5pt}{\color{flavour0}\underline{5}}\hspace{0.5pt}{\color{flavour0}\overline{6}}\hspace{0.5pt}{\color{flavour0}\overline{10}}){+}{A}({\color{flavour0}\underline{1}}\hspace{0.5pt}{\color{flavour0}\underline{2}}\hspace{0.5pt}{\color{flavour0}\underline{5}}\hspace{0.5pt}{\color{flavour0}\underline{4}}\hspace{0.5pt}{\color{flavour0}\underline{3}}\hspace{0.5pt}{\color{flavour0}\overline{8}}\hspace{0.5pt}{\color{flavour0}\overline{7}}\hspace{0.5pt}{\color{flavour0}\overline{9}}\hspace{0.5pt}{\color{flavour0}\overline{6}}\hspace{0.5pt}{\color{flavour0}\overline{10}}){+}}\fwboxL{90pt}{\underbrace{\rule[-2pt]{0pt}{15pt}\smash{\mathcal{A}({\color{flavour1}\underline{1}}\hspace{0.5pt}{\color{flavour2}\underline{2}}\hspace{0.5pt}{\color{flavour3}\underline{4}}\hspace{0.5pt}{\color{flavour4}\underline{3}}\hspace{0.5pt}{\color{flavour4}\overline{8}}\hspace{0.5pt}{\color{flavour3}\overline{7}}\hspace{0.5pt}{\color{flavour5}\underline{5}}\hspace{0.5pt}{\color{flavour5}\overline{9}}\hspace{0.5pt}{\color{flavour2}\overline{6}}\hspace{0.5pt}{\color{flavour1}\overline{10}})}}}$};
\node[rotate=-90](m1)at($(a2.270)+(94.5pt,-1.5pt)$){$\to$};
\node[anchor=-90](a3)at($(a2)-(-95.pt,54pt)$) {$\fwboxL{0pt}{\,\,\raisebox{9.pt}{$\overbrace{\hspace{278pt}}$}}\fwboxL{193pt}{{A}({\color{flavour0}\underline{1}}\hspace{0.5pt}{\color{flavour0}\underline{2}}\hspace{0.5pt}{\color{flavour0}\underline{4}}\hspace{0.5pt}{\color{flavour0}\underline{3}}\hspace{0.5pt}{\color{flavour0}\overline{8}}\hspace{0.5pt}{\color{flavour0}\overline{7}}\hspace{0.5pt}{\color{flavour0}\underline{5}}\hspace{0.5pt}{\color{flavour0}\overline{9}}\hspace{0.5pt}{\color{flavour0}\overline{6}}\hspace{0.5pt}{\color{flavour0}\overline{10}}){+}{A}({\color{flavour0}\underline{1}}\hspace{0.5pt}{\color{flavour0}\underline{2}}\hspace{0.5pt}{\color{flavour0}\underline{4}}\hspace{0.5pt}{\color{flavour0}\underline{5}}\hspace{0.5pt}{\color{flavour0}\underline{3}}\hspace{0.5pt}{\color{flavour0}\overline{8}}\hspace{0.5pt}{\color{flavour0}\overline{7}}\hspace{0.5pt}{\color{flavour0}\overline{9}}\hspace{0.5pt}{\color{flavour0}\overline{6}}\hspace{0.5pt}{\color{flavour0}\overline{10}}){+}}\fwboxL{90pt}{\underbrace{\rule[-2pt]{0pt}{15pt}\smash{\mathcal{A}({\color{flavour1}\underline{1}}\hspace{0.5pt}{\color{flavour2}\underline{2}}\hspace{0.5pt}{\color{flavour3}\underline{4}}\hspace{0.5pt}{\color{flavour4}\underline{3}}\hspace{0.5pt}{\color{flavour4}\overline{8}}\hspace{0.5pt}{\color{flavour5}\underline{5}}\hspace{0.5pt}{\color{flavour5}\overline{7}}\hspace{0.5pt}{\color{flavour3}\overline{9}}\hspace{0.5pt}{\color{flavour2}\overline{6}}\hspace{0.5pt}{\color{flavour1}\overline{10}})}}}$};
\node[]at($(a3.90)+(0,74.5pt)$){\includegraphics[scale=0.8]{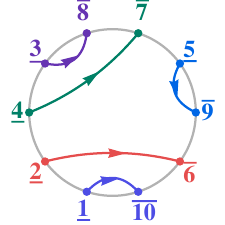}};
\node[rotate=-90](m3)at($(a3.270)+(94.5pt,-1.5pt)$){$\to$};
\node[anchor=-90](a4)at($(a3)-(-47.5pt,48pt)$) {$\fwboxL{0pt}{\,\,\raisebox{9.pt}{$\overbrace{\hspace{180pt}}$}}\fwboxL{90pt}{{A}({\color{flavour0}\underline{1}}\hspace{0.5pt}{\color{flavour0}\underline{2}}\hspace{0.5pt}{\color{flavour0}\underline{4}}\hspace{0.5pt}{\color{flavour0}\underline{3}}\hspace{0.5pt}{\color{flavour0}\overline{8}}\hspace{0.5pt}{\color{flavour0}\underline{5}}\hspace{0.5pt}{\color{flavour0}\overline{7}}\hspace{0.5pt}{\color{flavour0}\overline{9}}\hspace{0.5pt}{\color{flavour0}\overline{6}}\hspace{0.5pt}{\color{flavour0}\overline{10}}){+}{A}({\color{flavour0}\underline{1}}\hspace{0.5pt}{\color{flavour0}\underline{2}}\hspace{0.5pt}{\color{flavour0}\underline{4}}\hspace{0.5pt}{\color{flavour0}\underline{3}}\hspace{0.5pt}{\color{flavour0}\underline{5}}\hspace{0.5pt}{\color{flavour0}\overline{8}}\hspace{0.5pt}{\color{flavour0}\overline{7}}\hspace{0.5pt}{\color{flavour0}\overline{9}}\hspace{0.5pt}{\color{flavour0}\overline{6}}\hspace{0.5pt}{\color{flavour0}\overline{10}})}$};
\node[]at($(m3.90)+(-5pt,72.5pt)$){\includegraphics[scale=0.8]{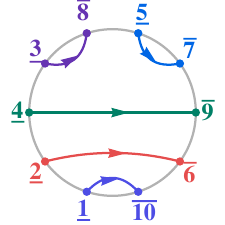}};
\end{tikzpicture}\vspace{-15pt}$$
\vspace{-04pt}\end{figure}

A less trivial example of how this algorithm proceeds would be for the partial amplitude $\mathcal{A}({\color{flavour1}\underline{1}}\hspace{1.02pt}{\color{flavour2}\underline{2}}\hspace{1.02pt}{\color{flavour4}\underline{4}}\hspace{1.02pt}{\color{flavour3}\underline{3}}\hspace{1.02pt}{\color{flavour3}\overline{8}}\hspace{1.02pt}{\color{flavour4}\overline{7}}\hspace{1.02pt}{\color{flavour2}\overline{9}}\hspace{1.02pt}{\color{flavour5}\underline{5}}\hspace{1.02pt}{\color{flavour5}\overline{6}}\hspace{1.02pt}{\color{flavour1}\overline{10}})$. The iterated use of the reduction identity (\ref{flavour_reduction_identity}) is shown in \mbox{Figure~\ref{flavour_reduction_example_figure}}, where we have written all multi-flavour partial amplitudes of highest maturity directly as single-flavour partial amplitudes.\\[-6pt]

\newpage
The result of the sequence of reductions illustrated in \mbox{Figure~\ref{flavour_reduction_example_figure}} is the identity:
\eq{\begin{split}\hspace{-75pt}\mathcal{A}({\color{flavour1}\underline{1}}\hspace{1.5pt}{\color{flavour2}\underline{2}}\hspace{1.02pt}{\color{flavour4}\underline{4}}\hspace{1.02pt}{\color{flavour3}\underline{3}}\hspace{1.02pt}{\color{flavour3}\overline{8}}\hspace{1.02pt}{\color{flavour4}\overline{7}}\hspace{1.02pt}{\color{flavour2}\overline{9}}\hspace{1.02pt}{\color{flavour5}\underline{5}}\hspace{1.02pt}{\color{flavour5}\overline{6}}\hspace{1.02pt}{\color{flavour1}\overline{10}})=&\phantom{{+}}
{A}({\color{flavour0}\underline{1}}\hspace{0.5pt}{\color{flavour0}\underline{2}}\hspace{0.5pt}{\color{flavour0}\underline{4}}\hspace{0.5pt}{\color{flavour0}\underline{3}}\hspace{0.5pt}{\color{flavour0}\overline{8}}\hspace{0.5pt}{\color{flavour0}\overline{7}}\hspace{0.5pt}{\color{flavour0}\overline{9}}\hspace{0.5pt}{\color{flavour0}\underline{5}}\hspace{0.5pt}{\color{flavour0}\overline{6}}\hspace{0.5pt}{\color{flavour0}\overline{10}}){+}{A}({\color{flavour0}\underline{1}}\hspace{0.5pt}{\color{flavour0}\underline{2}}\hspace{0.5pt}{\color{flavour0}\underline{5}}\hspace{0.5pt}{\color{flavour0}\underline{4}}\hspace{0.5pt}{\color{flavour0}\underline{3}}\hspace{0.5pt}{\color{flavour0}\overline{8}}\hspace{0.5pt}{\color{flavour0}\overline{7}}\hspace{0.5pt}{\color{flavour0}\overline{9}}\hspace{0.5pt}{\color{flavour0}\overline{6}}\hspace{0.5pt}{\color{flavour0}\overline{10}}){+}{A}({\color{flavour0}\underline{1}}\hspace{0.5pt}{\color{flavour0}\underline{2}}\hspace{0.5pt}{\color{flavour0}\underline{4}}\hspace{0.5pt}{\color{flavour0}\underline{3}}\hspace{0.5pt}{\color{flavour0}\overline{8}}\hspace{0.5pt}{\color{flavour0}\overline{7}}\hspace{0.5pt}{\color{flavour0}\underline{5}}\hspace{0.5pt}{\color{flavour0}\overline{9}}\hspace{0.5pt}{\color{flavour0}\overline{6}}\hspace{0.5pt}{\color{flavour0}\overline{10}})\hspace{-50pt}\\
&{+}{A}({\color{flavour0}\underline{1}}\hspace{0.5pt}{\color{flavour0}\underline{2}}\hspace{0.5pt}{\color{flavour0}\underline{4}}\hspace{0.5pt}{\color{flavour0}\underline{5}}\hspace{0.5pt}{\color{flavour0}\underline{3}}\hspace{0.5pt}{\color{flavour0}\overline{8}}\hspace{0.5pt}{\color{flavour0}\overline{7}}\hspace{0.5pt}{\color{flavour0}\overline{9}}\hspace{0.5pt}{\color{flavour0}\overline{6}}\hspace{0.5pt}{\color{flavour0}\overline{10}}){+}{A}({\color{flavour0}\underline{1}}\hspace{0.5pt}{\color{flavour0}\underline{2}}\hspace{0.5pt}{\color{flavour0}\underline{4}}\hspace{0.5pt}{\color{flavour0}\underline{3}}\hspace{0.5pt}{\color{flavour0}\overline{8}}\hspace{0.5pt}{\color{flavour0}\underline{5}}\hspace{0.5pt}{\color{flavour0}\overline{7}}\hspace{0.5pt}{\color{flavour0}\overline{9}}\hspace{0.5pt}{\color{flavour0}\overline{6}}\hspace{0.5pt}{\color{flavour0}\overline{10}}){+}{A}({\color{flavour0}\underline{1}}\hspace{0.5pt}{\color{flavour0}\underline{2}}\hspace{0.5pt}{\color{flavour0}\underline{4}}\hspace{0.5pt}{\color{flavour0}\underline{3}}\hspace{0.5pt}{\color{flavour0}\underline{5}}\hspace{0.5pt}{\color{flavour0}\overline{8}}\hspace{0.5pt}{\color{flavour0}\overline{7}}\hspace{0.5pt}{\color{flavour0}\overline{9}}\hspace{0.5pt}{\color{flavour0}\overline{6}}\hspace{0.5pt}{\color{flavour0}\overline{10}}).\!\!\hspace{-50pt}\end{split}\label{example_flavour_reduction_sequence_identity}}

As emphasized in the introduction, any single-flavour partial amplitude is \emph{automatically} equal to a component amplitude of $\mathcal{N}{=}1$ supersymmetric Yang-Mills theory (which are in turn equal to component amplitudes of the maximally ($\mathcal{N}{=}4$) supersymmetric Yang-Mills theory at tree-level). Thus, flavour-reduction allows all multi-flavoured partial amplitudes of pure gauge theory coupled to charged fermions to be efficiently encoded by, extracted from, and evaluated in terms of the partial amplitudes of sYM. Because these amplitudes are widely known and publicly available through computer packages such as \cite{Bourjaily:2023uln} (see also~\cite{Dixon:2010ik,Bourjaily:2010wh}), this allows for all the partial amplitudes required for pure gauge theory coupled to charged fermions (whether of the same or distinct flavours!) to be effectively evaluated.\\

As described in the section~\ref{section:fermionic_amplitudes_package}, the package \fpackage~implements all the ingredients required to express partial amplitudes in pure gauge theory as those of (maximally) supersymmetric Yang-Mills theory---and, if the user also has access to the package \package\ \cite{Bourjaily:2023uln}, then to convert these partial amplitudes into analytic expressions involving spinors.

\newpage
\subsection{Colour-Dependence of Charged Matter Amplitudes in Gauge Theory}\label{colour_tensors}

Melia's seminal work on the partial amplitudes of pure gauge theory coupled to charged matter left open the question of their colour dependence. While for some representations of charged matter and some gauge groups, colour-decompositions have been known for some time (see \emph{e.g.}~\cite{Cvitanovic:1976am,Berends:1987cv,Mangano:1987xk,Kosower:1987ic,Kosower:1988kh,Bern:1990ux,Mangano:1990by,Maltoni:2002mq}), the precise colour tensors required for the decomposition of amplitudes into Melia's all-plus, minimal basis of linearly independent partial amplitudes (\ref{all_plus_basis_of_partial_amps}) were first determined via unitarity by Johansson and Ochirov in \cite{Johansson:2015oia}. These colour tensors were subsequently proven correct by Melia in \cite{Melia:2015ika}; and this work was further clarified and extended beyond tree-level in \cite{Johansson:2014zca,Kalin:2017oqr,Ochirov:2019mtf}. We review the structure of these tensors here.

\subsubsection{A Brief Description of Johansson-Ochirov's Colour Tensors}\label{jo_tensors_review}

The precise colour tensors appearing in the decomposition (\ref{colour_decomposition_of_fermionic_amplitudes}) of amplitudes for Melia's all-plus basis of partial amplitudes are unusual in several respects. Some of this is best understood by comparison against more familiar cases of colour-decomposition. 

Consider first the case of amplitudes involving only gluons. One colour decomposition is the familiar trace expansion (see \emph{e.g.}~\cite{Dixon:1996wi}), which (at tree-level) reads
\eq{\begin{split}\mathcal{A}(\{g_1,\ldots,g_n\})=&\frac{1}{T(\mathbf{\b{R}})}\sum_{\sigma\in\mathfrak{S}([2,n])\hspace{-16pt}}\mathrm{tr}(\mathbf{T}_{\!\mathbf{\b{R}}}^{1}.\mathbf{T}_{\!\mathbf{\b{R}}}^{\sigma_1}.\mathbf{T}_{\!\mathbf{\b{R}}}^{\sigma_2}.\cdots.\mathbf{T}_{\!\mathbf{\b{R}}}^{\sigma_{\text{-}1}}){A}(1\,\vec{\sigma})\\
\equivL&\frac{1}{T(\mathbf{\b{R}})}\sum_{\sigma\in\mathfrak{S}([2,n])\hspace{-16pt}}\mathrm{tr}_{\mathbf{\b{R}}}(1\,\vec{\sigma})\,{A}(1\,\vec{\sigma})\\[-8pt]\end{split}\label{trace_expansion_for_glue}}
for which the set $\{(\mathbf{T}_{\!\mathbf{\b{R}}}^{\r{a}})\}_{\r{a}\in[\mathrm{dim}(\mathfrak{\r{g}})]}$ are the $\mathrm{dim}(\mathbf{\b{R}})\!\times\!\mathrm{dim}(\mathbf{\b{R}})$ matrices which encode the \emph{generators} of \emph{any} representation `$\mathbf{\b{R}}$' of a Lie algebra $\mathfrak{g}$ corresponding to the gauge group, and $T(\mathbf{\b{R}})$ is the representation's \emph{Dynkin index} defined by its connection to the Killing metric: $\mathrm{tr}_{\mathbf{\b{R}}}(1\,2)\equivL T(\mathbf{\b{R}})g^{\r{[\mathfrak{\r{g}}]\,[\mathfrak{g}]}}_{\mathbf{\r{ad}}}$. (See \emph{e.g.}~\cite{Bourjaily:2025hvq,Bourjaily:2024jbt} for more details.) 

The trace expansion follows from the representation of the structure constants of a Lie algebra $\r{f}\indices{\r{a\,b\,c}}{}$ in terms of traces over the generators of \emph{any} representation---namely, 
\eq{\r{f}\indices{\r{a\,b\,c}}{}\equivR\sum_{\r{d}\in\r{[\mathfrak{g}]}} \r{f}\indices{\r{a\,b}}{\r{d}}\,g_{\mathbf{\r{ad}}}^{\r{d\,c}}=\frac{1}{T(\mathbf{\b{R}})}\big(\mathrm{tr}_{\mathbf{\b{R}}}(\r{a\,b\,c}){-}\mathrm{tr}_{\mathbf{\b{R}}}(\r{b\,a\,c})\big)}
and using this to decompose all colour tensors appearing in the tree-level Feynman rules into (single-)traces over the generators of the representation $\mathbf{\b{R}}$. \emph{This representation need not have anything to do with the charge generators of matter}: indeed, the expression (\ref{trace_expansion_for_glue}) is for an amplitude involving only particles charged under $\mathbf{\r{ad}}(\mathbf{\b{R}})$---which is generated by the $\mathrm{dim}(\mathfrak{g})\!\times\!\mathrm{dim}(\mathfrak{g})$ matrices $(\mathbf{T}_{\mathbf{\r{ad}}}^{\r{b}})\indices{\r{a}}{\r{c}}\equivL\r{f\indices{ab}{c}}$.\\[-6pt]

Of course, the partial amplitudes appearing in (\ref{trace_expansion_for_glue}) are not linearly independent: they satisfy the KK relations (\ref{kk_relation_among_partials}). Re-writing this in terms of a KK basis of partial amplitudes results in the expansion found by Del Duca, Dixon, and Maltoni in \cite{DelDuca:1999rs}:
\eq{\mathcal{A}(\{g_1,\ldots,g_n\})=\sum_{\vec{\sigma}\in\mathfrak{S}([n]\backslash\{\r{\alpha},\b{\beta}\})\hspace{-30pt}}\r{f}^{\r{\alpha}\,\vec{\sigma}\,\b{\beta}}A(\r{\alpha}\,\vec{\sigma}\,\b{\beta})\label{ddm_expansion}}
with colour tensors defined via
\eq{\r{f}^{\r{\alpha}\,\vec{\sigma}\,\b{\beta}}\equivR\sum_{\g{d_i}\in\g{[\mathfrak{g}]}}\big(\r{f}\indices{\r{\alpha}\,\sigma_1}{\g{d_1}}\r{f}\indices{\g{d_1}\,\sigma_2}{\g{d_2}}\cdots\r{f}\indices{\g{d_{\text{-}2}}\,\sigma_{\text{-}1}}{\g{d_{\text{-}1}}}\big)g_{\mathbf{\r{ad}}}^{\g{d_{\text{-}1}}\,\b{\beta}}\,.\label{ddm_tensors}}
This set of colour tensors can be seen as linearly independent under the Jacobi relations among tree-graphs of structure constants (and spanning the set of colour-tensors relevant to tree-level Feynman diagrams). (It is worth remarking that the form (\ref{ddm_expansion}) follows immediately from on-shell recursion (see \emph{e.g.}~\cite{Bourjaily:2023ycy}).)\\[-6pt]

Extending this to amplitudes involving a single fermion line charged under the representation $\mathbf{\b{R}}$ of some Lie algebra $\mathfrak{g}$, results in the colour-decomposition \cite{Mangano:1987xk}
\eq{\mathcal{A}(\{{\color{flavour1}\psi_1},g_2,\ldots,g_{n\,\text{-}1},{\color{flavour1}\bar{\psi}_{n}}\})=\sum_{\vec{\sigma}\in\mathfrak{S}([2,n\,\text{-}1])\hspace{-30pt}}
{\color{flavour1}\{\underline{1}|}\vec{\sigma}{\color{flavour1}|\bar{n}\}}
A({\color{flavour1}\underline{1}}\,\vec{\sigma}\,{\color{flavour1}\overline{n}})\label{one_fermion_amp}}
where the colour tensors involved are defined as
\eq{{\color{flavour1}\{\underline{1}|}\vec{\sigma}{\color{flavour1}|\bar{n}\}}\equivR\sum_{\b{d_i}\in\b{[r]}}\big(\mathbf{T}\indices{\b{1}\,\sigma_1}{\b{d_1}}\mathbf{T}\indices{\b{d_1}\,\sigma_2}{\b{d_2}}\cdots\mathbf{T}\indices{\b{d_{\text{-}1}}\,\sigma_{\text{-}1}}{\b{\overline{n}}}\big)\,\label{one_flavour_colour_tensors}}
and the $\mathrm{dim}(\mathbf{\b{R}})\!\times\!\mathrm{dim}(\mathbf{\b{R}})$ matrices $\mathbf{T}$ are the generators of the \emph{representation} $\mathbf{\b{R}}$ encoding the `charge' of the fermion. (It is worth noting that, unlike the DDM expansion (\ref{ddm_expansion}), the colour decomposition (\ref{one_fermion_amp}) does not follow directly from on-shell recursion: shifting the momenta of any fermion/anti-fermion pair of the same flavour always involves contributions from poles at infinity \cite{ArkaniHamed:2008yf,Cohen:2010mi}.)\\[-6pt]

Like the trace (\ref{trace_expansion_for_glue}) and DDM (\ref{ddm_expansion}) expansions, the one-fermion tensors (\ref{one_flavour_colour_tensors}) are all essentially identical: they differ only in their `slot-sequences' which encode the ordering of external gauge bosons. This is not the case for the colour tensors that appear in the expansion of multi-flavoured amplitudes (\ref{colour_decomposition_of_fermionic_amplitudes}) as found by Johansson and Ochirov in \cite{Johansson:2015oia}, which are more involved. 

The colour tensors that encode the colour-dependence of the partial amplitudes appearing in Melia's all-plus basis (\ref{all_plus_basis_of_partial_amps}),
\eq{\mathcal{A}(\{{\color{flavour1}\underline{1}},{\color{flavour2}\underline{2}},\ldots,{\color{flavour1}\overline{n}}\})=\sum_{\vec{\sigma}}\mathcal{C}[{\color{flavour1}\underline{1}}\,\vec{\sigma}\,\,{\color{flavour1}\overline{n}}]\,\mathcal{A}({\color{flavour1}\underline{1}}\,\vec{\sigma}\,\,{\color{flavour1}\overline{n}})\,,}
are somewhat difficult to describe. Despite their generality (being defined for fermions transforming under \emph{any} charge representation of \emph{any} Lie algebra), they are much less familiar to physicists than the tensors appearing in (\ref{trace_expansion_for_glue}), (\ref{ddm_tensors}), or (\ref{one_flavour_colour_tensors}) above. We briefly review these tensors here, but refer the reader to ref.~\cite{Johansson:2015oia} for a more thorough discussion (see also~\cite{Melia:2015ika,Johansson:2014zca,Kalin:2017oqr,Ochirov:2019mtf}).

Roughly speaking, the colour tensors are constructed as sums of products of the generators of the fermions' charge, where some sets of adjoint indices have been contracted via the Killing metric. For $n_f$ fermion lines, these involve $n_f$ `curly bracket' tensors of the form (\ref{one_flavour_colour_tensors}) with some indices contracted between brackets. 

The general rule for their construction is perhaps easiest to understand by example. Consider the colour tensor coefficient $\mathcal{C}[{\color{flavour1}\underline{1}}\hspace{1.02pt}{\color{flavour3}\underline{3}}\hspace{1.02pt}{\color{flavour4}\underline{4}}\hspace{1.02pt}6\hspace{1.02pt}{\color{flavour4}\overline{7}}\hspace{1.02pt}{\color{flavour3}\overline{8}}\hspace{1.02pt}5\hspace{1.02pt}{\color{flavour2}\underline{2}}\hspace{1.02pt}{\color{flavour2}\overline{9}}\hspace{1.02pt}{\color{flavour1}\overline{10}}]$ of the partial amplitude $\mathcal{A}({\color{flavour1}\underline{1}}\hspace{1.02pt}{\color{flavour3}\underline{3}}\hspace{1.02pt}{\color{flavour4}\underline{4}}\hspace{1.02pt}6\hspace{1.02pt}{\color{flavour4}\overline{7}}\hspace{1.02pt}{\color{flavour3}\overline{8}}\hspace{1.02pt}5\hspace{1.02pt}{\color{flavour2}\underline{2}}\hspace{1.02pt}{\color{flavour2}\overline{9}}\hspace{1.02pt}{\color{flavour1}\overline{10}})$. To construct this tensor, we start from the chord diagram
\vspace{-2pt}\eq{\mathcal{A}({\color{flavour1}\underline{1}}\hspace{1.02pt}{\color{flavour3}\underline{3}}\hspace{1.02pt}{\color{flavour4}\underline{4}}\hspace{1.02pt}6\hspace{1.02pt}{\color{flavour4}\overline{7}}\hspace{1.02pt}{\color{flavour3}\overline{8}}\hspace{1.02pt}5\hspace{1.02pt}{\color{flavour2}\underline{2}}\hspace{1.02pt}{\color{flavour2}\overline{9}}\hspace{1.02pt}{\color{flavour1}\overline{10}})\;\;\;\;\bigger{\Leftrightarrow}\;\;\;\;\begin{array}{@{}c@{}}\includegraphics[scale=0.7]{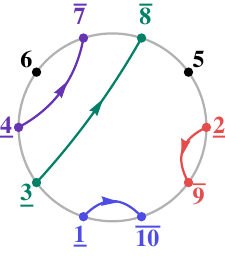}\end{array}\vspace{-5pt}}
and draw lines adjoint-lines from each fermion line and each external gluon to the base-line $({\color{flavour1}\underline{1},\overline{10}})$, crossing any other fermion lines along the way (but never crossing any of the newly added lines):
\vspace{-2pt}\eq{\begin{array}{@{}c@{}}\includegraphics[scale=0.7]{chords_for_colour_tensor_example}\end{array}\;\;\;\;\bigger{\Rightarrow}\;\;\;\;\begin{array}{@{}c@{}}\includegraphics[scale=0.7]{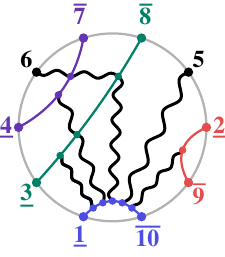}\end{array}\,.\vspace{-5pt}}
The colour tensor we need would be encoded by a graph with exactly the same topology as this decorated chord diagram:
\vspace{16pt}\eq{\hspace{40pt}\fwboxR{0pt}{\mathcal{C}[{\color{flavour1}\underline{1}}\hspace{1.02pt}{\color{flavour3}\underline{3}}\hspace{1.02pt}{\color{flavour4}\underline{4}}\hspace{1.02pt}6\hspace{1.02pt}{\color{flavour4}\overline{7}}\hspace{1.02pt}{\color{flavour3}\overline{8}}\hspace{1.02pt}5\hspace{1.02pt}{\color{flavour2}\underline{2}}\hspace{1.02pt}{\color{flavour2}\overline{9}}\hspace{1.02pt}{\color{flavour1}\overline{10}}]}\bigger{\Leftrightarrow\;}\begin{array}{@{}c@{}}\\[-39pt]\includegraphics[scale=1]{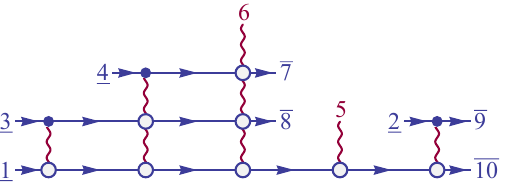}\end{array}\,.\label{fig:egcolourdiagram}}
(Such a graphical representation of the colour tensor can be obtained using the package \fpackage\ via the function \funL{drawColourTensors}.)

\begin{figure}[t!]\caption{Defining the tensor $\Xi_{\ell{=}4}^{\t{a}}$ (\ref{defn_of_xi}) as it would appear in diagrams such as (\ref{fig:egcolourdiagram}).}\label{fig:Xipic}\vspace{-15pt}$$\begin{array}{@{}c@{}}\includegraphics[scale=0.9]{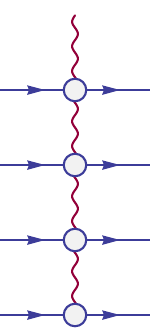}\\[-5.45pt]\end{array}\;=\;\begin{array}{@{}c@{}}\includegraphics[scale=.9]{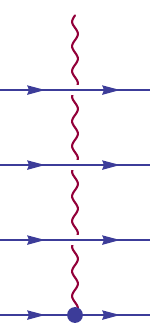}\\[-5.45pt]\end{array}{\;+\;}\begin{array}{@{}c@{}}\includegraphics[scale=.9]{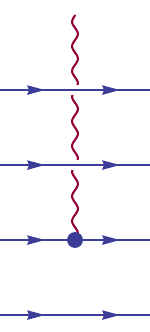}\\[-5.45pt]\end{array}{\;+\;}\begin{array}{@{}c@{}}\includegraphics[scale=.9]{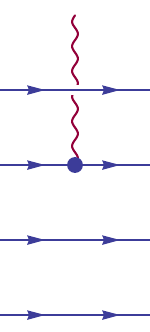}\\[-5.45pt]\end{array}{\;+\;}\begin{array}{@{}c@{}}\includegraphics[scale=.9]{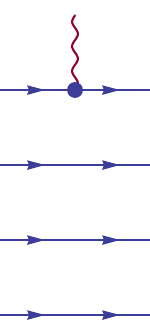}\\[-5.45pt]\end{array}\vspace{-14pt}$$
\end{figure}

The vertices appearing in (\ref{fig:egcolourdiagram}) are defined in \mbox{Figure~\ref{fig:Xipic}}; upon expanding each of these, we find $\mathcal{C}[{\color{flavour1}\underline{1}}\hspace{1.02pt}{\color{flavour3}\underline{3}}\hspace{1.02pt}{\color{flavour4}\underline{4}}\hspace{1.02pt}6\hspace{1.02pt}{\color{flavour4}\overline{7}}\hspace{1.02pt}{\color{flavour3}\overline{8}}\hspace{1.02pt}5\hspace{1.02pt}{\color{flavour2}\underline{2}}\hspace{1.02pt}{\color{flavour2}\overline{9}}\hspace{1.02pt}{\color{flavour1}\overline{10}}]$ to be given by the sum of contractions:
\vspace{-20pt}\eq{\begin{split}\\[20pt]&\hspace{-95pt}\phantom{+\;}\begin{array}{@{}c@{}}\\[-34pt]\includegraphics[scale=0.85]{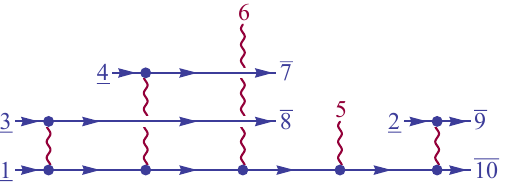}\end{array}{\hspace{-5pt}+\;}\begin{array}{@{}c@{}}\\[-34pt]\includegraphics[scale=.85]{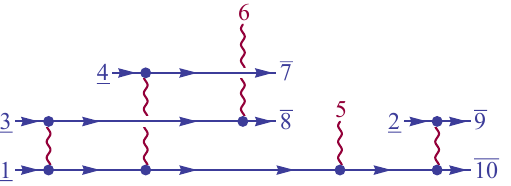}\end{array}\hspace{-100pt}\\[10pt]
&\hspace{-95pt}{+\;}\begin{array}{@{}c@{}}\\[-34pt]\includegraphics[scale=0.85]{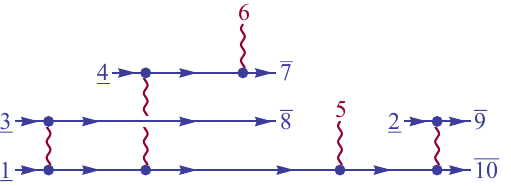}\end{array}{\hspace{-5pt}+\;}\begin{array}{@{}c@{}}\\[-34pt]\includegraphics[scale=.85]{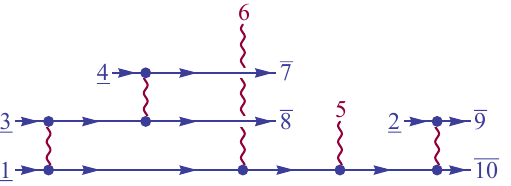}\end{array}\hspace{-100pt}\\[10pt]
&\hspace{-95pt}{+\;}\begin{array}{@{}c@{}}\\[-34pt]\includegraphics[scale=0.85]{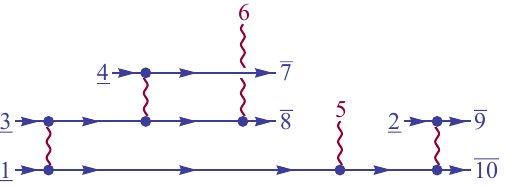}\end{array}{\hspace{-5pt}+\;}\begin{array}{@{}c@{}}\\[-34pt]\includegraphics[scale=.85]{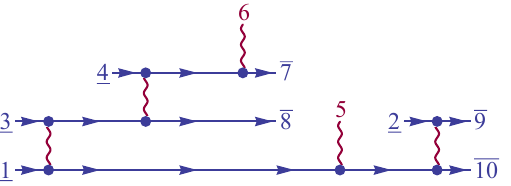}\end{array}\hspace{-100pt}\\[-45pt]
\end{split}\label{fig:egcolourdiagramExpanded}
}
(These figures are generated using the function \funL{drawColourTensorsExpanded}.)

More constructively (and less graphically), we can encode the colour tensor via the iterated replacements
\eq{C[{\color{flavour1}\underline{1}}\,\vec{\sigma}\,{\color{flavour1}\overline{n}}]\equivR{\color{flavour1}\{\underline{1}|}\vec{\sigma}{\color{flavour1}|\overline{n}\}}/\!\!/.\left\{\begin{array}{lll}{\color{flavour2}\underline{a}}\text{ (at level $\ell$)}&\mapsto&{\color{flavour2}\{\underline{a}|}\mathbf{\overline{T}}_{\t{a}}\!\otimes\!\Xi_{\ell{-}1}^{\t{a}},\\
{\color{flavour3}\overline{b}}\text{ (at level $\ell$)}&\mapsto&{\color{flavour3}|\overline{b}\}},\\
c\text{ (at level $\ell$)}&\mapsto&\Xi_{\ell}^{c}\end{array}\right\}\,.\label{colour_tensor_replacement_rule}}
where $\mathbf{\overline{T}}\equivR\text{-}\mathbf{T}^{T}$ denotes the generators of the conjugate representation and the `level' of a line/particle denotes the number of fermion lines between it and a point below the line $({\color{flavour1}\underline{1}},{\color{flavour1}\overline{n}})$, and\\[-10pt]
\eq{\Xi_\ell^{\t{a}}\equivR\sum_{s=1}^{\ell}\underbrace{\mathbf{1}\!\otimes\!\cdots\!\otimes\!\overbrace{\mathbf{{T}}_{\!\mathbf{\!\b{R}}}^{\t{a}}\!\otimes\!\cdots\!\otimes\!\mathbf{1}}^{s}}_{\ell}\,.\label{defn_of_xi}}
Applying the replacements (\ref{colour_tensor_replacement_rule}) would result in a representation of the tensor (\ref{fig:egcolourdiagram}) given by 
\eq{{\color{flavour1}\{\underline{1}|}{\color{flavour3}\{\underline{3}|}(\mathbf{\bar{T}}_{\t{\!a_1}}\!\!\otimes\!\Xi_{1}^{\t{a_1}}\!){\color{flavour4}\{\underline{4}|}(\mathbf{\bar{T}}_{\!\t{a_2}}\!\!\otimes\!\Xi_{2}^{\t{a_2}}\!)\Xi_{3}^{\r{6}}{\color{flavour4}|\overline{7}\}}{\color{flavour3}|\overline{8}\}}\Xi_{1}^{\r{5}}{\color{flavour2}\{\underline{2}|}(\mathbf{\bar{T}}_{\!\t{a_3}}\!\!\otimes\!\Xi_{1}^{\t{a_3}}\!){\color{flavour2}|\overline{9}\}}{\color{flavour1}|\overline{10}\}}\,.}
(This representation of colour-tensors can be obtained using \fpackage\ via the function \funL{writeAsCurlyBrackets}.)

The expanded form comes from applying the definition of the $\Xi_{\ell_i}$ tensors, resulting in a sum over $\prod_{i}(\ell_i{-}1)$ terms as in (\ref{fig:egcolourdiagramExpanded}). Each of these terms is given simply by the product of $n_f$ curly brackets defined in (\ref{one_flavour_colour_tensors}), with various contractions between them. For example: 
\eq{\hspace{-400pt}\begin{array}{@{}c@{}}\\[-10pt]\includegraphics[scale=0.65]{colourTensor_exempli_6}\end{array}\hspace{-5pt}\raisebox{-16pt}{$\bigger{\Leftrightarrow}\;{\color{flavour1}\{\underline{1}|}\mathbf{{T}}_{\!\t{a_1}}\!\mathbf{{T}}_{\!\t{a_2}}\!\mathbf{T}^{\r{6}}\mathbf{T}^{\r{5}}\mathbf{{T}}_{\!\t{a_3}}{\color{flavour1}|\overline{10}\}}{\color{flavour2}\{\underline{2}|}\mathbf{\overline{T}}^{\t{a_3}}\!{\color{flavour2}|\overline{9}\}}{\color{flavour3}\{\underline{3}|}\mathbf{\overline{T}}^{\t{a_1}}\!{\color{flavour3}|\overline{8}\}}{\color{flavour4}\{\underline{4}|}\mathbf{\overline{T}}^{\t{a_2}}\!{\color{flavour4}|\overline{7}\}}$}\hspace{-400pt}}
and similarly, 
\eq{\hspace{-400pt}\begin{array}{@{}c@{}}\\[-10pt]\includegraphics[scale=0.65]{colourTensor_exempli_1}\end{array}\hspace{-5pt}\raisebox{-16pt}{$\bigger{\Leftrightarrow}\;{\color{flavour1}\{\underline{1}|}\mathbf{{T}}_{\!\t{a_1}}\!\mathbf{T}^{\r{5}}\mathbf{{T}}_{\!\t{a_3}}{\color{flavour1}|\overline{10}\}}{\color{flavour2}\{\underline{2}|}\mathbf{\overline{T}}^{\t{a_3}}\!{\color{flavour2}|\overline{9}\}}{\color{flavour3}\{\underline{3}|}\mathbf{\overline{T}}^{\t{a_1}}\!\mathbf{{T}}_{\!\t{a_2}}\!{\color{flavour3}|\overline{8}\}}{\color{flavour4}\{\underline{4}|}\mathbf{\overline{T}}^{\t{a_2}}\!\mathbf{T}^{\r{6}}\!{\color{flavour4}|\overline{7}\}}$}\,.\hspace{-400pt}}
(The expansion into products of curly-brackets can be generated via the function \funL{writeAsCurlyBracketsExpanded}.)\\[-6pt]

In the package \fpackage, we not only allow for the systematic expression of these tensors in terms of nested (or expanded) curly brackets, but also allow the user to directly \emph{build} these tensors as explicit, numerical \built{SparseArray} objects given any set of charge generators---for \emph{any} representation of \emph{any} Lie algebra. These arrays allow for the rapid computation of arbitrary contractions/traces over this colour dependence relevant for cross sections, say. 

Although the construction of such tensors as literal arrays may itself seem rather peculiar, we find that virtually all representation-theoretic quantities relevant to generic amplitudes involving arbitrarily-coloured particles can be determined sufficiently rapidly in this way---without any recourse to the peculiarities of particular representations of particular gauge groups (such as Fierz identities or large-$N_c$ expansions). We build upon these ideas in the forthcoming \textsc{Mathematica} package for the construction and manipulation of colour tensors in \cite{lie_algebra_representation_tensor_tools}. The comparatively limited functionality provided by \fpackage\ for the colour tensors relevant to charged-matter amplitudes described in \cite{Johansson:2015oia} is largely motivated by a desire to make these tensors \emph{useful} without requiring users build an understanding of representation theory for themselves as described in \emph{e.g.}~\cite{birdtracks} (see also \cite{Zeppenfeld:1988bz,Bourjaily:2024jbt,Bourjaily:2025hvq}).\\[-6pt]

One final point is worth mentioning before we move on: as of today, there is no direct definition of the colour tensors relevant to \emph{single-flavoured} amplitudes. While these can always be obtained by summing over distinctly-flavoured amplitudes as in 
\eq{{A}(\{{\color{flavour0}\psi_1},{\color{flavour0}\psi_2},\ldots\})=\sum_{\vec{\sigma}\in\mathfrak{S}([n_f])}\mathcal{A}(\{{\color{flavour1}\psi_1},{\color{flavour2}\psi_2},{\color{flavour3}\psi_3},\ldots,{\color{flavour3}\bar{\psi}_{\text{-}\sigma_3}},{\color{flavour2}\bar{\psi}_{\text{-}\sigma_2}},{\color{flavour1}\bar{\psi}_{\text{-}\sigma_1}}\})\,,\label{same_flavour_amp_expansion_yet_again}}
the colour tensors appearing in this expansion are always simply those which appear in the \emph{distinctly}-flavoured partial amplitudes of (\ref{colour_decomposition_of_fermionic_amplitudes}).\\[-6pt]

\subsubsection{Bookkeeping: Conventions for Labeling Tensors}\label{tensor_labeling_conventions}

Especially when viewed as an explicit array of numbers, it is important to organize the \built{Sequence} of arguments of a colour tensor uniformly across all diagrams. To do this, we conventionally organize the \built{Slot}-\built{Sequence} of any \built{SparseArray} object encoding a colour-tensor according to the momentum-ordering of states. Following our conventions for labeling partial amplitudes in (\ref{momentum_ordering_of_states}), the sequence of colour-slots of every tensor will be the same---namely, 
\eq{\big\{\overbrace{\phantom{\fwbox{0pt}{n_f{+}1)}}{\color{flavour0}\underline{1}},{\color{flavour0}\underline{\ldots}}\,,{\color{flavour0}\underline{\smash{n_f}}}}^{\fwbox{10pt}{n_f\!\!\times\!\text{$\mathbf{\b{R}}$-slots}}},\overbrace{(n_f{+}1),\ldots,(n_f\text{+}n_g)}^{\fwbox{10pt}{n_g\!\!\times\!\text{$\mathbf{\r{ad}}(\mathbf{\b{R}})$-slots}}},\overbrace{{\color{flavour0}\overline{\phantom{\fwbox{0pt}{1}}{\text{-}n_f}}},{\color{flavour0}\overline{\phantom{\fwbox{0pt}{1}}\ldots\,}},{\color{flavour0}\overline{\phantom{\fwbox{0pt}{1}}\text{-}1}}}^{\fwbox{10pt}{n_f\!\!\times\!\text{$\mathbf{\b{\bar{R}}}$-slots}}\hspace{-0pt}}\big\}.\label{slot_ordering_of_tensors}}
Thus, all colour tensors will be of rank-$(n_f{+}n_g,n_f)$, with indices of the form 
\eq{\mathcal{C}\indices{{\color{flavour1}c_1}{\color{flavour2}c_2}\cdots\,\mathfrak{g}_{n_{\!\!f}\!{+}\!1}\cdots\mathfrak{g}_{n_{\!\!f}\!{+}\!n_g}}{\cdots\,{\color{flavour2}\bar{c}_{\text{-}2}}{\color{flavour1}\bar{c}_{\text{-}1}}}\,.}
This slot sequence is independent of the ordering sequence of states in partial amplitudes (like the association between states and the list of external momenta), and the same regardless of the flavour-pairing of states.\\

\subsubsection{\emph{Exempli Gratia}: Interesting Cases of Charged Matter Tensors}\label{eg_charged_matter}

It is worth reiterating an important feature of the tensors described by Johansson and Ochirov: they are defined \emph{without reference} to any peculiar simplifications available for particular charge-representations of particular gauge groups. As such, they are equally useful for `fundamentally' charged matter, adjoint-charged matter (as in supersymmetric Yang-Mills theory), or even particles charged under the reducible `$\mathbf{30380}\!\otimes\!\mathbf{248}$' representation of $\mathfrak{e}_8$ gauge theory. Especially for the sake of applications to `real-world' physics, there are at least two interesting cases to note.\\[-6pt]

\paragraph{Amplitudes in Massless QED}~\\[-34pt]

One especially interesting case to consider is that of massless, charged fermions in $\mathfrak{u}_1$ gauge theory. In this case, the charge generators would be given by the single $1\!\times\!1$ matrix $(1)$, with all structure constants vanishing. Nevertheless, the charge tensors described above correctly encode the `colour dependence' of amplitudes---all tensors being rank-$n$ tensors of one index each. The only colourful-content of these tensors is an integer counting the number of terms appearing in the expanded form such as in (\ref{fig:egcolourdiagramExpanded}). Thus, for the example of (\ref{fig:egcolourdiagramExpanded}) above, we'd have 
\vspace{4pt}\eq{\fwboxR{00pt}{(\mathfrak{u}_1)\hspace{198pt}}\fwbox{0pt}{\mathcal{C}[{\color{flavour1}\underline{1}}\hspace{1.02pt}{\color{flavour3}\underline{3}}\hspace{1.02pt}{\color{flavour4}\underline{4}}\hspace{1.02pt}6\hspace{1.02pt}{\color{flavour4}\overline{7}}\hspace{1.02pt}{\color{flavour3}\overline{8}}\hspace{1.02pt}5\hspace{1.02pt}{\color{flavour2}\underline{2}}\hspace{1.02pt}{\color{flavour2}\overline{9}}\hspace{1.02pt}{\color{flavour1}\overline{10}}]=6}\vspace{4pt}}
where, by this, we mean that it is a rank-$(6,4)$, $1\!\times\!1\!\times\!\cdots\!\times\!1$ tensor whose only component equals 6.

These integers can be seen to exactly compensate for the multiplicity of $\mathfrak{u}_1$-theory Feynman diagrams that contribute to each partial amplitude. Consider, for example, an amplitude involving 2 \emph{distinguishable} fermions and one photon:
\eq{\begin{split}\fwboxR{00pt}{(\mathfrak{u}_1)\hspace{198pt}}\fwboxR{0pt}{A(\{{\color{flavour1}\underline{1}},{\color{flavour2}\underline{2}},3,{\color{flavour2}\overline{4}},{\color{flavour1}\overline{5}}\})=\hspace{100pt}}\fwboxL{0pt}{\hspace{-100pt}{-}\mathcal{A}({\color{flavour1}\underline{1}}\hspace{1.02pt}3\hspace{1.02pt}{\color{flavour2}\underline{2}}\hspace{1.02pt}{\color{flavour2}\overline{4}}\hspace{1.02pt}{\color{flavour1}\overline{5}}){-}2\mathcal{A}({\color{flavour1}\underline{1}}\hspace{1.02pt}{\color{flavour2}\underline{2}}\hspace{1.02pt}3\hspace{1.02pt}{\color{flavour2}\overline{4}}\hspace{1.02pt}{\color{flavour1}\overline{5}}){-}\mathcal{A}({\color{flavour1}\underline{1}}\hspace{1.02pt}{\color{flavour2}\underline{2}}\hspace{1.02pt}{\color{flavour2}\overline{4}}\hspace{1.02pt}3\hspace{1.02pt}{\color{flavour1}\overline{5}})\,.}
\end{split}\label{u1_21_amp_example}}
Replacing each partial amplitude by the Feynman diagrams that contribute, and noting the sign conventions for partial amplitudes in gauge theory (for which each vertex is antisymmetric with respect to plane ordering of its edges), we see that
\eq{\begin{split}\hspace{-1000pt}A(\{{\color{flavour1}\underline{1}},{\color{flavour2}\underline{2}},3,{\color{flavour2}\overline{4}},{\color{flavour1}\overline{5}}\})=&{-}\Bigg[\fdiagramMaa\hspace{-5pt}{+}\fdiagramMab\hspace{-5pt}{+}\fdiagramMac\hspace{-0pt}\Bigg]{-}2\Bigg[\hspace{-0pt}\fdiagramMbb\hspace{-5pt}{+}\hspace{-5pt}\fdiagramMba\hspace{-0pt}\Bigg]\hspace{-500pt}\\[-10pt]
\hspace{-70pt}&{-}\Bigg[\hspace{-0pt}\fdiagramMca{+}\hspace{-5pt}\fdiagramMcb{+}\hspace{-5pt}\fdiagramMcc\Bigg]\hspace{-50pt}\\[-12pt]
\hspace{-70pt}=&\phantom{{-}\hspace{7pt}}\fdiagramMab\hspace{-2pt}{+}\hspace{-3pt}\fdiagramMacc\hspace{-2pt}{+}\hspace{-3pt}\fdiagramMcb\hspace{4pt}{+}\hspace{-4pt}\fdiagramMccc\,.\hspace{-50pt}\end{split}}
(Note: partial amplitudes change sign for odd permutations of their embedded vertices (exactly to compensate the antisymmetry of their colour-dependence).)\\[-6pt]

Alternatively, consider the case of 2 \emph{indistinguishable} fermion lines interacting with one photon:
\eq{\begin{split}\fwboxR{00pt}{(\mathfrak{u}_1)\hspace{198pt}}\fwboxR{0pt}{A(\{{\color{flavour0}\underline{1}},{\color{flavour0}\underline{2}},3,{\color{flavour0}\overline{4}},{\color{flavour0}\overline{5}}\})=\hspace{100pt}}\fwboxL{0pt}{\hspace{-100pt}{-}\mathcal{A}({\color{flavour1}\underline{1}}\hspace{1.02pt}3\hspace{1.02pt}{\color{flavour2}\underline{2}}\hspace{1.02pt}{\color{flavour2}\overline{4}}\hspace{1.02pt}{\color{flavour1}\overline{5}}){-}2\mathcal{A}({\color{flavour1}\underline{1}}\hspace{1.02pt}{\color{flavour2}\underline{2}}\hspace{1.02pt}3\hspace{1.02pt}{\color{flavour2}\overline{4}}\hspace{1.02pt}{\color{flavour1}\overline{5}}){-}\mathcal{A}({\color{flavour1}\underline{1}}\hspace{1.02pt}{\color{flavour2}\underline{2}}\hspace{1.02pt}{\color{flavour2}\overline{4}}\hspace{1.02pt}3\hspace{1.02pt}{\color{flavour1}\overline{5}})}\\[-20pt]
\\\fwboxL{0pt}{\hspace{-100pt}{-}\mathcal{A}({\color{flavour1}\underline{1}}\hspace{1.02pt}3\hspace{1.02pt}{\color{flavour2}\underline{2}}\hspace{1.02pt}{\color{flavour2}\overline{5}}\hspace{1.02pt}{\color{flavour1}\overline{4}}){-}2\mathcal{A}({\color{flavour1}\underline{1}}\hspace{1.02pt}{\color{flavour2}\underline{2}}\hspace{1.02pt}3\hspace{1.02pt}{\color{flavour2}\overline{5}}\hspace{1.02pt}{\color{flavour1}\overline{4}}){-}\mathcal{A}({\color{flavour1}\underline{1}}\hspace{1.02pt}{\color{flavour2}\underline{2}}\hspace{1.02pt}{\color{flavour2}\overline{5}}\hspace{1.02pt}3\hspace{1.02pt}{\color{flavour1}\overline{4}})}\\[-0pt]
\fwboxR{0pt}{=\hspace{100pt}}\fwboxL{0pt}{\hspace{-100pt}{\phantom{{-}}{A}({\color{flavour0}\underline{1}}\hspace{1.02pt}3\hspace{1.02pt}{\color{flavour0}\overline{4}}\hspace{1.02pt}{\color{flavour0}\underline{2}}\hspace{1.02pt}{\color{flavour0}\overline{5}}){+}{A}({\color{flavour0}\underline{1}}\hspace{1.02pt}{\color{flavour0}\overline{4}}\hspace{1.02pt}{\color{flavour0}\underline{2}}\hspace{1.02pt}3\hspace{1.02pt}{\color{flavour0}\overline{5}})\,.}}
\\[-10pt]\end{split}\label{u1_sf_21_amp_example}}
Writing the Feynman diagrams consistent with each of the two single-flavour partial amplitudes in the last line, we see that the amplitude is given by the terms
\eq{\begin{split}
&\phantom{{+}}\fwboxL{65pt}{\fdiagramSFba}{+}\fwboxL{66pt}{\fdiagramSFbb}{+}\fwboxL{65pt}{\fdiagramSFbc}{+}\fwboxL{50pt}{\fdiagramSFaa}{+}\fwboxL{40pt}{\fdiagramSFab}\\[-0pt]
&{+}\fwboxL{65pt}{\fdiagramSFca}{+}\fwboxL{65pt}{\fdiagramSFcb}{+}\fwboxL{65pt}{\fdiagramSFcc}{+}\fwboxL{50pt}{\fdiagramSFda}\hspace{0pt}{+}\hspace{-0pt}\fwboxL{40pt}{\fdiagramSFdb}\fwbox{0pt}{\,.}\\[-10pt]
\end{split}}
Notice that the internal three-`gluon' vertex cancels from the sum. Upon rearranging the vertices (with appropriate signs) and reorganizing these terms, we find that
\begin{align}
\hspace{-20pt}A(\{{\color{flavour0}\underline{1}},{\color{flavour0}\underline{2}},3,{\color{flavour0}\overline{4}},{\color{flavour0}\overline{5}}\})=&\phantom{{-}}{A}({\color{flavour0}\underline{1}}\hspace{1.02pt}3\hspace{1.02pt}{\color{flavour0}\overline{4}}\hspace{1.02pt}{\color{flavour0}\underline{2}}\hspace{1.02pt}{\color{flavour0}\overline{5}}){+}{A}({\color{flavour0}\underline{1}}\hspace{1.02pt}{\color{flavour0}\overline{4}}\hspace{1.02pt}{\color{flavour0}\underline{2}}\hspace{1.02pt}3\hspace{1.02pt}{\color{flavour0}\overline{5}})\\
=&\phantom{{-}}\Bigg[\sfFinalA{+}\sfFinalB{+}\sfFinalC{+}\sfFinalD\Bigg]\nonumber\\
&{-}\Bigg[\sfFinalAb{+}\sfFinalBb{+}\sfFinalCb{+}\sfFinalDb\Bigg]\fwbox{0pt}{\,.}\nonumber
\end{align}
This exactly matches the Feynman diagrams of QED for an amplitude involving two indistinguishable fermions and one photon. 

\newpage
\paragraph{\textbf{Fundamental} Matter of $\mathfrak{su}_{N}$ Gauge Theory}~\\[-36pt]

Another case of exceptional phenomenological interest and considerable simplicity is that of \emph{fundamental} (`$\mathbf{\b{N}}$'-)charged matter of $\mathfrak{su}_{N}$ gauge theory. In this case, the Fierz identity
\eq{\sum_{\t{a},\t{b}\in[\mathfrak{g}]}g^{\mathbf{\r{ad}}}_{\t{a\,b}}(\mathbf{T}_{\!\!\mathbf{\b{N}}}^{\t{a}}\!)\indices{c_1}{\,\bar{c}_2}(\mathbf{T}_{\!\!\mathbf{\b{N}}}^{\t{b}}\!)\indices{c_3}{\,\bar{c}_4}=\delta\indices{c_1}{\bar{c}_4}\delta\indices{c_3}{\bar{c}_2}{-}\frac{1}{N}\delta\indices{c_1}{\bar{c}_2}\delta\indices{c_3}{\bar{c}_4}\,,\label{fundamental_fierz_for_sun}}
allows all the tensors appearing in the expansion of multi-flavoured amplitudes into sums over products of Kronecker $\delta$'s and generator matrices $\mathbf{T}$ weighted by inverse powers of $N$. 

Consider, for example, the colour tensor $\mathcal{C}[{\color{flavour1}\underline{1}}\hspace{0.5pt}5\hspace{0.5pt}4\hspace{0.5pt}{\color{flavour3}\underline{3}}\hspace{0.5pt}6\hspace{0.5pt}{\color{flavour3}\overline{7}}\hspace{0.5pt}{\color{flavour2}\underline{2}}\hspace{0.5pt}{\color{flavour2}\overline{8}}\hspace{0.5pt}{\color{flavour1}\overline{9}}]$. For fermions which all transform under the fundamental, `$\mathbf{\b{N_c}}$' representation of $\mathfrak{su}_{N_c}$ gauge theory, we could use (\ref{fundamental_fierz_for_sun}) to express it as
\eq{\begin{split}
\hspace{-100pt}\mathcal{C}[{\color{flavour1}\underline{1}}\hspace{0.5pt}5\hspace{0.5pt}4\hspace{0.5pt}{\color{flavour3}\underline{3}}\hspace{0.5pt}6\hspace{0.5pt}{\color{flavour3}\overline{7}}\hspace{0.5pt}{\color{flavour2}\underline{2}}\hspace{0.5pt}{\color{flavour2}\overline{8}}\hspace{0.5pt}{\color{flavour1}\overline{9}}]\bigger{\Leftrightarrow}\hspace{-10pt}&\raisebox{10pt}{$\begin{array}{@{}c@{}}\includegraphics[scale=0.75]{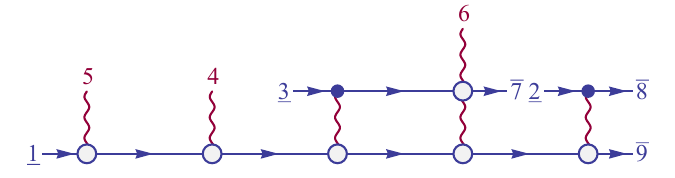}\end{array}$}\hspace{20pt}\\
&\hspace{-69pt}{=}\hspace{31pt}(\!\mathbf{T}^{{\color{hred}5}}\mathbf{T}^{{\color{hred}4}}\!)\!\indices{{\color{flavour1}{1}}}{{\color{flavour3}\overline{7}}}\delta\indices{{\color{flavour2}{2}}}{{\color{flavour1}\overline{9}}}(\!\mathbf{T}^{{\color{hred}6}}\!)\!\indices{{\color{flavour3}{3}}}{{\color{flavour2}\overline{8}}}{+}(\!\mathbf{T}^{{\color{hred}5}}\mathbf{T}^{{\color{hred}4}}\mathbf{T}^{{\color{hred}6}}\!)\!\indices{{\color{flavour1}{1}}}{{\color{flavour3}\overline{7}}}\delta\indices{{\color{flavour2}{2}}}{{\color{flavour1}\overline{9}}}\delta\indices{{\color{flavour3}{3}}}{{\color{flavour2}\overline{8}}}\\
&\hspace{-60pt}{-}\frac{1}{N_c}\Big[(\!\mathbf{T}^{{\color{hred}5}}\mathbf{T}^{{\color{hred}4}}\!)\!\indices{{\color{flavour1}{1}}}{{\color{flavour3}\overline{7}}}\delta\indices{{\color{flavour2}{2}}}{{\color{flavour2}\overline{8}}}(\!\mathbf{T}^{{\color{hred}6}}\!)\!\indices{{\color{flavour3}{3}}}{{\color{flavour1}\overline{9}}}{+}(\!\mathbf{T}^{{\color{hred}5}}\mathbf{T}^{{\color{hred}4}}\!)\!\indices{{\color{flavour1}{1}}}{{\color{flavour2}\overline{8}}}\delta\indices{{\color{flavour2}{2}}}{{\color{flavour1}\overline{9}}}(\!\mathbf{T}^{{\color{hred}6}}\!)\!\indices{{\color{flavour3}{3}}}{{\color{flavour3}\overline{7}}}{+}(\!\mathbf{T}^{{\color{hred}5}}\mathbf{T}^{{\color{hred}4}}\mathbf{T}^{{\color{hred}6}}\!)\!\indices{{\color{flavour1}{1}}}{{\color{flavour3}\overline{7}}}\delta\indices{{\color{flavour2}{2}}}{{\color{flavour2}\overline{8}}}\delta\indices{{\color{flavour3}{3}}}{{\color{flavour1}\overline{9}}}{+}(\!\mathbf{T}^{{\color{hred}5}}\mathbf{T}^{{\color{hred}4}}\mathbf{T}^{{\color{hred}6}}\!)\!\indices{{\color{flavour1}{1}}}{{\color{flavour2}\overline{8}}}\delta\indices{{\color{flavour2}{2}}}{{\color{flavour1}\overline{9}}}\delta\indices{{\color{flavour3}{3}}}{{\color{flavour3}\overline{7}}}\Big]\hspace{-100pt}\\
&\hspace{-60pt}{+}\frac{1}{N_c^2}\Big[(\!\mathbf{T}^{{\color{hred}5}}\mathbf{T}^{{\color{hred}4}}\!)\!\indices{{\color{flavour1}{1}}}{{\color{flavour1}\overline{9}}}\delta\indices{{\color{flavour2}{2}}}{{\color{flavour2}\overline{8}}}(\!\mathbf{T}^{{\color{hred}6}}\!)\!\indices{{\color{flavour3}{3}}}{{\color{flavour3}\overline{7}}}{+}(\!\mathbf{T}^{{\color{hred}5}}\mathbf{T}^{{\color{hred}4}}\mathbf{T}^{{\color{hred}6}}\!)\!\indices{{\color{flavour1}{1}}}{{\color{flavour1}\overline{9}}}\delta\indices{{\color{flavour2}{2}}}{{\color{flavour2}\overline{8}}}\delta\indices{{\color{flavour3}{3}}}{{\color{flavour3}\overline{7}}}\Big].\fwboxL{0pt}{\fwboxR{145pt}{(\text{$\mathbf{\b{R}}{=}\mathbf{\b{N_c}}$ of }\mathfrak{su}_{N_c}}\text{ theory})}\end{split}\label{example_sun_expansion_of_colour_tensors}}
These expansions have been automated in \fpackage\ in the function \funL{writeAsSUNcFundamentalTensors}.

The form of colour tensors in (\ref{example_sun_expansion_of_colour_tensors}) should be very familiar to most researchers and similar to those of `colour-flow' described in \cite{Maltoni:2002mq}.\\[-6pt]

In the package \fpackage, we have implemented the representation of colour tensors graphically or symbolically in terms of curly-brackets (nested as well as expanded), included the Fierz-expanded form as in (\ref{example_sun_expansion_of_colour_tensors}) for fundamentally-charged matter of $\mathfrak{su}_N$ gauge theory, and also implemented the explicit construction and contraction of these tensors given any user-provided charge generators.

\newpage
\section{Using the \fpackage\ Package}\label{section:fermionic_amplitudes_package}

The complete source code of the \textsc{Mathematica} package \fpackage\ together with a demonstration notebook walking the user through its primary functionality with illustrative examples and internal checks are available from the abstract page of this work on the \texttt{arXiv}. Specifically, these files may be found under the heading of `\textbf{Ancillary files}' in the upper-right-hand panel of the abstract page.\\[-6pt]

\subsection{Initializing and Installing the Package}\label{subsec:obtaining_and_installing}

To load the package, simply make sure that the file \fpackage\built{.m} is within the same directory as any (\emph{saved}) \built{Notebook} that you are using, and evaluate the following:
\mathematicaBox{
\mathematicaSequence[1]{\smash{\built{SetDirectory}\brace{\built{NotebookDirectory}\brace{}\,};}\\
\texttt{<\hspace{-1pt}<\,\fpackage\built{.m}}}{$\rule{0pt}{80pt}\begin{array}{@{}c@{}}\includegraphics[scale=0.6]{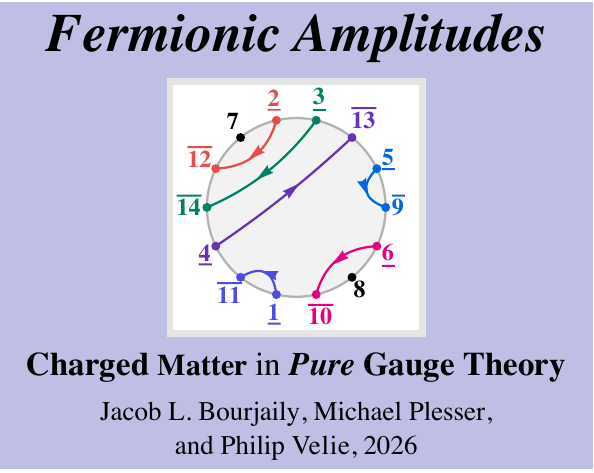}\end{array}$}
}
(These are the first lines of the demonstration notebook.)

If the user would like to make this package available to other \built{Notebooks}---those unsaved or located in other directories---\fpackage~can be copied to the \built{\$Path} directories using the function \funL{installFermionicAmplitudesPackage}.\\

\textbf{Note}: There are some features of \fpackage~that are designed for use with the package \package~\cite{Bourjaily:2023uln}. Upon initialization, the package \fpackage\ checks to see if \package~can be found in the \built{NotebookDirectory} (or elsewhere within the directories of \built{\$Path}); if it can be found, \package~will be loaded upon initialization of \fpackage.\\[-6pt]

\subsection{Basic Syntax and Conventions}\label{subsec:syntax_and_conventions_for_the_package}

The primary symbols used in the representation of colour-dressed amplitudes in gauge theory with charged fermions are `\funL{amp}'---used for both single-flavour and multi-flavour partial amplitudes---and `\funL{colourTensor}'. 

Single-flavour partial amplitudes are expressed in the form \fun{amp}\brace{\var{states}\pattern\built{List}} where \var{states} is an \emph{ordered} \built{List} of external states. See, for example, equation (\ref{example_encoding_of_sf_amp}).

External states each have a (protected) \built{Head} identifying the type of particle and its momentum-indexing. Following the conventions of (\ref{tex_conventions_for_labeling_states}) and (\ref{slot_ordering_of_tensors}) described above, these states may be specified according to \mbox{Table~\ref{syntax_for_states_table}}. 

Notice that the momentum index of anti-fermions is taken to be \emph{negative}---indicating that they are assigned the last $[n_f]$ indices of the range $[n]$. Finally, it is important that the number of anti-fermions be equal to the number of fermions.\\[-6pt]
\begin{table}[t]\caption{Conventions for labeling states of partial amplitudes.}\label{syntax_for_states_table}\vspace{-20pt}$$\begin{array}{@{}c@{$\;\;$}l@{$\;\;$}l@{$\;\;\;$}l@{}}\text{State}&\text{Syntax}&\text{(Momentum) Indices}&\multicolumn{1}{c}{\text{Notes}}\\\hline
\text{`${\color{black}{a}}$'}&\text{\fun{g}\brace{\var{$a$}\pattern}}&\var{a}\in\{n_f\text{+}1,\ldots,n_f\text{+}n_g\}&\text{indefinite-helicity gauge boson; momentum: $p_{\var{a}}$}\\
\text{\phantom{${}^{+}$}`${\color{black}{a{}^{+}}}$'}&\text{\fun{p}\brace{\var{$a$}\pattern}}&\var{a}\in\{n_f\text{+}1,\ldots,n_f\text{+}n_g\}&\text{a \emph{positive}-helicity gauge-boson; momentum: $p_{\var{a}}$}\\
\text{\phantom{${}^{-}$}`${\color{black}{a{}^{-}}}$'}&\text{\fun{m}\brace{\var{$a$}\pattern}}&\var{a}\in\{n_f\text{+}1,\ldots,n_f\text{+}n_g\}&\text{a \hspace{-2pt}\emph{negative}-helicity gauge-boson; momentum: $p_{\var{a}}$}\\~\\[-10pt]
\text{`${\color{flavour0}\underline{a}}$'}&\text{\fun{f}\brace{\var{$a$}\pattern}}&\var{a}\in\{1,\ldots,n_f\}&\text{a \emph{positive}-helicity fermion $\psi$; momentum: $p_{\var{a}}$}\\
\text{`${\color{flavour0}\overline{a}}$'}&\text{\fun{fb}\brace{\var{$a$}\pattern}}&\var{a}\in\{1,\ldots,n_f\}&\text{a \hspace{-2pt}\emph{negative}-helicity fermion $\bar{\psi}$; momentum: $p_{\text{-}\var{a}}$}
\end{array}$$\vspace{-20pt}\end{table}

To denote the partial amplitude involving some ordered set of \var{states} but with \emph{distinguishable} fermions, the \var{flavourPairing} between fermions and anti-fermions is encoded by a second argument \fun{amp}\brace{\var{states}\pattern\built{List},\var{flavourPairing}\pattern\built{List}} that follows precisely the conventions outlined in (\ref{flavour_pairing_convention}). As discussed in \mbox{section~\ref{state_labeling_conventions}}, \emph{fermions} are conventionally assigned flavour indices equal to their momentum indices, but the momentum-ordering of anti-fermions does not imply anything about their flavour assignments. 

All \fun{colourTensor} objects require flavour-pairing\footnote{Unless no fermions are involved, in which case these tensors are simply defined to be those of DDM \cite{DelDuca:1999rs}---which are defined in (\ref{ddm_tensors}).} and follow the same labeling conventions as multi-flavour partial amplitudes.\\

To specify the \funL[1]{chargeGenerators} of fermions, \funL{setChargeGenerators} should be used. This establishes a number of \built{Global} variables which are used by the function  \funL{buildColourTensors}, which converts any symbolic \fun{colourTensor} objects into explicit, numeric \built{SparseArray} objects. The \built{Slot}-\built{Sequence} of these arrays is always taken to follow momentum-index ordering, as described in \mbox{section~\ref{tensor_labeling_conventions}} above.

Although the user may specify any \funL[1]{chargeGenerators} she pleases, we have stored the familiar (unitary) generators of the fundamental representations of $\mathfrak{u}_1,\mathfrak{su}_{2},\mathfrak{su}_3,$ and $\mathfrak{su}_4$---which can be loaded via the arguments \fun{u}\brace{1}\,,\fun{su}\brace{2}\,,\fun{su}\brace{3}\,, and \fun{su}\brace{4}\, with the function \funL{setChargeGenerators}. (We have also included these generators---together with the $\mathbf{\b{7}}$-dimensional representation of $\mathfrak{g}_2$---in the Chevalley basis.)

\newpage
\subsection{Primary Functionality Provided by the Package}\label{subsec:primary_functionality_of_package}

The primary functionality of \fpackage\ is outlined through a number of illustrative examples in the demonstration notebook included with this work's submission to the \texttt{arXiv}. Principle among the functions defined are:
\begin{itemize}
\item for the symbolic representation of colour-dressed amplitudes involving same-flavoured or distinctly-flavoured charged fermions in gauge theory:
\begin{itemize}
\item \funL{distinctFlavourAmp}
\item \funL{distinctFlavourAmpSquared}
\item \funL{sameFlavourAmp}
\item \funL{sameFlavourAmpSquared}\\[-20pt]
\end{itemize}
\item for manipulations of partial amplitudes and their relations---including the representation of multi-flavour partial amplitudes in terms of single-flavoured ones, and vice-versa:
\begin{itemize}
\item\funL{partialAmpBasis}
\item\funL{randomPartialAmp}
\item\funL{toDistinctFlavours}
\item\funL{toSingleFlavour}
\item\funL{kkProjectionRule}
\item\funL{toAllPlus}
\item\funL{toAllPlusBasis}\\[-20pt]
\end{itemize}
\item for the graphical representation of partial amplitudes' chord diagrams and colour tensors, and related symbolic forms:
\begin{itemize}
\item\funL{drawChordDiagrams}
\item\funL{drawColourTensors}
\item\funL{drawColourTensorsExpanded}
\item\funL{writeAsCurlyBrackets}
\item\funL{writeAsCurlyBracketsExpanded}
\item\funL{writeAsSUNcFundamentalTensors}\\[-20pt]
\end{itemize}
\item for the direct construction of explicit colour tensors for any (user-provided or pre-defined) set of charge generators, and the computation of their contractions---useful for the construction of (colour-summed) squared amplitudes, and for identifying any linear relations satisfied by a set of colour tensors:
\begin{itemize}
\item\funL{setChargeGenerators}
\item\funL{buildColourTensors}
\item\funL{colourTensorRelations}\\[-20pt]
\end{itemize}
\item and for the translation to forms used by the \package\ package:
\begin{itemize}
\item\funL{toAnalytic}
\item\funL{toNumeric}\\[-26pt]
\end{itemize}
\end{itemize}
Our implementation of the many algorithms involved may be verified in a number of ways. In particular, in the walkthrough notebook we demonstrate the equivalence between colour-dressed amplitudes represented in different bases, and (for adjoint-charged fermions) against DDM. 

The primary functions provided by the package are documented in \mbox{Appendix~\ref{appendix:protected_symbols}}, including a number of illustrations of their usage.\\[16pt]

\section{Conclusions and Outlook}\label{sec:conclusion}

The implementation of Melia's flavour-reduction algorithm combined with the colour-tensors of Johansson and Ochirov provide an effective way to determine amplitudes involving charged matter in gauge theory. In the case of \emph{massless} charged matter, our ability to determine partial amplitudes in terms of those of supersymmetric Yang-Mills theory proves especially powerful. Even for these amplitudes, there remain important open questions. 

For example, consider the case of amplitudes involving only a single flavour of fermion. There must exist a colour decomposition of the form 
\eq{A(\{{\color{flavour0}\psi_1},\ldots,{\color{flavour0}\bar{\psi}_{\text{-}1}}\})=\sum_{\hspace{-5pt}\vec{\sigma}\in\mathfrak{S}([n\text{-}2])\hspace{-10pt}}\mathcal{C}(\vec{\sigma})\,A({\color{flavour0}\psi_1},\vec{\sigma},{\color{flavour0}\bar{\psi}_{\text{-}1}})\,,}
which is \emph{directly} expressed in terms of single-flavoured partial amplitudes; but as of today---at least for general charge representations of general gauge groups---the colour tensors which decorate each partial amplitude are not known. 

For the partial amplitudes involving distinct-flavoured fermions, the colour tensors of Johansson and Ochirov (like those of DDM) are arguably too broad: for any fixed gauge group or representation of charged matter, these tensors grow factorially with multiplicity; and yet, we know on general grounds that the number of \emph{independent} colour tensors which exist must grow at most exponentially \cite{Bourjaily:2024jbt}. (This is visible already for relatively low multiplicity in the case of $\mathfrak{su}_2$ gauge theory (and for $\mathfrak{u}_1$ gauge theory, obviously).)

Moreover, the colour tensors involved in charged-matter amplitudes exhibit relatively dense overlap in colour-space---resulting in virtually all partial amplitudes interfering with each other in the computation of colour-summed squared amplitudes. This interference does \emph{not} diminish with increasing numbers of independent colours---making it hard to see amplitudes' large-$N_c$ behavior. (This complaint applies equally to the colour tensors of DDM (\ref{ddm_tensors}), but is unlike the traces of (\ref{trace_expansion_for_glue}), which become non-interfering in the large-$N_c$ limit of $\mathfrak{su}_{N_c}$ gauge theory.)

\newpage
The construction of a manifestly colour-orthogonal, linearly independent set of colour tensors for any process involving arbitrarily-coloured particles (at arbitrary orders of perturbation theory) was described in ref.\ \cite{Bourjaily:2025hvq}. It would be worthwhile to explore the representation of amplitudes involving charged matter and gauge bosons in terms of such tensors.\\[-6pt]

Finally, there are a number of interesting questions and relevant applications involving charged fermions of non-vanishing mass. In many such cases, distinguishable fermions would have distinct masses, rendering much of the technology described here moot.  For degenerate (or indistinguishable) massive fermions, however, virtually nothing about the reduction of partial amplitudes will change---only their connection to amplitudes of sYM would require modification. Amplitudes involving degenerate massive fermions may be computable on the Higgs branch of sYM, or may be effectively obtained using the formalism of \cite{Arkani-Hamed:2017jhn}. But we must leave such questions to future work.

\vspace{\fill}\vspace{-4pt}
\section*{Acknowledgments}%
\vspace{-4pt}
\noindent The authors gratefully acknowledge fruitful conversations with Henrik Johansson. One of the authors (JB) would like to thank the Erwin Schr\"odinger International Institute for Mathematics and Physics (ESI), University of Vienna (Austria), for the opportunity to participate in the Thematic Programme ``Amplitudes and Algebraic Geometry'' where some of this work was completed. This work was supported in part by a grant from the US Department of Energy (No.\ DE-SC00019066).

\newpage
\addtocontents{toc}{\protect~\\[-10pt]\mbox{\vspace{0pt}}\protect\hrulefill\par}
\appendix
%
\setcounter{section}{-1}
\hypertarget{context_organization_of_appendix}{}\vspace{-0pt}\section[\mbox{\hspace{-18pt}Context-Organized List of Functions Provided by the Package}]{\hspace{-17.5pt}Context-Organized List of Functions Provided}\vspace{-0pt}\label{function_appendix}

\vspace{5pt}\sectionAppendix{Protected Symbols and Abstract Expressions}{appendix:protected_symbols}\vspace{-10pt}

\defnBox[5]{f}{\var{momentumIndex}\pattern}{labels a fermion (with `charge' $\mathbf{\b{R}}$), with momentum indexed by \var{momentumIndex}\,$\in[n_f]\!\subset\![n]$. In the context of distinct-flavoured partial amplitudes, we take by convention that \fun{f}\brace{\var{$a$}} has `flavour' index \var{$a$}.}

\defnBox[5]{fb}{\var{momentumIndex}\pattern}{labels an anti-fermion (with `charge' $\mathbf{\b{\bar{R}}}$), with momentum indexed by ({-}\var{momentumIndex})$\in\![n_f]\subset\![n]$. To be clear, there is no presumption of flavour indexing. So `$\fun{fb}\brace{1}$\,' is simply the anti-fermion associated with the `$(\text{-}1)$th' (or \built{Last}) momentum index---namely, $p_n$.}

\defnBox[5]{g}{\var{momentumIndex}\pattern}{labels a gauge boson (with `charge' $\mathbf{\r{ad}}(\mathbf{\b{R}})$) associated with the momentum indexed by \var{momentumIndex}\,$\in([n_g]{+}n_f)$.}

\defnBox[5]{p}{\var{momentumIndex}\pattern}{labels a \emph{positive-helicity} gauge boson (with `charge' $\mathbf{\r{ad}}(\mathbf{\b{R}})$) associated with the momentum indexed by \var{momentumIndex}\,$\in([n_g]{+}n_f)$.}

\defnBox[5]{m}{\var{momentumIndex}\pattern}{labels a \emph{negative-helicity} gauge boson (`charge' $\mathbf{\r{ad}}(\mathbf{\b{R}})$) associated with the momentum indexed by \var{momentumIndex}\,$\in([n_g]{+}n_f)$.}

\defnBox{colourTensor}{\{\var{stateSequence}\patternTwo\},\{\var{flavourPairingSequence}\patternTwo\}}{abstractly denotes the colour tensor defined by \cite{Johansson:2015oia} that dresses the partial amplitude \mbox{\fun{amp}\brace{\{\var{stateSequence}\},\{\var{flavourPairingSequence}\}}} appearing in the \emph{all-plus} basis of partial amplitudes (\ref{all_plus_basis_of_partial_amps}).\\[10pt]
To be clear, although a wider variety of partial amplitude bases may be defined, \fun{colourTensor} objects are only defined for those in this specific basis of partial amplitudes (or trivial permutations thereof).
}

\defnBox{colourTensorOverlapMatrix}{\{\var{tensorSequence}\patternTwo\}}{abstractly represents the self-overlap matrix between a \built{List} of \fun{colourTensor} (or \fun{colorFactor}) objects and their conjugate tensors. The function \funL{buildColourTensors} will convert this to a numeric array for the \funL[3]{chargeGenerators} \built{Set} by \funL{setChargeGenerators}.}

\defnBox{amp}{\{\var{stateSequence}\patternTwo\}}{abstractly represents an ordered partial amplitude involving the states $\{\var{stateSequence}\}$ for which all fermions have the \emph{same} flavour---and therefore are indistinguishable.}

\defnBox[2]{amp}{\{\var{stateSequence}\patternTwo\},\{\var{flavourPairingSequence}\patternTwo\}}{denotes an ordered partial amplitude involving the states $\{\var{stateSequence}\}$ for which each fermion \fun{f}\brace{\var{$a$}} is assigned a \emph{distinct} flavour, shared by exactly one anti-fermion \mbox{\fun{fb}\brace{\var{$b$}}}. If the flavour-pair \{\fun{f}\brace{\var{$a$}},\fun{fb}\brace{\var{$b$}}\,\} appear in \built{Position} $\{i,j\}\!\subset\![n]$ of \{\var{stateSequence}\}, then this flavour-pairing is encoded by $\{i,j\}\in\!\{$\var{flavourPairingSequence}$\}$.\\[-10pt]

\textbf{Note}: the \built{Ordering} of the pairs listed in $\{$\var{flavourPairingSequence}$\}$ encodes each fermion-line's orientation. To be in the all-plus basis of partial amplitudes, these pairs must all be ordered---that is the \built{First} element of each flavour-pair appearing in \{\var{flavourPairingSequence}\} must denote the \built{Position} of a \emph{fermion} \mbox{(\fun{f}\brace{\var{$a$}}\,)} and the \built{Last} must denote that of an \emph{anti}-fermion (\fun{fb}\brace{\var{$b$}}\,).\\[-10pt]

In addition to fermions/anti-fermions, \{\var{stateSequence}\} can include (helicity-assignment agnostic) gauge bosons encoded by \fun{g}\brace{\var{$c$}}, or helicity-specific gauge bosons---with plus/minus helicities encoded by \fun{p}\brace{\var{$c$}}\,, and \fun{m}\brace{\var{$c$}}\,, respectively. 
}

\vspace{8pt}\sectionAppendix{Formatted and Graphical Representations of Objects}{appendix:formatted_output_and_graphics}\vspace{-10pt}

\defnBox{drawChords}{\var{expression}\pattern}{draws all chord diagrams associated with a single-flavour or multi-flavour partial amplitude.
\mathematicaBox{
\mathematicaSequence[1]{egPartialAmp=\funL[1]{randomPartialAmp}\brace{3,4};\\
\funL[1]{nice}@\%}{
$\rule{0pt}{18pt}\mathcal{A}_{3}^{4}({\color{flavour1}\mathbf{\underline{1}}}\hspace{0.5pt},\hspace{-1.5pt}\mathbf{5}\hspace{0.5pt},\hspace{-1.5pt}{\color{flavour3}\mathbf{\overline{10}}}\hspace{0.5pt},\hspace{-1.5pt}{\color{flavour3}\mathbf{\underline{3}}}\hspace{0.5pt},\hspace{-1.5pt}{\color{flavour2}\mathbf{\underline{2}}}\hspace{0.5pt},\hspace{-1.5pt}\mathbf{6}\hspace{0.5pt},\hspace{-1.5pt}{\color{flavour2}\mathbf{\overline{9}}}\hspace{0.5pt},\hspace{-1.5pt}\mathbf{7}\hspace{0.5pt},\hspace{-1.5pt}{\color{flavour1}\mathbf{\overline{8}}}\hspace{0.5pt},\hspace{-1.5pt}\mathbf{4})$}\\[-18pt]
\mathematicaSequence{\funL[1]{drawChords}@egPartialAmp}{
$\begin{array}{@{}c@{}}\includegraphics[scale=0.75]{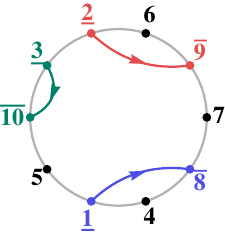}\end{array}$}\\[-20pt]
\mathematicaSequence[1]{egPartialAmp=\fun{amp}@\built{First}\brace{egPartialAmp};\\\funL[1]{nice}@\%}{
$\rule{0pt}{18pt}{A}_{3}^{4}({\color{flavour0}\mathbf{\underline{1}}}\hspace{0.5pt},\hspace{-1.5pt}\mathbf{5}\hspace{0.5pt},\hspace{-1.5pt}{\color{flavour0}\mathbf{\overline{10}}}\hspace{0.5pt},\hspace{-1.5pt}{\color{flavour0}\mathbf{\underline{3}}}\hspace{0.5pt},\hspace{-1.5pt}{\color{flavour0}\mathbf{\underline{2}}}\hspace{0.5pt},\hspace{-1.5pt}\mathbf{6}\hspace{0.5pt},\hspace{-1.5pt}{\color{flavour0}\mathbf{\overline{9}}}\hspace{0.5pt},\hspace{-1.5pt}\mathbf{7}\hspace{0.5pt},\hspace{-1.5pt}{\color{flavour0}\mathbf{\overline{8}}}\hspace{0.5pt},\hspace{-1.5pt}\mathbf{4})$}
\mathematicaSequence{\funL[1]{drawChords}@egPartialAmp}{
$\begin{array}{@{}c@{}}\includegraphics[scale=0.75]{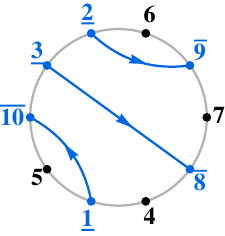}\end{array}{+}\begin{array}{@{}c@{}}\includegraphics[scale=0.75]{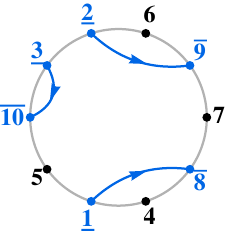}\end{array}$}
}
\textbf{Note}: if no flavour pairings exist, \fun{drawChords} will return 0:
\mathematicaBox{
\mathematicaSequence{\funL[1]{drawChords}\brace{\fun{amp}\brace{\{\fun{f}\brace{1},\fun{f}\brace{2},\fun{fb}\brace{2},\fun{fb}\brace{1}\,\},\!\{\!\{1,3\},\{2,4\}\!\}}\,}}{0
}}
\vspace{-50pt}}

\defnBox{drawColourTensors}{\var{expression}\pattern}{converts any \fun{colourTensor} objects (associated with an element of an all-plus partial amplitude basis) appearing in \var{expression} into a \emph{compact} graphical illustration of the tensor. See \mbox{section~\ref{colour_tensors}} for more details.
\mathematicaBox{
\mathematicaSequence[1]{egTensor=\fun{colourTensor}@@\built{RandomChoice}\brace{\funL[1]{partialAmpBasis}\brace{4,2}\,};\\
\funL[1]{nice}@\%}{$\rule{0pt}{18pt}\mathcal{C}[{\color{flavour1}\mathbf{\underline{1}}}\hspace{0.5pt},\hspace{-1.5pt}\mathbf{6}\hspace{0.5pt},\hspace{-1.5pt}{\color{flavour3}\mathbf{\underline{3}}}\hspace{0.5pt},\hspace{-1.5pt}\mathbf{5}\hspace{0.5pt},\hspace{-1.5pt}{\color{flavour2}\mathbf{\underline{2}}}\hspace{0.5pt},\hspace{-1.5pt}{\color{flavour2}\mathbf{\overline{9}}}\hspace{0.5pt},\hspace{-1.5pt}{\color{flavour3}\mathbf{\overline{8}}}\hspace{0.5pt},\hspace{-1.5pt}{\color{flavour4}\mathbf{\underline{4}}}\hspace{0.5pt},\hspace{-1.5pt}{\color{flavour4}\mathbf{\overline{7}}}\hspace{0.5pt},\hspace{-1.5pt}{\color{flavour1}\mathbf{\overline{10}}}]\indices{{\color{flavour1}c_1}{\color{flavour2}c_2}{\color{flavour3}c_3}{\color{flavour4}c_4}\,{\color{black}\mathfrak{g}_5}{\color{black}\mathfrak{g}_6}}{{\color{flavour4}c_7}{\color{flavour3}c_8}{\color{flavour2}c_9}{\color{flavour1}c_{10}}}$}
\mathematicaSequence{\funL[1]{drawColourTensors}\brace{egTensor}}{$\begin{array}{@{}c@{}}\includegraphics[scale=0.75]{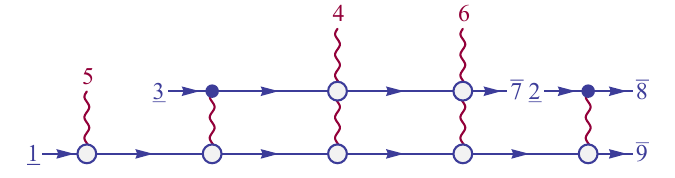}\end{array}$}
}
}

\defnBox{drawColourTensorsExpanded}{\var{expression}\pattern}{converts any \fun{colourTensor} objects (associated with an element of an all-plus partial amplitude basis) appearing in \var{expression} into a graphical illustration of the (`curly-\mbox{bracket'-)}\emph{expanded} contributions to the tensor. See \mbox{section~\ref{colour_tensors}} for more details.
\mathematicaBox{
\mathematicaSequence[1]{egTensor=\fun{colourTensor}@@\built{RandomChoice}\brace{\funL[1]{partialAmpBasis}\brace{4,2}\,};\\
\funL[1]{nice}@\%}{$\rule{0pt}{18pt}\mathcal{C}[{\color{flavour1}\mathbf{\underline{1}}}\hspace{0.5pt},\hspace{-1.5pt}\mathbf{6}\hspace{0.5pt},\hspace{-1.5pt}{\color{flavour3}\mathbf{\underline{3}}}\hspace{0.5pt},\hspace{-1.5pt}\mathbf{5}\hspace{0.5pt},\hspace{-1.5pt}{\color{flavour2}\mathbf{\underline{2}}}\hspace{0.5pt},\hspace{-1.5pt}{\color{flavour2}\mathbf{\overline{9}}}\hspace{0.5pt},\hspace{-1.5pt}{\color{flavour3}\mathbf{\overline{8}}}\hspace{0.5pt},\hspace{-1.5pt}{\color{flavour4}\mathbf{\underline{4}}}\hspace{0.5pt},\hspace{-1.5pt}{\color{flavour4}\mathbf{\overline{7}}}\hspace{0.5pt},\hspace{-1.5pt}{\color{flavour1}\mathbf{\overline{10}}}]\indices{{\color{flavour1}c_1}{\color{flavour2}c_2}{\color{flavour3}c_3}{\color{flavour4}c_4}\,{\color{black}\mathfrak{g}_5}{\color{black}\mathfrak{g}_6}}{{\color{flavour4}c_7}{\color{flavour3}c_8}{\color{flavour2}c_9}{\color{flavour1}c_{10}}}$}
\mathematicaSequence{\funL[1]{drawColourTensors}\brace{egTensor}}{$\begin{array}{@{}c@{}}\includegraphics[scale=0.75]{draw_colour_tensors_doc_eg}\end{array}$}
\mathematicaSequence{\funL[1]{drawColourTensorsExpanded}\brace{egTensor}}{$\hspace{-10pt}\begin{array}{@{}c@{}}\includegraphics[scale=0.6]{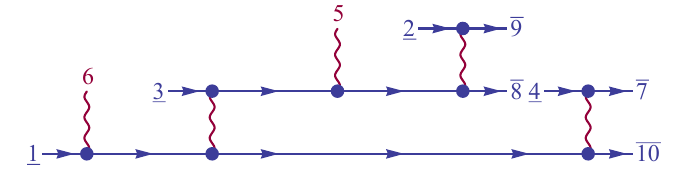}\end{array}\hspace{-2pt}{+}\hspace{-7.5pt}\begin{array}{@{}c@{}}\includegraphics[scale=0.6]{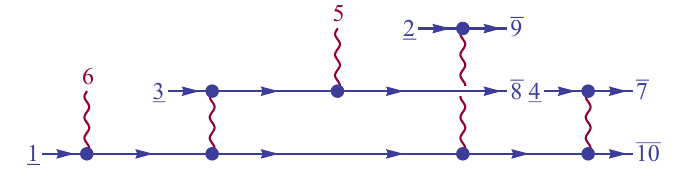}\end{array}$\\
\mbox{}$\hspace{-10pt}{+}\hspace{-10pt}\begin{array}{@{}c@{}}\includegraphics[scale=0.6]{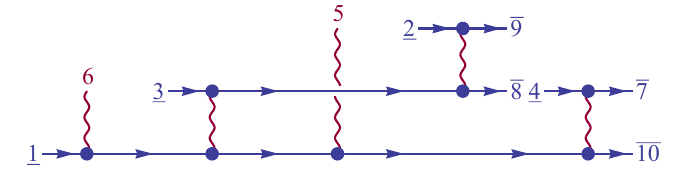}\end{array}\hspace{-2pt}{+}\hspace{-7.5pt}\begin{array}{@{}c@{}}\includegraphics[scale=0.6]{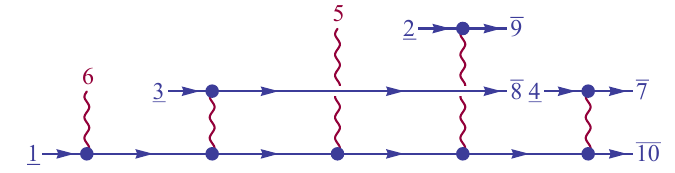}\end{array}$\\
}
}
}

\defnBox{nice}{\var{expression}\pattern}{formats expressions---especially those involving protected symbols---in a way that is more human-readable.}

\defnBox{writeAsCurlyBrackets}{\var{expression}\pattern}{converts any \fun{colourTensor} objects (associated with an element of an all-plus partial amplitude basis) appearing in \var{expression} into a symbolic (machine-readable) expression involving \emph{nested} `curly-brackets'. See \mbox{section~\ref{colour_tensors}} for more details.
\mathematicaBox{
\mathematicaSequence[1]{egTensor=\mbox{\fun{colourTensor}[\{\fun{f}\brace{\hspace{-1pt}1\hspace{-1pt}},\fun{g}\brace{\hspace{-1pt}5\hspace{-1pt}},\fun{f}\brace{\hspace{-1pt}3\hspace{-1pt}},\fun{g}\brace{\hspace{-1pt}4\hspace{-1pt}},\fun{g}\brace{\hspace{-1pt}6\hspace{-1pt}},\fun{fb}\brace{\hspace{-1pt}3\hspace{-1pt}},\fun{f}\brace{\hspace{-1pt}2\hspace{-1pt}},\fun{fb}\brace{\hspace{-1pt}2\hspace{-1pt}},\fun{fb}\brace{\hspace{-1pt}1\hspace{-1pt}}\},}\\\mbox{}\hspace{130pt}\;\;\!\{\!\{1\!,\!9\}\!,\!\{7\!,\!8\}\!,\!\{3\!,\!6\}\!\}];\\
\funL[]{drawColourTensors}@egTensor}{$\begin{array}{@{}c@{}}\includegraphics[scale=0.8]{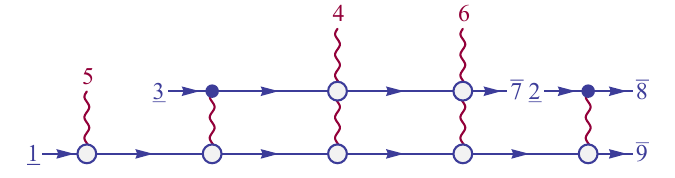}\end{array}\vspace{-10pt}$}
\mathematicaSequence{
\funL[1]{nice}@\funL[1]{writeAsCurlyBrackets}\brace{egTensor}}{\rule{0pt}{15pt}\raisebox{-2pt}{$\begin{array}{@{}c@{}}\includegraphics[scale=1]{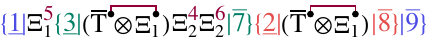}\end{array}$}}
}
}

\defnBox{writeAsCurlyBracketsExpanded}{\var{expression}\pattern}{converts any \fun{colourTensor} objects (associated with an element of an all-plus partial amplitude basis) appearing in \var{expression} into a symbolic (machine-readable) expression involving \emph{expanded} `curly-brackets'. See \mbox{section~\ref{colour_tensors}} for more details.
\mathematicaBox{
\mathematicaSequence[1]{egTensor=\mbox{\fun{colourTensor}[\{\fun{f}\brace{\hspace{-1pt}1\hspace{-1pt}},\fun{g}\brace{\hspace{-1pt}5\hspace{-1pt}},\fun{f}\brace{\hspace{-1pt}3\hspace{-1pt}},\fun{g}\brace{\hspace{-1pt}4\hspace{-1pt}},\fun{g}\brace{\hspace{-1pt}6\hspace{-1pt}},\fun{fb}\brace{\hspace{-1pt}3\hspace{-1pt}},\fun{f}\brace{\hspace{-1pt}2\hspace{-1pt}},\fun{fb}\brace{\hspace{-1pt}2\hspace{-1pt}},\fun{fb}\brace{\hspace{-1pt}1\hspace{-1pt}}\},}\\\mbox{}\hspace{130pt}\;\;\!\{\!\{1\!,\!9\}\!,\!\{7\!,\!8\}\!,\!\{3\!,\!6\}\!\}];\\
\funL[]{drawColourTensors}@egTensor}{$\begin{array}{@{}c@{}}\includegraphics[scale=0.8]{eg_colour_tensor_1}\end{array}\vspace{-10pt}$}
\mathematicaSequence{\funL[1]{nice}@\funL[1]{writeAsCurlyBracketsExpanded}\brace{egTensor}}{$\begin{array}{@{}c@{}}\includegraphics[scale=0.8]{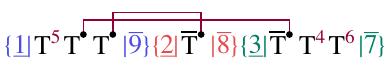}\end{array}{+}\begin{array}{@{}c@{}}\includegraphics[scale=0.8]{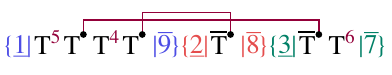}\end{array}$\\[-10pt]
$\fwboxR{0pt}{{+}}\begin{array}{@{}c@{}}\includegraphics[scale=0.8]{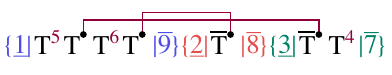}\end{array}{+}\begin{array}{@{}c@{}}\includegraphics[scale=0.8]{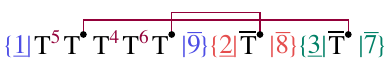}\end{array}$}
}
}

\defnBox{writeAsSUNcFundamentalTensors}{\var{expression}\pattern}{writes any \fun{colourTensor} objects as products of generators and $\delta$ tensors.
\mathematicaBox{
\mathematicaSequence[1]{egTensor=\mbox{\fun{colourTensor}[\{\fun{f}\brace{\hspace{-1pt}1\hspace{-1pt}},\fun{g}\brace{\hspace{-1pt}5\hspace{-1pt}},\fun{f}\brace{\hspace{-1pt}3\hspace{-1pt}},\fun{g}\brace{\hspace{-1pt}4\hspace{-1pt}},\fun{g}\brace{\hspace{-1pt}6\hspace{-1pt}},\fun{fb}\brace{\hspace{-1pt}3\hspace{-1pt}},\fun{f}\brace{\hspace{-1pt}2\hspace{-1pt}},\fun{fb}\brace{\hspace{-1pt}2\hspace{-1pt}},\fun{fb}\brace{\hspace{-1pt}1\hspace{-1pt}}\},}\\\mbox{}\hspace{130pt}\;\;\!\{\!\{1\!,\!9\}\!,\!\{7\!,\!8\}\!,\!\{3\!,\!6\}\!\}];\\
\funL[]{drawColourTensors}@egTensor}{$\begin{array}{@{}c@{}}\includegraphics[scale=0.8]{eg_colour_tensor_1}\end{array}\vspace{-10pt}$}
\mathematicaSequence{\funL[]{nice}@\funL[1]{writeAsSUNcFundamentalTensors}\brace{egTensor}}{$\rule{0pt}{10pt}
\hspace{0pt}(\!\mathbf{T}^{{\color{hred}5}}\!)\!\indices{{\color{flavour1}{1}}}{{\color{flavour3}\overline{7}}}\delta\indices{{\color{flavour2}{2}}}{{\color{flavour1}\overline{9}}}(\!\mathbf{T}^{{\color{hred}4}}\mathbf{T}^{{\color{hred}6}}\!)\!\indices{{\color{flavour3}{3}}}{{\color{flavour2}\overline{8}}}{+}(\!\mathbf{T}^{{\color{hred}5}}\mathbf{T}^{{\color{hred}4}}\!)\!\indices{{\color{flavour1}{1}}}{{\color{flavour3}\overline{7}}}\delta\indices{{\color{flavour2}{2}}}{{\color{flavour1}\overline{9}}}(\!\mathbf{T}^{{\color{hred}6}}\!)\!\indices{{\color{flavour3}{3}}}{{\color{flavour2}\overline{8}}}{+}(\!\mathbf{T}^{{\color{hred}5}}\mathbf{T}^{{\color{hred}6}}\!)\!\indices{{\color{flavour1}{1}}}{{\color{flavour3}\overline{7}}}\delta\indices{{\color{flavour2}{2}}}{{\color{flavour1}\overline{9}}}(\!\mathbf{T}^{{\color{hred}4}}\!)\!\indices{{\color{flavour3}{3}}}{{\color{flavour2}\overline{8}}}{+}(\!\mathbf{T}^{{\color{hred}5}}\mathbf{T}^{{\color{hred}4}}\mathbf{T}^{{\color{hred}6}}\!)\!\indices{{\color{flavour1}{1}}}{{\color{flavour3}\overline{7}}}\delta\indices{{\color{flavour2}{2}}}{{\color{flavour1}\overline{9}}}\delta\indices{{\color{flavour3}{3}}}{{\color{flavour2}\overline{8}}}$\\[2pt]
\mbox{}\hspace{-19pt}${-}\frac{1}{N_c}\Big[(\!\mathbf{T}^{{\color{hred}5}}\!)\!\indices{{\color{flavour1}{1}}}{{\color{flavour3}\overline{7}}}\delta\indices{{\color{flavour2}{2}}}{{\color{flavour2}\overline{8}}}(\!\mathbf{T}^{{\color{hred}4}}\mathbf{T}^{{\color{hred}6}}\!)\!\indices{{\color{flavour3}{3}}}{{\color{flavour1}\overline{9}}}{+}(\!\mathbf{T}^{{\color{hred}5}}\!)\!\indices{{\color{flavour1}{1}}}{{\color{flavour2}\overline{8}}}\delta\indices{{\color{flavour2}{2}}}{{\color{flavour1}\overline{9}}}(\!\mathbf{T}^{{\color{hred}4}}\mathbf{T}^{{\color{hred}6}}\!)\!\indices{{\color{flavour3}{3}}}{{\color{flavour3}\overline{7}}}{+}(\!\mathbf{T}^{{\color{hred}5}}\mathbf{T}^{{\color{hred}4}}\!)\!\indices{{\color{flavour1}{1}}}{{\color{flavour3}\overline{7}}}\delta\indices{{\color{flavour2}{2}}}{{\color{flavour2}\overline{8}}}(\!\mathbf{T}^{{\color{hred}6}}\!)\!\indices{{\color{flavour3}{3}}}{{\color{flavour1}\overline{9}}}{+}(\!\mathbf{T}^{{\color{hred}5}}\mathbf{T}^{{\color{hred}4}}\!)\!\indices{{\color{flavour1}{1}}}{{\color{flavour2}\overline{8}}}\delta\indices{{\color{flavour2}{2}}}{{\color{flavour1}\overline{9}}}(\!\mathbf{T}^{{\color{hred}6}}\!)\!\indices{{\color{flavour3}{3}}}{{\color{flavour3}\overline{7}}}$\\
\mbox{}\hspace{-19pt}$\mbox{}\hspace{24pt}{+}(\!\mathbf{T}^{{\color{hred}5}}\mathbf{T}^{{\color{hred}6}}\!)\!\indices{{\color{flavour1}{1}}}{{\color{flavour3}\overline{7}}}\delta\indices{{\color{flavour2}{2}}}{{\color{flavour2}\overline{8}}}(\!\mathbf{T}^{{\color{hred}4}}\!)\!\indices{{\color{flavour3}{3}}}{{\color{flavour1}\overline{9}}}{+}(\!\mathbf{T}^{{\color{hred}5}}\mathbf{T}^{{\color{hred}6}}\!)\!\indices{{\color{flavour1}{1}}}{{\color{flavour2}\overline{8}}}\delta\indices{{\color{flavour2}{2}}}{{\color{flavour1}\overline{9}}}(\!\mathbf{T}^{{\color{hred}4}}\!)\!\indices{{\color{flavour3}{3}}}{{\color{flavour3}\overline{7}}}{+}(\!\mathbf{T}^{{\color{hred}5}}\mathbf{T}^{{\color{hred}4}}\mathbf{T}^{{\color{hred}6}}\!)\!\indices{{\color{flavour1}{1}}}{{\color{flavour3}\overline{7}}}\delta\indices{{\color{flavour2}{2}}}{{\color{flavour2}\overline{8}}}\delta\indices{{\color{flavour3}{3}}}{{\color{flavour1}\overline{9}}}{+}(\!\mathbf{T}^{{\color{hred}5}}\mathbf{T}^{{\color{hred}4}}\mathbf{T}^{{\color{hred}6}}\!)\!\indices{{\color{flavour1}{1}}}{{\color{flavour2}\overline{8}}}\delta\indices{{\color{flavour2}{2}}}{{\color{flavour1}\overline{9}}}\delta\indices{{\color{flavour3}{3}}}{{\color{flavour3}\overline{7}}}\Big]$\\
\mbox{}\hspace{-19pt}${+}\frac{1}{N_c^2}\Big[(\!\mathbf{T}^{{\color{hred}5}}\!)\!\indices{{\color{flavour1}{1}}}{{\color{flavour1}\overline{9}}}\delta\indices{{\color{flavour2}{2}}}{{\color{flavour2}\overline{8}}}(\!\mathbf{T}^{{\color{hred}4}}\mathbf{T}^{{\color{hred}6}}\!)\!\indices{{\color{flavour3}{3}}}{{\color{flavour3}\overline{7}}}{+}(\!\mathbf{T}^{{\color{hred}5}}\mathbf{T}^{{\color{hred}4}}\!)\!\indices{{\color{flavour1}{1}}}{{\color{flavour1}\overline{9}}}\delta\indices{{\color{flavour2}{2}}}{{\color{flavour2}\overline{8}}}(\!\mathbf{T}^{{\color{hred}6}}\!)\!\indices{{\color{flavour3}{3}}}{{\color{flavour3}\overline{7}}}{+}(\!\mathbf{T}^{{\color{hred}5}}\mathbf{T}^{{\color{hred}6}}\!)\!\indices{{\color{flavour1}{1}}}{{\color{flavour1}\overline{9}}}\delta\indices{{\color{flavour2}{2}}}{{\color{flavour2}\overline{8}}}(\!\mathbf{T}^{{\color{hred}4}}\!)\!\indices{{\color{flavour3}{3}}}{{\color{flavour3}\overline{7}}}{+}(\!\mathbf{T}^{{\color{hred}5}}\mathbf{T}^{{\color{hred}4}}\mathbf{T}^{{\color{hred}6}}\!)\!\indices{{\color{flavour1}{1}}}{{\color{flavour1}\overline{9}}}\delta\indices{{\color{flavour2}{2}}}{{\color{flavour2}\overline{8}}}\delta\indices{{\color{flavour3}{3}}}{{\color{flavour3}\overline{7}}}\Big]$}
}
}

\vspace{10pt}\sectionAppendix{Amplitudes for Charged Fermions and Gauge Bosons}{appendix:partial_amplitude_bases}\vspace{-10pt}

\vspace{10pt}\subsectionAppendix{Partial Amplitudes Involving Charged Matter and Gauge Bosons}{appendix:partial_amplitude_manipulations}\vspace{-10pt}

\defnBox[2]{partialAmpBasis}{\var{nFermionLines}\pattern}{returns the set of partial amplitudes of the \emph{all-plus} Melia basis (\ref{all_plus_basis_of_partial_amps}) for $n_f{=}$\var{nFermionLines} and $n_g{=}0$.
\mathematicaBox{
\mathematicaSequence{\funL[1]{nice}@\funL[1]{partialAmpBasis}\brace{3}}{
$\hspace{-6pt}\big\{\!\mathcal{A}_{3}({\color{flavour1}\mathbf{\underline{1}}}\hspace{0.5pt},\hspace{-1.5pt}{\color{flavour2}\mathbf{\underline{2}}}\hspace{0.5pt},\hspace{-1.5pt}{\color{flavour3}\mathbf{\underline{3}}}\hspace{0.5pt},\hspace{-1.5pt}{\color{flavour3}\mathbf{\overline{4}}}\hspace{0.5pt},\hspace{-1.5pt}{\color{flavour2}\mathbf{\overline{5}}}\hspace{0.5pt},\hspace{-1.5pt}{\color{flavour1}\mathbf{\overline{6}}}),\mathcal{A}_{3}({\color{flavour1}\mathbf{\underline{1}}}\hspace{0.5pt},\hspace{-1.5pt}{\color{flavour2}\mathbf{\underline{2}}}\hspace{0.5pt},\hspace{-1.5pt}{\color{flavour2}\mathbf{\overline{5}}}\hspace{0.5pt},\hspace{-1.5pt}{\color{flavour3}\mathbf{\underline{3}}}\hspace{0.5pt},\hspace{-1.5pt}{\color{flavour3}\mathbf{\overline{4}}}\hspace{0.5pt},\hspace{-1.5pt}{\color{flavour1}\mathbf{\overline{6}}}),\mathcal{A}_{3}({\color{flavour1}\mathbf{\underline{1}}}\hspace{0.5pt},\hspace{-1.5pt}{\color{flavour3}\mathbf{\underline{3}}}\hspace{0.5pt},\hspace{-1.5pt}{\color{flavour2}\mathbf{\underline{2}}}\hspace{0.5pt},\hspace{-1.5pt}{\color{flavour2}\mathbf{\overline{5}}}\hspace{0.5pt},\hspace{-1.5pt}{\color{flavour3}\mathbf{\overline{4}}}\hspace{0.5pt},\hspace{-1.5pt}{\color{flavour1}\mathbf{\overline{6}}}),\mathcal{A}_{3}({\color{flavour1}\mathbf{\underline{1}}}\hspace{0.5pt},\hspace{-1.5pt}{\color{flavour3}\mathbf{\underline{3}}}\hspace{0.5pt},\hspace{-1.5pt}{\color{flavour3}\mathbf{\overline{4}}}\hspace{0.5pt},\hspace{-1.5pt}{\color{flavour2}\mathbf{\underline{2}}}\hspace{0.5pt},\hspace{-1.5pt}{\color{flavour2}\mathbf{\overline{5}}}\hspace{0.5pt},\hspace{-1.5pt}{\color{flavour1}\mathbf{\overline{6}}})\big\}$
}}
}

\defnBox[3]{partialAmpBasis}{\var{nFermionLines}\pattern,\var{nGaugeBosons}\pattern}{returns the set of partial amplitudes of the \emph{all-plus} Melia basis (\ref{all_plus_basis_of_partial_amps}) for $n_f{=}$\var{nFermionLines} and $n_g{=}$\var{nGaugeBosons}.
\mathematicaBox{
\mathematicaSequence{\funL[1]{nice}@\funL[1]{partialAmpBasis}\brace{2,2}}{
$\hspace{-6pt}\big\{\!\mathcal{A}_{2}^{2}({\color{flavour1}\mathbf{\underline{1}}}\hspace{0.5pt},\hspace{-1.5pt}\mathbf{3}\hspace{0.5pt},\hspace{-1.5pt}\mathbf{4}\hspace{0.5pt},\hspace{-1.5pt}{\color{flavour2}\mathbf{\underline{2}}}\hspace{0.5pt},\hspace{-1.5pt}{\color{flavour2}\mathbf{\overline{5}}}\hspace{0.5pt},\hspace{-1.5pt}{\color{flavour1}\mathbf{\overline{6}}}),\mathcal{A}_{2}^{2}({\color{flavour1}\mathbf{\underline{1}}}\hspace{0.5pt},\hspace{-1.5pt}\mathbf{3}\hspace{0.5pt},\hspace{-1.5pt}{\color{flavour2}\mathbf{\underline{2}}}\hspace{0.5pt},\hspace{-1.5pt}\mathbf{4}\hspace{0.5pt},\hspace{-1.5pt}{\color{flavour2}\mathbf{\overline{5}}}\hspace{0.5pt},\hspace{-1.5pt}{\color{flavour1}\mathbf{\overline{6}}}),\mathcal{A}_{2}^{2}({\color{flavour1}\mathbf{\underline{1}}}\hspace{0.5pt},\hspace{-1.5pt}\mathbf{3}\hspace{0.5pt},\hspace{-1.5pt}{\color{flavour2}\mathbf{\underline{2}}}\hspace{0.5pt},\hspace{-1.5pt}{\color{flavour2}\mathbf{\overline{5}}}\hspace{0.5pt},\hspace{-1.5pt}\mathbf{4}\hspace{0.5pt},\hspace{-1.5pt}{\color{flavour1}\mathbf{\overline{6}}}),\mathcal{A}_{2}^{2}({\color{flavour1}\mathbf{\underline{1}}}\hspace{0.5pt},\hspace{-1.5pt}{\color{flavour2}\mathbf{\underline{2}}}\hspace{0.5pt},\hspace{-1.5pt}\mathbf{3}\hspace{0.5pt},\hspace{-1.5pt}\mathbf{4}\hspace{0.5pt},\hspace{-1.5pt}{\color{flavour2}\mathbf{\overline{5}}}\hspace{0.5pt},\hspace{-1.5pt}{\color{flavour1}\mathbf{\overline{6}}}),\\
\mathcal{A}_{2}^{2}({\color{flavour1}\mathbf{\underline{1}}}\hspace{0.5pt},\hspace{-1.5pt}{\color{flavour2}\mathbf{\underline{2}}}\hspace{0.5pt},\hspace{-1.5pt}\mathbf{3}\hspace{0.5pt},\hspace{-1.5pt}{\color{flavour2}\mathbf{\overline{5}}}\hspace{0.5pt},\hspace{-1.5pt}\mathbf{4}\hspace{0.5pt},\hspace{-1.5pt}{\color{flavour1}\mathbf{\overline{6}}}),\mathcal{A}_{2}^{2}({\color{flavour1}\mathbf{\underline{1}}}\hspace{0.5pt},\hspace{-1.5pt}{\color{flavour2}\mathbf{\underline{2}}}\hspace{0.5pt},\hspace{-1.5pt}{\color{flavour2}\mathbf{\overline{5}}}\hspace{0.5pt},\hspace{-1.5pt}\mathbf{3}\hspace{0.5pt},\hspace{-1.5pt}\mathbf{4}\hspace{0.5pt},\hspace{-1.5pt}{\color{flavour1}\mathbf{\overline{6}}}),\mathcal{A}_{2}^{2}({\color{flavour1}\mathbf{\underline{1}}}\hspace{0.5pt},\hspace{-1.5pt}\mathbf{4}\hspace{0.5pt},\hspace{-1.5pt}\mathbf{3}\hspace{0.5pt},\hspace{-1.5pt}{\color{flavour2}\mathbf{\underline{2}}}\hspace{0.5pt},\hspace{-1.5pt}{\color{flavour2}\mathbf{\overline{5}}}\hspace{0.5pt},\hspace{-1.5pt}{\color{flavour1}\mathbf{\overline{6}}}),\mathcal{A}_{2}^{2}({\color{flavour1}\mathbf{\underline{1}}}\hspace{0.5pt},\hspace{-1.5pt}\mathbf{4}\hspace{0.5pt},\hspace{-1.5pt}{\color{flavour2}\mathbf{\underline{2}}}\hspace{0.5pt},\hspace{-1.5pt}\mathbf{3}\hspace{0.5pt},\hspace{-1.5pt}{\color{flavour2}\mathbf{\overline{5}}}\hspace{0.5pt},\hspace{-1.5pt}{\color{flavour1}\mathbf{\overline{6}}}),\\
\mathcal{A}_{2}^{2}({\color{flavour1}\mathbf{\underline{1}}}\hspace{0.5pt},\hspace{-1.5pt}\mathbf{4}\hspace{0.5pt},\hspace{-1.5pt}{\color{flavour2}\mathbf{\underline{2}}}\hspace{0.5pt},\hspace{-1.5pt}{\color{flavour2}\mathbf{\overline{5}}}\hspace{0.5pt},\hspace{-1.5pt}\mathbf{3}\hspace{0.5pt},\hspace{-1.5pt}{\color{flavour1}\mathbf{\overline{6}}}),\mathcal{A}_{2}^{2}({\color{flavour1}\mathbf{\underline{1}}}\hspace{0.5pt},\hspace{-1.5pt}{\color{flavour2}\mathbf{\underline{2}}}\hspace{0.5pt},\hspace{-1.5pt}\mathbf{4}\hspace{0.5pt},\hspace{-1.5pt}\mathbf{3}\hspace{0.5pt},\hspace{-1.5pt}{\color{flavour2}\mathbf{\overline{5}}}\hspace{0.5pt},\hspace{-1.5pt}{\color{flavour1}\mathbf{\overline{6}}}),\mathcal{A}_{2}^{2}({\color{flavour1}\mathbf{\underline{1}}}\hspace{0.5pt},\hspace{-1.5pt}{\color{flavour2}\mathbf{\underline{2}}}\hspace{0.5pt},\hspace{-1.5pt}\mathbf{4}\hspace{0.5pt},\hspace{-1.5pt}{\color{flavour2}\mathbf{\overline{5}}}\hspace{0.5pt},\hspace{-1.5pt}\mathbf{3}\hspace{0.5pt},\hspace{-1.5pt}{\color{flavour1}\mathbf{\overline{6}}}),\mathcal{A}_{2}^{2}({\color{flavour1}\mathbf{\underline{1}}}\hspace{0.5pt},\hspace{-1.5pt}{\color{flavour2}\mathbf{\underline{2}}}\hspace{0.5pt},\hspace{-1.5pt}{\color{flavour2}\mathbf{\overline{5}}}\hspace{0.5pt},\hspace{-1.5pt}\mathbf{4}\hspace{0.5pt},\hspace{-1.5pt}\mathbf{3}\hspace{0.5pt},\hspace{-1.5pt}{\color{flavour1}\mathbf{\overline{6}}})\big\}$
}}
}

\defnBox[3]{partialAmpBasis}{\var{nFermionLines}\pattern,\var{nPlus}\pattern,\var{nMinus}\pattern}{returns the set of partial amplitudes of the \emph{all-plus} Melia basis (\ref{all_plus_basis_of_partial_amps}) for $n_f{=}$\var{nFermionLines} and $n_g{=}$(\var{nPlus}{+}\var{nMinus}). By convention, the plus-helicity gauge bosons are associated with momentum indices $([\var{\textsl{nPlus}}]{+}n_f)$ while those with minus-helicity are assigned momentum indices $([\var{\textsl{nMinus}}]{+}n_f{+}\var{\textsl{nPlus}})$.
\mathematicaBox{
\mathematicaSequence{\funL[1]{nice}@\funL[1]{partialAmpBasis}\brace{1,2,1}}{
$\hspace{-6pt}\big\{\!\mathcal{A}_{1}^{(2,1)}({\color{flavour1}\mathbf{\underline{1}}}\hspace{0.5pt},\hspace{-1.5pt}\mathbf{2^{{+}}}\hspace{-3pt}\!\hspace{0.5pt},\hspace{-1.5pt}\mathbf{3^{{+}}}\hspace{-3pt}\!\hspace{0.5pt},\hspace{-1.5pt}\mathbf{4^{{-}}}\!\hspace{-3pt}\hspace{0.5pt},\hspace{-1.5pt}{\color{flavour1}\mathbf{\overline{5}}}),\mathcal{A}_{1}^{(2,1)}({\color{flavour1}\mathbf{\underline{1}}}\hspace{0.5pt},\hspace{-1.5pt}\mathbf{2^{{+}}}\hspace{-3pt}\!\hspace{0.5pt},\hspace{-1.5pt}\mathbf{4^{{-}}}\!\hspace{-3pt}\hspace{0.5pt},\hspace{-1.5pt}\mathbf{3^{{+}}}\hspace{-3pt}\!\hspace{0.5pt},\hspace{-1.5pt}{\color{flavour1}\mathbf{\overline{5}}}),\mathcal{A}_{1}^{(2,1)}({\color{flavour1}\mathbf{\underline{1}}}\hspace{0.5pt},\hspace{-1.5pt}\mathbf{3^{{+}}}\hspace{-3pt}\!\hspace{0.5pt},\hspace{-1.5pt}\mathbf{2^{{+}}}\hspace{-3pt}\!\hspace{0.5pt},\hspace{-1.5pt}\mathbf{4^{{-}}}\!\hspace{-3pt}\hspace{0.5pt},\hspace{-1.5pt}{\color{flavour1}\mathbf{\overline{5}}}),\\
\phantom{\big\{}\hspace{-7pt}\mathcal{A}_{1}^{(2,1)}({\color{flavour1}\mathbf{\underline{1}}}\hspace{0.5pt},\hspace{-1.5pt}\mathbf{3^{{+}}}\hspace{-3pt}\!\hspace{0.5pt},\hspace{-1.5pt}\mathbf{4^{{-}}}\!\hspace{-3pt}\hspace{0.5pt},\hspace{-1.5pt}\mathbf{2^{{+}}}\hspace{-3pt}\!\hspace{0.5pt},\hspace{-1.5pt}{\color{flavour1}\mathbf{\overline{5}}}),\mathcal{A}_{1}^{(2,1)}({\color{flavour1}\mathbf{\underline{1}}}\hspace{0.5pt},\hspace{-1.5pt}\mathbf{4^{{-}}}\!\hspace{-3pt}\hspace{0.5pt},\hspace{-1.5pt}\mathbf{2^{{+}}}\hspace{-3pt}\!\hspace{0.5pt},\hspace{-1.5pt}\mathbf{3^{{+}}}\hspace{-3pt}\!\hspace{0.5pt},\hspace{-1.5pt}{\color{flavour1}\mathbf{\overline{5}}}),\mathcal{A}_{1}^{(2,1)}({\color{flavour1}\mathbf{\underline{1}}}\hspace{0.5pt},\hspace{-1.5pt}\mathbf{4^{{-}}}\!\hspace{-3pt}\hspace{0.5pt},\hspace{-1.5pt}\mathbf{3^{{+}}}\hspace{-3pt}\!\hspace{0.5pt},\hspace{-1.5pt}\mathbf{2^{{+}}}\hspace{-3pt}\!\hspace{0.5pt},\hspace{-1.5pt}{\color{flavour1}\mathbf{\overline{5}}})\big\}$
}}
}

\defnBox[3]{partialAmpBasis}{\var{nFermionLines}\pattern,\{\var{bosonHelicitySequence}\patternTwo\}}{returns the set of partial amplitudes of the \emph{all-plus} Melia basis (\ref{all_plus_basis_of_partial_amps}) for $n_f{=}$\var{nFermionLines} and $n_g{=}$\built{Length}\brace{\var{bosonHelicitySequence}}\,. In contrast to simply specifying the number of plus/minus-helicity bosons, the momentum indices of the bosons are ordered according to \{\var{bosonHelicitySequence}\}.
\mathematicaBox{
\mathematicaSequence{\funL[1]{nice}@\funL[1]{partialAmpBasis}\brace{2,\{\fun{m},\fun{p}\}}}{$\hspace{-6pt}\big\{\!\mathcal{A}_{2}^{(1,1)}({\color{flavour1}\mathbf{\underline{1}}}\hspace{0.5pt},\hspace{-1.5pt}\mathbf{3^{{-}}}\!\hspace{-3pt}\hspace{0.5pt},\hspace{-1.5pt}\mathbf{4^{{+}}}\hspace{-3pt}\!\hspace{0.5pt},\hspace{-1.5pt}{\color{flavour2}\mathbf{\underline{2}}}\hspace{0.5pt},\hspace{-1.5pt}{\color{flavour2}\mathbf{\overline{5}}}\hspace{0.5pt},\hspace{-1.5pt}{\color{flavour1}\mathbf{\overline{6}}}),\mathcal{A}_{2}^{(1,1)}({\color{flavour1}\mathbf{\underline{1}}}\hspace{0.5pt},\hspace{-1.5pt}\mathbf{3^{{-}}}\!\hspace{-3pt}\hspace{0.5pt},\hspace{-1.5pt}{\color{flavour2}\mathbf{\underline{2}}}\hspace{0.5pt},\hspace{-1.5pt}\mathbf{4^{{+}}}\hspace{-3pt}\!\hspace{0.5pt},\hspace{-1.5pt}{\color{flavour2}\mathbf{\overline{5}}}\hspace{0.5pt},\hspace{-1.5pt}{\color{flavour1}\mathbf{\overline{6}}}),\mathcal{A}_{2}^{(1,1)}({\color{flavour1}\mathbf{\underline{1}}}\hspace{0.5pt},\hspace{-1.5pt}\mathbf{3^{{-}}}\!\hspace{-3pt}\hspace{0.5pt},\hspace{-1.5pt}{\color{flavour2}\mathbf{\underline{2}}}\hspace{0.5pt},\hspace{-1.5pt}{\color{flavour2}\mathbf{\overline{5}}}\hspace{0.5pt},\hspace{-1.5pt}\mathbf{4^{{+}}}\hspace{-3pt}\!\hspace{0.5pt},\hspace{-1.5pt}{\color{flavour1}\mathbf{\overline{6}}}),\\\mathcal{A}_{2}^{(1,1)}({\color{flavour1}\mathbf{\underline{1}}}\hspace{0.5pt},\hspace{-1.5pt}{\color{flavour2}\mathbf{\underline{2}}}\hspace{0.5pt},\hspace{-1.5pt}\mathbf{3^{{-}}}\!\hspace{-3pt}\hspace{0.5pt},\hspace{-1.5pt}\mathbf{4^{{+}}}\hspace{-3pt}\!\hspace{0.5pt},\hspace{-1.5pt}{\color{flavour2}\mathbf{\overline{5}}}\hspace{0.5pt},\hspace{-1.5pt}{\color{flavour1}\mathbf{\overline{6}}}),\mathcal{A}_{2}^{(1,1)}({\color{flavour1}\mathbf{\underline{1}}}\hspace{0.5pt},\hspace{-1.5pt}{\color{flavour2}\mathbf{\underline{2}}}\hspace{0.5pt},\hspace{-1.5pt}\mathbf{3^{{-}}}\!\hspace{-3pt}\hspace{0.5pt},\hspace{-1.5pt}{\color{flavour2}\mathbf{\overline{5}}}\hspace{0.5pt},\hspace{-1.5pt}\mathbf{4^{{+}}}\hspace{-3pt}\!\hspace{0.5pt},\hspace{-1.5pt}{\color{flavour1}\mathbf{\overline{6}}}),\mathcal{A}_{2}^{(1,1)}({\color{flavour1}\mathbf{\underline{1}}}\hspace{0.5pt},\hspace{-1.5pt}{\color{flavour2}\mathbf{\underline{2}}}\hspace{0.5pt},\hspace{-1.5pt}{\color{flavour2}\mathbf{\overline{5}}}\hspace{0.5pt},\hspace{-1.5pt}\mathbf{3^{{-}}}\!\hspace{-3pt}\hspace{0.5pt},\hspace{-1.5pt}\mathbf{4^{{+}}}\hspace{-3pt}\!\hspace{0.5pt},\hspace{-1.5pt}{\color{flavour1}\mathbf{\overline{6}}}),\\[3pt]\mathcal{A}_{2}^{(1,1)}({\color{flavour1}\mathbf{\underline{1}}}\hspace{0.5pt},\hspace{-1.5pt}\mathbf{4^{{+}}}\hspace{-3pt}\!\hspace{0.5pt},\hspace{-1.5pt}\mathbf{3^{{-}}}\!\hspace{-3pt}\hspace{0.5pt},\hspace{-1.5pt}{\color{flavour2}\mathbf{\underline{2}}}\hspace{0.5pt},\hspace{-1.5pt}{\color{flavour2}\mathbf{\overline{5}}}\hspace{0.5pt},\hspace{-1.5pt}{\color{flavour1}\mathbf{\overline{6}}}),\mathcal{A}_{2}^{(1,1)}({\color{flavour1}\mathbf{\underline{1}}}\hspace{0.5pt},\hspace{-1.5pt}\mathbf{4^{{+}}}\hspace{-3pt}\!\hspace{0.5pt},\hspace{-1.5pt}{\color{flavour2}\mathbf{\underline{2}}}\hspace{0.5pt},\hspace{-1.5pt}\mathbf{3^{{-}}}\!\hspace{-3pt}\hspace{0.5pt},\hspace{-1.5pt}{\color{flavour2}\mathbf{\overline{5}}}\hspace{0.5pt},\hspace{-1.5pt}{\color{flavour1}\mathbf{\overline{6}}}),\mathcal{A}_{2}^{(1,1)}({\color{flavour1}\mathbf{\underline{1}}}\hspace{0.5pt},\hspace{-1.5pt}\mathbf{4^{{+}}}\hspace{-3pt}\!\hspace{0.5pt},\hspace{-1.5pt}{\color{flavour2}\mathbf{\underline{2}}}\hspace{0.5pt},\hspace{-1.5pt}{\color{flavour2}\mathbf{\overline{5}}}\hspace{0.5pt},\hspace{-1.5pt}\mathbf{3^{{-}}}\!\hspace{-3pt}\hspace{0.5pt},\hspace{-1.5pt}{\color{flavour1}\mathbf{\overline{6}}}),\\[3pt]\mathcal{A}_{2}^{(1,1)}({\color{flavour1}\mathbf{\underline{1}}}\hspace{0.5pt},\hspace{-1.5pt}{\color{flavour2}\mathbf{\underline{2}}}\hspace{0.5pt},\hspace{-1.5pt}\mathbf{4^{{+}}}\hspace{-3pt}\!\hspace{0.5pt},\hspace{-1.5pt}\mathbf{3^{{-}}}\!\hspace{-3pt}\hspace{0.5pt},\hspace{-1.5pt}{\color{flavour2}\mathbf{\overline{5}}}\hspace{0.5pt},\hspace{-1.5pt}{\color{flavour1}\mathbf{\overline{6}}}),\mathcal{A}_{2}^{(1,1)}({\color{flavour1}\mathbf{\underline{1}}}\hspace{0.5pt},\hspace{-1.5pt}{\color{flavour2}\mathbf{\underline{2}}}\hspace{0.5pt},\hspace{-1.5pt}\mathbf{4^{{+}}}\hspace{-3pt}\!\hspace{0.5pt},\hspace{-1.5pt}{\color{flavour2}\mathbf{\overline{5}}}\hspace{0.5pt},\hspace{-1.5pt}\mathbf{3^{{-}}}\!\hspace{-3pt}\hspace{0.5pt},\hspace{-1.5pt}{\color{flavour1}\mathbf{\overline{6}}}),\mathcal{A}_{2}^{(1,1)}({\color{flavour1}\mathbf{\underline{1}}}\hspace{0.5pt},\hspace{-1.5pt}{\color{flavour2}\mathbf{\underline{2}}}\hspace{0.5pt},\hspace{-1.5pt}{\color{flavour2}\mathbf{\overline{5}}}\hspace{0.5pt},\hspace{-1.5pt}\mathbf{4^{{+}}}\hspace{-3pt}\!\hspace{0.5pt},\hspace{-1.5pt}\mathbf{3^{{-}}}\!\hspace{-3pt}\hspace{0.5pt},\hspace{-1.5pt}{\color{flavour1}\mathbf{\overline{6}}})\big\}$}
}
}

\defnBox[2]{randomPartialAmp}{\var{nFermionLines}\pattern}{
\mathematicaBox{
\mathematicaSequence{\funL[1]{nice}@\funL[1]{randomPartialAmp}\brace{5}\,}{
$\mathcal{A}_{5}({\color{flavour4}\mathbf{\underline{4}}}\hspace{0.5pt},\hspace{-1.5pt}{\color{flavour4}\mathbf{\overline{6}}}\hspace{0.5pt},\hspace{-1.5pt}{\color{flavour2}\mathbf{\underline{2}}}\hspace{0.5pt},\hspace{-1.5pt}{\color{flavour5}\mathbf{\underline{5}}}\hspace{0.5pt},\hspace{-1.5pt}{\color{flavour5}\mathbf{\overline{9}}}\hspace{0.5pt},\hspace{-1.5pt}{\color{flavour2}\mathbf{\overline{7}}}\hspace{0.5pt},\hspace{-1.5pt}{\color{flavour3}\mathbf{\overline{8}}}\hspace{0.5pt},\hspace{-1.5pt}{\color{flavour3}\mathbf{\underline{3}}}\hspace{0.5pt},\hspace{-1.5pt}{\color{flavour1}\mathbf{\underline{1}}}\hspace{0.5pt},\hspace{-1.5pt}{\color{flavour1}\mathbf{\overline{10}}})$}
}
}

\defnBox[3]{randomPartialAmp}{\var{nFermionLines}\pattern,\var{nGaugeBosons}\pattern}{
\mathematicaBox{
\mathematicaSequence{\funL[1]{nice}@\funL[1]{randomPartialAmp}\brace{5,2}\,}{
$\mathcal{A}_{5}^{2}({\color{flavour4}\mathbf{\overline{12}}}\hspace{0.5pt},\hspace{-1.5pt}{\color{flavour2}\mathbf{\overline{11}}}\hspace{0.5pt},\hspace{-1.5pt}{\color{flavour1}\mathbf{\underline{1}}}\hspace{0.5pt},\hspace{-1.5pt}{\color{flavour5}\mathbf{\underline{5}}}\hspace{0.5pt},\hspace{-1.5pt}{\color{flavour5}\mathbf{\overline{8}}}\hspace{0.5pt},\hspace{-1.5pt}\mathbf{6}\hspace{0.5pt},\hspace{-1.5pt}{\color{flavour1}\mathbf{\overline{10}}}\hspace{0.5pt},\hspace{-1.5pt}\mathbf{7}\hspace{0.5pt},\hspace{-1.5pt}{\color{flavour2}\mathbf{\underline{2}}}\hspace{0.5pt},\hspace{-1.5pt}{\color{flavour4}\mathbf{\underline{4}}}\hspace{0.5pt},\hspace{-1.5pt}{\color{flavour3}\mathbf{\underline{3}}}\hspace{0.5pt},\hspace{-1.5pt}{\color{flavour3}\mathbf{\overline{9}}})$}
}
}

\defnBox[3]{randomPartialAmp}{\var{nFermionLines}\pattern,\var{nPlus}\pattern,\var{nMinus}\pattern}{
\mathematicaBox{
\mathematicaSequence{\funL[]{nice}@\funL[1]{randomPartialAmp}\brace{4,3,2}\,}{
$\mathcal{A}_{4}^{(3,2)}(\mathbf{5^{{+}}}\hspace{-3pt}\!\hspace{0.5pt},\hspace{-1.5pt}{\color{flavour2}\mathbf{\underline{2}}}\hspace{0.5pt},\hspace{-1.5pt}{\color{flavour1}\mathbf{\overline{13}}}\hspace{0.5pt},\hspace{-1.5pt}{\color{flavour3}\mathbf{\overline{12}}}\hspace{0.5pt},\hspace{-1.5pt}{\color{flavour4}\mathbf{\overline{11}}}\hspace{0.5pt},\hspace{-1.5pt}{\color{flavour4}\mathbf{\underline{4}}}\hspace{0.5pt},\hspace{-1.5pt}{\color{flavour3}\mathbf{\underline{3}}}\hspace{0.5pt},\hspace{-1.5pt}\mathbf{7^{{+}}}\hspace{-3pt}\!\hspace{0.5pt},\hspace{-1.5pt}\mathbf{9^{{-}}}\!\hspace{-3pt}\hspace{0.5pt},\hspace{-1.5pt}{\color{flavour1}\mathbf{\underline{1}}}\hspace{0.5pt},\hspace{-1.5pt}\mathbf{6^{{+}}}\hspace{-3pt}\!\hspace{0.5pt},\hspace{-1.5pt}\mathbf{8^{{-}}}\!\hspace{-3pt}\hspace{0.5pt},\hspace{-1.5pt}{\color{flavour2}\mathbf{\overline{10}}})$}
}
}

\defnBox[3]{randomPartialAmp}{\var{nFermionLines}\pattern,\{\var{bosonHelicitySequence}\patternTwo\}}{
\mathematicaBox{
\mathematicaSequence{\funL[]{nice}@\funL[1]{randomPartialAmp}\brace{5,\{\fun{m},\fun{p},\fun{m},\fun{p}\}}\,}{
$\mathcal{A}_{4}^{(2,2)}({\color{flavour4}\mathbf{\underline{4}}}\hspace{0.5pt},\hspace{-1.5pt}{\color{flavour4}\mathbf{\overline{10}}}\hspace{0.5pt},\hspace{-1.5pt}{\color{flavour2}\mathbf{\underline{2}}}\hspace{0.5pt},\hspace{-1.5pt}\mathbf{8^{{+}}}\hspace{-3pt}\!\hspace{0.5pt},\hspace{-1.5pt}{\color{flavour3}\mathbf{\underline{3}}}\hspace{0.5pt},\hspace{-1.5pt}{\color{flavour3}\mathbf{\overline{9}}}\hspace{0.5pt},\hspace{-1.5pt}{\color{flavour2}\mathbf{\overline{11}}}\hspace{0.5pt},\hspace{-1.5pt}\mathbf{6^{{+}}}\hspace{-3pt}\!\hspace{0.5pt},\hspace{-1.5pt}{\color{flavour1}\mathbf{\overline{12}}}\hspace{0.5pt},\hspace{-1.5pt}{\color{flavour1}\mathbf{\underline{1}}}\hspace{0.5pt},\hspace{-1.5pt}\mathbf{7^{{-}}}\!\hspace{-3pt}\hspace{0.5pt},\hspace{-1.5pt}\mathbf{5^{{-}}}\!\hspace{-3pt})$}
}
}

\newpage
\vspace{0pt}\subsectionAppendix{Manipulations of Partial Amplitudes Involving Charged Matter}{appendix:partial_amplitude_manipulations}\vspace{-10pt}

\defnBox{toDistinctFlavours}{\var{expression}\pattern}{writes any single-flavour \fun{amp}\brace{} objects appearing in \var{expression} as a sum of distinct-flavour partial amplitudes. 
\mathematicaBox{
\mathematicaSequence[1]{egAmp=\fun{amp}\brace{\{\fun{fb}\brace{3},\fun{f}\brace{1},\fun{fb}\brace{2},\fun{f}\brace{3},\fun{fb}\brace{1},\fun{f}\brace{2}\,\}};\\
\funL[]{nice}@\%}{$\rule{0pt}{16pt}{A}_{3}({\color{flavour0}\mathbf{\overline{4}}}\hspace{0.5pt},\hspace{-1.5pt}{\color{flavour0}\mathbf{\underline{1}}}\hspace{0.5pt},\hspace{-1.5pt}{\color{flavour0}\mathbf{\overline{5}}}\hspace{0.5pt},\hspace{-1.5pt}{\color{flavour0}\mathbf{\underline{3}}}\hspace{0.5pt},\hspace{-1.5pt}{\color{flavour0}\mathbf{\overline{6}}}\hspace{0.5pt},\hspace{-1.5pt}{\color{flavour0}\mathbf{\underline{2}}})\vspace{-8pt}$}
\mathematicaSequence{\funL[]{drawChords}\brace{egAmp}}{$\begin{array}{@{}c@{}}\includegraphics[scale=0.6]{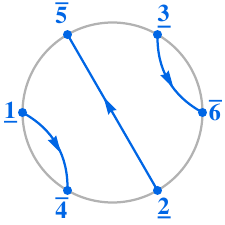}\end{array}{+}\begin{array}{@{}c@{}}\includegraphics[scale=0.6]{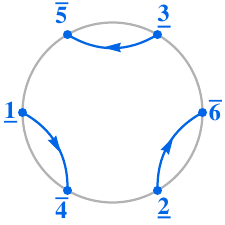}\end{array}{+}\begin{array}{@{}c@{}}\includegraphics[scale=0.6]{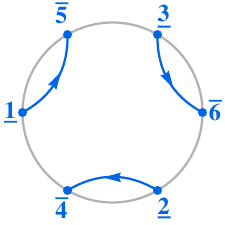}\end{array}{+}\begin{array}{@{}c@{}}\includegraphics[scale=0.6]{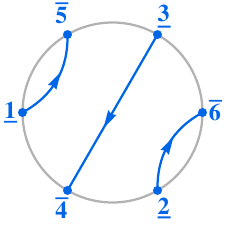}\end{array}{+}\begin{array}{@{}c@{}}\includegraphics[scale=0.6]{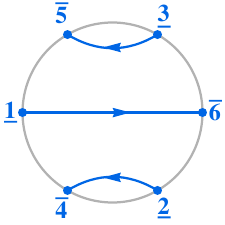}\end{array}\vspace{-8pt}$}
\mathematicaSequence{\funL[]{toDistinctFlavours}@egAmp}{
\phantom{{+}}\funL{amp}\brace{\{\fun{fb}\brace{3},\fun{f}\brace{1},\fun{fb}\brace{2},\fun{f}\brace{3},\fun{fb}\brace{1},\fun{f}\brace{2}\,\},\!\{\!\{2,1\},\{6,3\},\{4,5\}\!\}}\\
{+}\fun{amp}\brace{\{\fun{fb}\brace{3},\fun{f}\brace{1},\fun{fb}\brace{2},\fun{f}\brace{3},\fun{fb}\brace{1},\fun{f}\brace{2}\,\},\!\{\!\{2,1\},\{6,5\},\{4,3\}\!\}}\\
{+}\fun{amp}\brace{\{\fun{fb}\brace{3},\fun{f}\brace{1},\fun{fb}\brace{2},\fun{f}\brace{3},\fun{fb}\brace{1},\fun{f}\brace{2}\,\},\!\{\!\{2,3\},\{6,1\},\{4,5\}\!\}}\\
{+}\fun{amp}\brace{\{\fun{fb}\brace{3},\fun{f}\brace{1},\fun{fb}\brace{2},\fun{f}\brace{3},\fun{fb}\brace{1},\fun{f}\brace{2}\,\},\!\{\!\{2,3\},\{6,5\},\{4,1\}\!\}}\\
{+}\fun{amp}\brace{\{\fun{fb}\brace{3},\fun{f}\brace{1},\fun{fb}\brace{2},\fun{f}\brace{3},\fun{fb}\brace{1},\fun{f}\brace{2}\,\},\!\{\!\{2,5\},\{6,1\},\{4,3\}\!\}}}
\mathematicaSequence{\funL[]{nice}@\%}{$\mathcal{A}_{3}({\color{flavour1}\mathbf{\overline{4}}}\hspace{0.5pt},\hspace{-1.5pt}{\color{flavour1}\mathbf{\underline{1}}}\hspace{0.5pt},\hspace{-1.5pt}{\color{flavour2}\mathbf{\overline{5}}}\hspace{0.5pt},\hspace{-1.5pt}{\color{flavour3}\mathbf{\underline{3}}}\hspace{0.5pt},\hspace{-1.5pt}{\color{flavour3}\mathbf{\overline{6}}}\hspace{0.5pt},\hspace{-1.5pt}{\color{flavour2}\mathbf{\underline{2}}})%
{+}\mathcal{A}_{3}({\color{flavour1}\mathbf{\overline{4}}}\hspace{0.5pt},\hspace{-1.5pt}{\color{flavour1}\mathbf{\underline{1}}}\hspace{0.5pt},\hspace{-1.5pt}{\color{flavour3}\mathbf{\overline{5}}}\hspace{0.5pt},\hspace{-1.5pt}{\color{flavour3}\mathbf{\underline{3}}}\hspace{0.5pt},\hspace{-1.5pt}{\color{flavour2}\mathbf{\overline{6}}}\hspace{0.5pt},\hspace{-1.5pt}{\color{flavour2}\mathbf{\underline{2}}})%
{+}\mathcal{A}_{3}({\color{flavour2}\mathbf{\overline{4}}}\hspace{0.5pt},\hspace{-1.5pt}{\color{flavour1}\mathbf{\underline{1}}}\hspace{0.5pt},\hspace{-1.5pt}{\color{flavour1}\mathbf{\overline{5}}}\hspace{0.5pt},\hspace{-1.5pt}{\color{flavour3}\mathbf{\underline{3}}}\hspace{0.5pt},\hspace{-1.5pt}{\color{flavour3}\mathbf{\overline{6}}}\hspace{0.5pt},\hspace{-1.5pt}{\color{flavour2}\mathbf{\underline{2}}})\\
\mbox{}\hspace{-10pt}{+}\mathcal{A}_{3}({\color{flavour3}\mathbf{\overline{4}}}\hspace{0.5pt},\hspace{-1.5pt}{\color{flavour1}\mathbf{\underline{1}}}\hspace{0.5pt},\hspace{-1.5pt}{\color{flavour1}\mathbf{\overline{5}}}\hspace{0.5pt},\hspace{-1.5pt}{\color{flavour3}\mathbf{\underline{3}}}\hspace{0.5pt},\hspace{-1.5pt}{\color{flavour2}\mathbf{\overline{6}}}\hspace{0.5pt},\hspace{-1.5pt}{\color{flavour2}\mathbf{\underline{2}}}){+}\mathcal{A}_{3}({\color{flavour2}\mathbf{\overline{4}}}\hspace{0.5pt},\hspace{-1.5pt}{\color{flavour1}\mathbf{\underline{1}}}\hspace{0.5pt},\hspace{-1.5pt}{\color{flavour3}\mathbf{\overline{5}}}\hspace{0.5pt},\hspace{-1.5pt}{\color{flavour3}\mathbf{\underline{3}}}\hspace{0.5pt},\hspace{-1.5pt}{\color{flavour1}\mathbf{\overline{6}}}\hspace{0.5pt},\hspace{-1.5pt}{\color{flavour2}\mathbf{\underline{2}}})$}
\mathematicaSequence{\funL[]{drawChords}\brace{\funL[]{toDistinctFlavours}@egAmp}}{$\begin{array}{@{}c@{}}\includegraphics[scale=0.6]{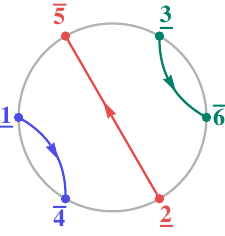}\end{array}{+}\begin{array}{@{}c@{}}\includegraphics[scale=0.6]{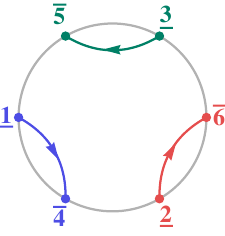}\end{array}{+}\begin{array}{@{}c@{}}\includegraphics[scale=0.6]{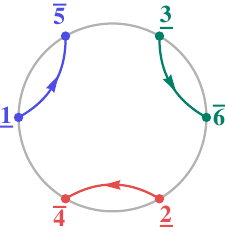}\end{array}{+}\begin{array}{@{}c@{}}\includegraphics[scale=0.6]{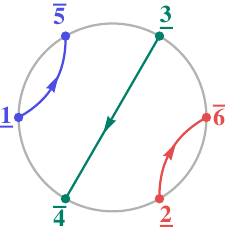}\end{array}{+}\begin{array}{@{}c@{}}\includegraphics[scale=0.6]{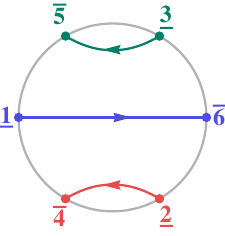}\end{array}\vspace{-8pt}$}
}
}

\defnBox{toAllPlus}{\var{expression}\pattern}{converts any distinct-flavour \fun{amp}\brace{} objects appearing in \var{expression} into those for which all fermion lines are oriented in the `plus' direction---that is, for which the \built{Position} of any fermion of some flavour is earlier than the corresponding anti-fermion of that flavour.\\[-8pt]

\textbf{Note}: The function \fun{toAllPlus} does \emph{not} \emph{necessarily} map a partial amplitudes into \text{the} \emph{all-plus basis} (\ref{all_plus_basis_of_partial_amps}) described in section~\ref{subsection_review_of_multiflavoured_partials}---merely into \emph{an} all-plus basis for some choice of KK-projection akin to those of  (\ref{kk_projected_set_of_distinct_flavour_partials}). 
\mathematicaBox{
\mathematicaSequence{egAmp=\funL[1]{randomPartialAmp}\brace{3,1}}{\funL{amp}\brace{\{\fun{fb}\brace{3},\fun{f}\brace{2},\fun{fb}\brace{2},\fun{f}\brace{3},\fun{fb}\brace{1},\fun{f}\brace{1},\fun{g}\brace{4}\,\},\!\{\!\{6,5\},\{2,3\},\{4,1\}\!\}}}
\vspace{-5pt}
\mathematicaSequence{
\funL[1]{nice}\brace{egAmp}}{$\mathcal{A}_{3}^{1}({\color{flavour3}\mathbf{\overline{5}}}\hspace{0.5pt},\hspace{-1.5pt}{\color{flavour2}\mathbf{\underline{2}}}\hspace{0.5pt},\hspace{-1.5pt}{\color{flavour2}\mathbf{\overline{6}}}\hspace{0.5pt},\hspace{-1.5pt}{\color{flavour3}\mathbf{\underline{3}}}\hspace{0.5pt},\hspace{-1.5pt}{\color{flavour1}\mathbf{\overline{7}}}\hspace{0.5pt},\hspace{-1.5pt}{\color{flavour1}\mathbf{\underline{1}}}\hspace{0.5pt},\hspace{-1.5pt}\mathbf{4})$}
\mathematicaSequence{\funL[1]{nice}@\funL[1]{toAllPlus}\brace{egAmp}}{$\rule{0pt}{10pt}\mathcal{A}_{3}^{1}({\color{flavour1}\mathbf{\underline{1}}}\hspace{0.5pt},\hspace{-1.5pt}{\color{flavour3}\mathbf{\underline{3}}}\hspace{0.5pt},\hspace{-1.5pt}{\color{flavour2}\mathbf{\underline{2}}}\hspace{0.5pt},\hspace{-1.5pt}{\color{flavour2}\mathbf{\overline{6}}}\hspace{0.5pt},\hspace{-1.5pt}\mathbf{4}\hspace{0.5pt},\hspace{-1.5pt}{\color{flavour3}\mathbf{\overline{5}}}\hspace{0.5pt},\hspace{-1.5pt}{\color{flavour1}\mathbf{\overline{7}}}){+}\mathcal{A}_{3}^{1}({\color{flavour1}\mathbf{\underline{1}}}\hspace{0.5pt},\hspace{-1.5pt}{\color{flavour3}\mathbf{\underline{3}}}\hspace{0.5pt},\hspace{-1.5pt}{\color{flavour2}\mathbf{\underline{2}}}\hspace{0.5pt},\hspace{-1.5pt}\mathbf{4}\hspace{0.5pt},\hspace{-1.5pt}{\color{flavour2}\mathbf{\overline{6}}}\hspace{0.5pt},\hspace{-1.5pt}{\color{flavour3}\mathbf{\overline{5}}}\hspace{0.5pt},\hspace{-1.5pt}{\color{flavour1}\mathbf{\overline{7}}}){+}\mathcal{A}_{3}^{1}({\color{flavour1}\mathbf{\underline{1}}}\hspace{0.5pt},\hspace{-1.5pt}{\color{flavour3}\mathbf{\underline{3}}}\hspace{0.5pt},\hspace{-1.5pt}\mathbf{4}\hspace{0.5pt},\hspace{-1.5pt}{\color{flavour2}\mathbf{\underline{2}}}\hspace{0.5pt},\hspace{-1.5pt}{\color{flavour2}\mathbf{\overline{6}}}\hspace{0.5pt},\hspace{-1.5pt}{\color{flavour3}\mathbf{\overline{5}}}\hspace{0.5pt},\hspace{-1.5pt}{\color{flavour1}\mathbf{\overline{7}}})$\\$\mbox{}\hspace{-10pt}\displaystyle{+}\mathcal{A}_{3}^{1}({\color{flavour1}\mathbf{\underline{1}}}\hspace{0.5pt},\hspace{-1.5pt}\mathbf{4}\hspace{0.5pt},\hspace{-1.5pt}{\color{flavour3}\mathbf{\underline{3}}}\hspace{0.5pt},\hspace{-1.5pt}{\color{flavour2}\mathbf{\underline{2}}}\hspace{0.5pt},\hspace{-1.5pt}{\color{flavour2}\mathbf{\overline{6}}}\hspace{0.5pt},\hspace{-1.5pt}{\color{flavour3}\mathbf{\overline{5}}}\hspace{0.5pt},\hspace{-1.5pt}{\color{flavour1}\mathbf{\overline{7}}})$}
}
\vspace{-50pt}}

\defnBox{toAllPlusBasis}{\var{expression}\pattern}{expands any distinctly-flavoured partial amplitude into the \emph{all-plus} basis of Melia of the form $\mathcal{A}({\color{flavour1}\underline{1}},\vec{\sigma},{\color{flavour1}\overline{a}})$ where ${\color{flavour1}\bar{\psi}_a}$ is the anti-fermion with the same flavour as fermion ${\color{flavour1}\psi_1}$.
\mathematicaBox{%
\mathematicaSequence[1]{egPartial=\funL[1]{randomPartialAmp}\brace{5,1};\\
\funL[1]{nice}@egPartial}{\rule{0pt}{14pt}$\mathcal{A}_{5}^{1}({\color{flavour4}\mathbf{\underline{4}}}\hspace{1.02pt},\hspace{-1.5pt}\mathbf{6}\hspace{1.02pt},\hspace{-1.5pt}{\color{flavour3}\mathbf{\underline{3}}}\hspace{1.02pt},\hspace{-1.5pt}{\color{flavour3}\mathbf{\overline{10}}}\hspace{1.02pt},\hspace{-1.5pt}{\color{flavour4}\mathbf{\overline{11}}}\hspace{1.02pt},\hspace{-1.5pt}{\color{flavour5}\mathbf{\overline{7}}}\hspace{1.02pt},\hspace{-1.5pt}{\color{flavour5}\mathbf{\underline{5}}}\hspace{1.02pt},\hspace{-1.5pt}{\color{flavour2}\mathbf{\overline{9}}}\hspace{1.02pt},\hspace{-1.5pt}{\color{flavour1}\mathbf{\overline{8}}}\hspace{1.02pt},\hspace{-1.5pt}{\color{flavour1}\mathbf{\underline{1}}}\hspace{1.02pt},\hspace{-1.5pt}{\color{flavour2}\mathbf{\underline{2}}})$}
\mathematicaSequence{egAllPlus=\funL[1]{toAllPlus}\brace{egPartial};\\
\funL[1]{nice}@egAllPlus}{$\rule{0pt}{14pt}{-}\mathcal{A}_{5}^{1}({\color{flavour3}\mathbf{\underline{3}}}\hspace{1.02pt},\hspace{-1.5pt}\mathbf{6}\hspace{1.02pt},\hspace{-1.5pt}{\color{flavour4}\mathbf{\underline{4}}}\hspace{1.02pt},\hspace{-1.5pt}{\color{flavour2}\mathbf{\underline{2}}}\hspace{1.02pt},\hspace{-1.5pt}{\color{flavour1}\mathbf{\underline{1}}}\hspace{1.02pt},\hspace{-1.5pt}{\color{flavour1}\mathbf{\overline{8}}}\hspace{1.02pt},\hspace{-1.5pt}{\color{flavour2}\mathbf{\overline{9}}}\hspace{1.02pt},\hspace{-1.5pt}{\color{flavour5}\mathbf{\underline{5}}}\hspace{1.02pt},\hspace{-1.5pt}{\color{flavour5}\mathbf{\overline{7}}}\hspace{1.02pt},\hspace{-1.5pt}{\color{flavour4}\mathbf{\overline{11}}}\hspace{1.02pt},\hspace{-1.5pt}{\color{flavour3}\mathbf{\overline{10}}})$}
\mathematicaSequence{egInBasis=\funL[1]{toAllPlusBasis}\brace{egPartial};\\
\funL[1]{nice}@egInBasis}{$\rule{0pt}{14pt}{-}\mathcal{A}_{5}^{1}({\color{flavour1}\mathbf{\underline{1}}}\hspace{1.02pt},\hspace{-1.5pt}{\color{flavour2}\mathbf{\underline{2}}}\hspace{1.02pt},\hspace{-1.5pt}{\color{flavour4}\mathbf{\underline{4}}}\hspace{1.02pt},\hspace{-1.5pt}\mathbf{6}\hspace{1.02pt},\hspace{-1.5pt}{\color{flavour3}\mathbf{\underline{3}}}\hspace{1.02pt},\hspace{-1.5pt}{\color{flavour3}\mathbf{\overline{10}}}\hspace{1.02pt},\hspace{-1.5pt}{\color{flavour4}\mathbf{\overline{11}}}\hspace{1.02pt},\hspace{-1.5pt}{\color{flavour5}\mathbf{\underline{5}}}\hspace{1.02pt},\hspace{-1.5pt}{\color{flavour5}\mathbf{\overline{7}}}\hspace{1.02pt},\hspace{-1.5pt}{\color{flavour2}\mathbf{\overline{9}}}\hspace{1.02pt},\hspace{-1.5pt}{\color{flavour1}\mathbf{\overline{8}}}){-}\mathcal{A}_{5}^{1}({\color{flavour1}\mathbf{\underline{1}}}\hspace{1.02pt},\hspace{-1.5pt}{\color{flavour2}\mathbf{\underline{2}}}\hspace{1.02pt},\hspace{-1.5pt}{\color{flavour5}\mathbf{\underline{5}}}\hspace{1.02pt},\hspace{-1.5pt}{\color{flavour4}\mathbf{\underline{4}}}\hspace{1.02pt},\hspace{-1.5pt}\mathbf{6}\hspace{1.02pt},\hspace{-1.5pt}{\color{flavour3}\mathbf{\underline{3}}}\hspace{1.02pt},\hspace{-1.5pt}{\color{flavour3}\mathbf{\overline{10}}}\hspace{1.02pt},\hspace{-1.5pt}{\color{flavour4}\mathbf{\overline{11}}}\hspace{1.02pt},\hspace{-1.5pt}{\color{flavour5}\mathbf{\overline{7}}}\hspace{1.02pt},\hspace{-1.5pt}{\color{flavour2}\mathbf{\overline{9}}}\hspace{1.02pt},\hspace{-1.5pt}{\color{flavour1}\mathbf{\overline{8}}})$}
}
}

\defnBox{toSingleFlavour}{\var{expression}\pattern}{expands any multi-flavoured partial amplitude to a sum of one-flavour partial amplitudes using Melia's flavour-reduction algorithm. See \mbox{section~\ref{section_flavour_reduction_algorithm}}.
\mathematicaBox{
\mathematicaSequence[1]{egPartial=\funL[1]{randomPartialAmp}\brace{4,2};\\
\funL[1]{nice}\brace{egPartial}}{$\rule{0pt}{16pt}\mathcal{A}_{4}^{2}({\color{flavour4}\mathbf{\overline{7}}}\hspace{0.5pt},\hspace{-1.5pt}{\color{flavour4}\mathbf{\underline{4}}}\hspace{0.5pt},\hspace{-1.5pt}{\color{flavour2}\mathbf{\overline{9}}}\hspace{0.5pt},\hspace{-1.5pt}{\color{flavour3}\mathbf{\underline{3}}}\hspace{0.5pt},\hspace{-1.5pt}\mathbf{6}\hspace{0.5pt},\hspace{-1.5pt}{\color{flavour3}\mathbf{\overline{8}}}\hspace{0.5pt},\hspace{-1.5pt}{\color{flavour2}\mathbf{\underline{2}}}\hspace{0.5pt},\hspace{-1.5pt}{\color{flavour1}\mathbf{\underline{1}}}\hspace{0.5pt},\hspace{-1.5pt}{\color{flavour1}\mathbf{\overline{10}}}\hspace{0.5pt},\hspace{-1.5pt}\mathbf{5})$}
\mathematicaSequence{\funL[1]{nice}@\funL[1]{toSingleFlavour}\brace{egPartial}}{$\rule{0pt}{10pt}\hspace{-8.5pt}{A}_{4}^{2}({\color{flavour0}\mathbf{\underline{1}}}\hspace{0.5pt},\hspace{-1.5pt}{\color{flavour0}\mathbf{\underline{2}}}\hspace{0.5pt},\hspace{-1.5pt}{\color{flavour0}\mathbf{\underline{3}}}\hspace{0.5pt},\hspace{-1.5pt}{\color{flavour0}\mathbf{\underline{4}}}\hspace{0.5pt},\hspace{-1.5pt}\mathbf{6}\hspace{0.5pt},\hspace{-1.5pt}{\color{flavour0}\mathbf{\overline{8}}}\hspace{0.5pt},\hspace{-1.5pt}{\color{flavour0}\mathbf{\overline{9}}}\hspace{0.5pt},\hspace{-1.5pt}{\color{flavour0}\mathbf{\overline{7}}}\hspace{-0.9pt},\hspace{-1.5pt}\mathbf{5}\hspace{0.5pt},\hspace{-1.5pt}{\color{flavour0}\mathbf{\overline{10}}})\!{+}\!{A}_{4}^{2}({\color{flavour0}\mathbf{\underline{1}}}\hspace{0.5pt},\hspace{-1.5pt}{\color{flavour0}\mathbf{\underline{2}}}\hspace{0.5pt},\hspace{-1.5pt}{\color{flavour0}\mathbf{\underline{3}}}\hspace{0.5pt},\hspace{-1.5pt}\mathbf{6}\hspace{0.5pt},\hspace{-1.5pt}{\color{flavour0}\mathbf{\underline{4}}}\hspace{0.5pt},\hspace{-1.5pt}{\color{flavour0}\mathbf{\overline{8}}}\hspace{0.5pt},\hspace{-1.5pt}{\color{flavour0}\mathbf{\overline{9}}}\hspace{0.5pt},\hspace{-1.5pt}{\color{flavour0}\mathbf{\overline{7}}}\hspace{-0.9pt},\hspace{-1.5pt}\mathbf{5}\hspace{0.5pt},\hspace{-1.5pt}{\color{flavour0}\mathbf{\overline{10}}})\!{+}\!{A}_{4}^{2}({\color{flavour0}\mathbf{\underline{1}}}\hspace{0.5pt},\hspace{-1.5pt}{\color{flavour0}\mathbf{\underline{2}}}\hspace{0.5pt},\hspace{-1.5pt}{\color{flavour0}\mathbf{\underline{3}}}\hspace{0.5pt},\hspace{-1.5pt}\mathbf{6}\hspace{0.5pt},\hspace{-1.5pt}{\color{flavour0}\mathbf{\overline{8}}}\hspace{0.5pt},\hspace{-1.5pt}{\color{flavour0}\mathbf{\underline{4}}}\hspace{0.5pt},\hspace{-1.5pt}{\color{flavour0}\mathbf{\overline{9}}}\hspace{0.5pt},\hspace{-1.5pt}{\color{flavour0}\mathbf{\overline{7}}}\hspace{-0.9pt},\hspace{-1.5pt}\mathbf{5}\hspace{0.5pt},\hspace{-1.5pt}{\color{flavour0}\mathbf{\overline{10}}})\\~\mbox{\hspace{-22.5pt}~}{+}{A}_{4}^{2}({\color{flavour0}\mathbf{\underline{1}}}\hspace{0.5pt},\hspace{-1.5pt}{\color{flavour0}\mathbf{\underline{2}}}\hspace{0.5pt},\hspace{-1.5pt}{\color{flavour0}\mathbf{\underline{3}}}\hspace{0.5pt},\hspace{-1.5pt}\mathbf{6}\hspace{0.5pt},\hspace{-1.5pt}{\color{flavour0}\mathbf{\overline{8}}}\hspace{0.5pt},\hspace{-1.5pt}{\color{flavour0}\mathbf{\overline{9}}}\hspace{0.5pt},\hspace{-1.5pt}{\color{flavour0}\mathbf{\underline{4}}}\hspace{0.5pt},\hspace{-1.5pt}{\color{flavour0}\mathbf{\overline{7}}}\hspace{-0.9pt},\hspace{-1.5pt}\mathbf{5}\hspace{0.5pt},\hspace{-1.5pt}{\color{flavour0}\mathbf{\overline{10}}})\!{+}\!{A}_{4}^{2}({\color{flavour0}\mathbf{\underline{1}}}\hspace{0.5pt},\hspace{-1.5pt}{\color{flavour0}\mathbf{\underline{2}}}\hspace{0.5pt},\hspace{-1.5pt}{\color{flavour0}\mathbf{\underline{4}}}\hspace{0.5pt},\hspace{-1.5pt}{\color{flavour0}\mathbf{\underline{3}}}\hspace{0.5pt},\hspace{-1.5pt}\mathbf{6}\hspace{0.5pt},\hspace{-1.5pt}{\color{flavour0}\mathbf{\overline{8}}}\hspace{0.5pt},\hspace{-1.5pt}{\color{flavour0}\mathbf{\overline{9}}}\hspace{0.5pt},\hspace{-1.5pt}{\color{flavour0}\mathbf{\overline{7}}}\hspace{-0.9pt},\hspace{-1.5pt}\mathbf{5}\hspace{0.5pt},\hspace{-1.5pt}{\color{flavour0}\mathbf{\overline{10}}})$}
}
\vspace{-0pt}}

\defnBox{kkProjectionRule}{\var{stateA}\pattern,\var{stateB}\pattern}{represents a \built{Rule} which maps any \fun{amp} objects (whether single-flavour or multi-flavour) into partial amps of the form $A(\var{stateA},\ldots,\var{stateB})$. \textbf{Note}: for multi-flavour partial amplitudes, this may \emph{not} result in a unique representation, as there exist fermion-line-orientation reversing identities such as (\ref{eqn:kkSwap}) among multi-flavoured partial amplitudes even with a choice of KK basis. See \mbox{section~\ref{relations_among_partial_amplitudes_general}} for further discussion.
\mathematicaBox{
\mathematicaSequence[1]{egPartial=\funL[1]{randomPartialAmp}\brace{4,1};\\
\funL[1]{nice}\brace{egPartial}}{$\rule{0pt}{16pt}\mathcal{A}_{4}^{1}({\color{flavour1}\mathbf{\underline{1}}}\hspace{0.5pt},\hspace{-1.5pt}{\color{flavour1}\mathbf{\overline{7}}}\hspace{0.5pt},\hspace{-1.5pt}{\color{flavour4}\mathbf{\overline{9}}}\hspace{0.5pt},\hspace{-1.5pt}{\color{flavour2}\mathbf{\overline{8}}}\hspace{0.5pt},\hspace{-1.5pt}{\color{flavour2}\mathbf{\underline{2}}}\hspace{0.5pt},\hspace{-1.5pt}{\color{flavour4}\mathbf{\underline{4}}}\hspace{0.5pt},\hspace{-1.5pt}{\color{flavour3}\mathbf{\underline{3}}}\hspace{0.5pt},\hspace{-1.5pt}\mathbf{5}\hspace{0.5pt},\hspace{-1.5pt}{\color{flavour3}\mathbf{\overline{6}}})$}
\mathematicaSequence{\funL[1]{nice}@\big(egPartial/.\funL[1]{kkProjectionRule}\brace{\fun{f}\brace{1},\fun{f}\brace{3}\,}\big)}{$\rule{0pt}{10pt}\hspace{-0pt}\mathcal{A}_{4}^{1}({\color{flavour1}\mathbf{\underline{1}}}\hspace{0.5pt},\hspace{-1.5pt}{\color{flavour1}\mathbf{\overline{7}}}\hspace{0.5pt},\hspace{-1.5pt}{\color{flavour4}\mathbf{\overline{9}}}\hspace{0.5pt},\hspace{-1.5pt}{\color{flavour2}\mathbf{\overline{8}}}\hspace{0.5pt},\hspace{-1.5pt}{\color{flavour2}\mathbf{\underline{2}}}\hspace{0.5pt},\hspace{-1.5pt}{\color{flavour4}\mathbf{\underline{4}}}\hspace{0.5pt},\hspace{-1.5pt}{\color{flavour3}\mathbf{\overline{6}}}\hspace{0.5pt},\hspace{-1.5pt}\mathbf{5}\hspace{0.5pt},\hspace{-1.5pt}{\color{flavour3}\mathbf{\underline{3}}}){+}\mathcal{A}_{4}^{1}({\color{flavour1}\mathbf{\underline{1}}}\hspace{0.5pt},\hspace{-1.5pt}{\color{flavour1}\mathbf{\overline{7}}}\hspace{0.5pt},\hspace{-1.5pt}{\color{flavour3}\mathbf{\overline{6}}}\hspace{0.5pt},\hspace{-1.5pt}{\color{flavour4}\mathbf{\overline{9}}}\hspace{0.5pt},\hspace{-1.5pt}{\color{flavour2}\mathbf{\overline{8}}}\hspace{0.5pt},\hspace{-1.5pt}{\color{flavour2}\mathbf{\underline{2}}}\hspace{0.5pt},\hspace{-1.5pt}{\color{flavour4}\mathbf{\underline{4}}}\hspace{0.5pt},\hspace{-1.5pt}\mathbf{5}\hspace{0.5pt},\hspace{-1.5pt}{\color{flavour3}\mathbf{\underline{3}}}){+}\mathcal{A}_{4}^{1}({\color{flavour1}\mathbf{\underline{1}}}\hspace{0.5pt},\hspace{-1.5pt}{\color{flavour1}\mathbf{\overline{7}}}\hspace{0.5pt},\hspace{-1.5pt}{\color{flavour3}\mathbf{\overline{6}}}\hspace{0.5pt},\hspace{-1.5pt}{\color{flavour4}\mathbf{\overline{9}}}\hspace{0.5pt},\hspace{-1.5pt}{\color{flavour2}\mathbf{\overline{8}}}\hspace{0.5pt},\hspace{-1.5pt}{\color{flavour2}\mathbf{\underline{2}}}\hspace{0.5pt},\hspace{-1.5pt}\mathbf{5}\hspace{0.5pt},\hspace{-1.5pt}{\color{flavour4}\mathbf{\underline{4}}}\hspace{0.5pt},\hspace{-1.5pt}{\color{flavour3}\mathbf{\underline{3}}})\\\mbox{}\hspace{-9pt}{+}\mathcal{A}_{4}^{1}({\color{flavour1}\mathbf{\underline{1}}}\hspace{0.5pt},\hspace{-1.5pt}{\color{flavour1}\mathbf{\overline{7}}}\hspace{0.5pt},\hspace{-1.5pt}{\color{flavour3}\mathbf{\overline{6}}}\hspace{0.5pt},\hspace{-1.5pt}{\color{flavour4}\mathbf{\overline{9}}}\hspace{0.5pt},\hspace{-1.5pt}{\color{flavour2}\mathbf{\overline{8}}}\hspace{0.5pt},\hspace{-1.5pt}\mathbf{5}\hspace{0.5pt},\hspace{-1.5pt}{\color{flavour2}\mathbf{\underline{2}}}\hspace{0.5pt},\hspace{-1.5pt}{\color{flavour4}\mathbf{\underline{4}}}\hspace{0.5pt},\hspace{-1.5pt}{\color{flavour3}\mathbf{\underline{3}}}){+}\mathcal{A}_{4}^{1}({\color{flavour1}\mathbf{\underline{1}}}\hspace{0.5pt},\hspace{-1.5pt}{\color{flavour1}\mathbf{\overline{7}}}\hspace{0.5pt},\hspace{-1.5pt}{\color{flavour3}\mathbf{\overline{6}}}\hspace{0.5pt},\hspace{-1.5pt}{\color{flavour4}\mathbf{\overline{9}}}\hspace{0.5pt},\hspace{-1.5pt}\mathbf{5}\hspace{0.5pt},\hspace{-1.5pt}{\color{flavour2}\mathbf{\overline{8}}}\hspace{0.5pt},\hspace{-1.5pt}{\color{flavour2}\mathbf{\underline{2}}}\hspace{0.5pt},\hspace{-1.5pt}{\color{flavour4}\mathbf{\underline{4}}}\hspace{0.5pt},\hspace{-1.5pt}{\color{flavour3}\mathbf{\underline{3}}}){+}\mathcal{A}_{4}^{1}({\color{flavour1}\mathbf{\underline{1}}}\hspace{0.5pt},\hspace{-1.5pt}{\color{flavour1}\mathbf{\overline{7}}}\hspace{0.5pt},\hspace{-1.5pt}{\color{flavour3}\mathbf{\overline{6}}}\hspace{0.5pt},\hspace{-1.5pt}\mathbf{5}\hspace{0.5pt},\hspace{-1.5pt}{\color{flavour4}\mathbf{\overline{9}}}\hspace{0.5pt},\hspace{-1.5pt}{\color{flavour2}\mathbf{\overline{8}}}\hspace{0.5pt},\hspace{-1.5pt}{\color{flavour2}\mathbf{\underline{2}}}\hspace{0.5pt},\hspace{-1.5pt}{\color{flavour4}\mathbf{\underline{4}}}\hspace{0.5pt},\hspace{-1.5pt}{\color{flavour3}\mathbf{\underline{3}}})$}
\mathematicaSequence{\funL[1]{nice}@(egPartial/.\funL[1]{kkProjectionRule}\brace{\fun{f}\brace{1},\fun{fb}\brace{2}\,})}{$\rule{0pt}{10pt}\hspace{-9pt}{-}\mathcal{A}_{4}^{1}({\color{flavour1}\mathbf{\underline{1}}}\hspace{0.5pt},\hspace{-1.5pt}{\color{flavour1}\mathbf{\overline{7}}}\hspace{0.5pt},\hspace{-1.5pt}{\color{flavour4}\mathbf{\overline{9}}}\hspace{0.5pt},\hspace{-1.5pt}{\color{flavour3}\mathbf{\overline{6}}}\hspace{0.5pt},\hspace{-1.5pt}\mathbf{5}\hspace{0.5pt},\hspace{-1.5pt}{\color{flavour3}\mathbf{\underline{3}}}\hspace{0.5pt},\hspace{-1.5pt}{\color{flavour4}\mathbf{\underline{4}}}\hspace{0.5pt},\hspace{-1.5pt}{\color{flavour2}\mathbf{\underline{2}}}\hspace{0.5pt},\hspace{-1.5pt}{\color{flavour2}\mathbf{\overline{8}}}){-}\mathcal{A}_{4}^{1}({\color{flavour1}\mathbf{\underline{1}}}\hspace{0.5pt},\hspace{-1.5pt}{\color{flavour1}\mathbf{\overline{7}}}\hspace{0.5pt},\hspace{-1.5pt}{\color{flavour3}\mathbf{\overline{6}}}\hspace{0.5pt},\hspace{-1.5pt}\mathbf{5}\hspace{0.5pt},\hspace{-1.5pt}{\color{flavour3}\mathbf{\underline{3}}}\hspace{0.5pt},\hspace{-1.5pt}{\color{flavour4}\mathbf{\underline{4}}}\hspace{0.5pt},\hspace{-1.5pt}{\color{flavour4}\mathbf{\overline{9}}}\hspace{0.5pt},\hspace{-1.5pt}{\color{flavour2}\mathbf{\underline{2}}}\hspace{0.5pt},\hspace{-1.5pt}{\color{flavour2}\mathbf{\overline{8}}}){-}\mathcal{A}_{4}^{1}({\color{flavour1}\mathbf{\underline{1}}}\hspace{0.5pt},\hspace{-1.5pt}{\color{flavour1}\mathbf{\overline{7}}}\hspace{0.5pt},\hspace{-1.5pt}{\color{flavour3}\mathbf{\overline{6}}}\hspace{0.5pt},\hspace{-1.5pt}\mathbf{5}\hspace{0.5pt},\hspace{-1.5pt}{\color{flavour3}\mathbf{\underline{3}}}\hspace{0.5pt},\hspace{-1.5pt}{\color{flavour4}\mathbf{\overline{9}}}\hspace{0.5pt},\hspace{-1.5pt}{\color{flavour4}\mathbf{\underline{4}}}\hspace{0.5pt},\hspace{-1.5pt}{\color{flavour2}\mathbf{\underline{2}}}\hspace{0.5pt},\hspace{-1.5pt}{\color{flavour2}\mathbf{\overline{8}}})\\\mbox{}\hspace{-9pt}{-}\mathcal{A}_{4}^{1}({\color{flavour1}\mathbf{\underline{1}}}\hspace{0.5pt},\hspace{-1.5pt}{\color{flavour3}\mathbf{\overline{6}}}\hspace{0.5pt},\hspace{-1.5pt}\mathbf{5}\hspace{0.5pt},\hspace{-1.5pt}{\color{flavour3}\mathbf{\underline{3}}}\hspace{0.5pt},\hspace{-1.5pt}{\color{flavour1}\mathbf{\overline{7}}}\hspace{0.5pt},\hspace{-1.5pt}{\color{flavour4}\mathbf{\underline{4}}}\hspace{0.5pt},\hspace{-1.5pt}{\color{flavour4}\mathbf{\overline{9}}}\hspace{0.5pt},\hspace{-1.5pt}{\color{flavour2}\mathbf{\underline{2}}}\hspace{0.5pt},\hspace{-1.5pt}{\color{flavour2}\mathbf{\overline{8}}}){-}\mathcal{A}_{4}^{1}({\color{flavour1}\mathbf{\underline{1}}}\hspace{0.5pt},\hspace{-1.5pt}{\color{flavour3}\mathbf{\overline{6}}}\hspace{0.5pt},\hspace{-1.5pt}\mathbf{5}\hspace{0.5pt},\hspace{-1.5pt}{\color{flavour3}\mathbf{\underline{3}}}\hspace{0.5pt},\hspace{-1.5pt}{\color{flavour1}\mathbf{\overline{7}}}\hspace{0.5pt},\hspace{-1.5pt}{\color{flavour4}\mathbf{\overline{9}}}\hspace{0.5pt},\hspace{-1.5pt}{\color{flavour4}\mathbf{\underline{4}}}\hspace{0.5pt},\hspace{-1.5pt}{\color{flavour2}\mathbf{\underline{2}}}\hspace{0.5pt},\hspace{-1.5pt}{\color{flavour2}\mathbf{\overline{8}}})$}
}
\vspace{-50pt}}

\vspace{5pt}\subsectionAppendix{Colour-Dressed Amplitudes Involving Charged Matter}{appendix:colour_dressed_amplitudes}\vspace{-10pt}

\defnBox{distinctFlavourAmp}{\var{input}\patternTwo}{is \built{Total}@\fun{distinctFlavourAmpTerms}\brace{\var{input}}\,.
}

\defnBox{distinctFlavourAmpTerms}{\var{input}\patternTwo}{returns a \built{List} of contributions to the colour-dressed amplitude with \var{input} syntax equivalent for \funL{partialAmpBasis}.\\[-12pt]
\mathematicaBox{
\mathematicaSequence{\funL[1]{distinctFlavourAmpTerms}\brace{2}}{\big\{\funL{amp}\brace{\{\fun{f}\brace{1},\fun{f}\brace{2},\fun{fb}\brace{2},\fun{fb}\brace{1}\,\},\!\{\!\{1,4\},\{2,3\}\!\}}\\
\mbox{}\hspace{5pt}\funL{colourTensor}\brace{\{\fun{f}\brace{1},\fun{f}\brace{2},\fun{fb}\brace{2},\fun{fb}\brace{1}\,\},\!\{\!\{1,4\},\{2,3\}\!\}}\big\}\\[-14pt]}
\mathematicaSequence{\funL[1]{nice}@\%}{$\big\{\mathcal{A}_{2}({\color{flavour1}\mathbf{\underline{1}}}\hspace{0.5pt},\hspace{-1.5pt}{\color{flavour2}\mathbf{\underline{2}}}\hspace{0.5pt},\hspace{-1.5pt}{\color{flavour2}\mathbf{\overline{3}}}\hspace{0.5pt},\hspace{-1.5pt}{\color{flavour1}\mathbf{\overline{4}}})\,\mathcal{C}[{\color{flavour1}\mathbf{\underline{1}}}\hspace{0.5pt},\hspace{-1.5pt}{\color{flavour2}\mathbf{\underline{2}}}\hspace{0.5pt},\hspace{-1.5pt}{\color{flavour2}\mathbf{\overline{3}}}\hspace{0.5pt},\hspace{-1.5pt}{\color{flavour1}\mathbf{\overline{4}}}]\indices{{\color{flavour1}c_1}{\color{flavour2}c_2}}{{\color{flavour2}c_3}{\color{flavour1}c_4}}\big\}$}
\mathematicaSequence{\funL[1]{nice}@\funL[1]{distinctFlavourAmpTerms}\brace{2,1}}{$\big\{\mathcal{A}_{2}^{1}({\color{flavour1}\mathbf{\underline{1}}}\hspace{0.5pt},\hspace{-1.5pt}\mathbf{3}\hspace{0.5pt},\hspace{-1.5pt}{\color{flavour2}\mathbf{\underline{2}}}\hspace{0.5pt},\hspace{-1.5pt}{\color{flavour2}\mathbf{\overline{4}}}\hspace{0.5pt},\hspace{-1.5pt}{\color{flavour1}\mathbf{\overline{5}}})\,\mathcal{C}[{\color{flavour1}\mathbf{\underline{1}}}\hspace{0.5pt},\hspace{-1.5pt}\mathbf{3}\hspace{0.5pt},\hspace{-1.5pt}{\color{flavour2}\mathbf{\underline{2}}}\hspace{0.5pt},\hspace{-1.5pt}{\color{flavour2}\mathbf{\overline{4}}}\hspace{0.5pt},\hspace{-1.5pt}{\color{flavour1}\mathbf{\overline{5}}}]\indices{{\color{flavour1}c_1}{\color{flavour2}c_2}\,\mathfrak{g}_3}{{\color{flavour2}c_4}{\color{flavour1}c_5}},\mathcal{A}_{2}^{1}({\color{flavour1}\mathbf{\underline{1}}}\hspace{0.5pt},\hspace{-1.5pt}{\color{flavour2}\mathbf{\underline{2}}}\hspace{0.5pt},\hspace{-1.5pt}\mathbf{3}\hspace{0.5pt},\hspace{-1.5pt}{\color{flavour2}\mathbf{\overline{4}}}\hspace{0.5pt},\hspace{-1.5pt}{\color{flavour1}\mathbf{\overline{5}}})\,\mathcal{C}[{\color{flavour1}\mathbf{\underline{1}}}\hspace{0.5pt},\hspace{-1.5pt}{\color{flavour2}\mathbf{\underline{2}}}\hspace{0.5pt},\hspace{-1.5pt}\mathbf{3}\hspace{0.5pt},\hspace{-1.5pt}{\color{flavour2}\mathbf{\overline{4}}}\hspace{0.5pt},\hspace{-1.5pt}{\color{flavour1}\mathbf{\overline{5}}}]\indices{{\color{flavour1}c_1}{\color{flavour2}c_2}\,\mathfrak{g}_3}{{\color{flavour2}c_4}{\color{flavour1}c_5}},\\\mbox{}\hspace{7pt}
\mathcal{A}_{2}^{1}({\color{flavour1}\mathbf{\underline{1}}}\hspace{0.5pt},\hspace{-1.5pt}{\color{flavour2}\mathbf{\underline{2}}}\hspace{0.5pt},\hspace{-1.5pt}{\color{flavour2}\mathbf{\overline{4}}}\hspace{0.5pt},\hspace{-1.5pt}\mathbf{3}\hspace{0.5pt},\hspace{-1.5pt}{\color{flavour1}\mathbf{\overline{5}}})\,\mathcal{C}[{\color{flavour1}\mathbf{\underline{1}}}\hspace{0.5pt},\hspace{-1.5pt}{\color{flavour2}\mathbf{\underline{2}}}\hspace{0.5pt},\hspace{-1.5pt}{\color{flavour2}\mathbf{\overline{4}}}\hspace{0.5pt},\hspace{-1.5pt}\mathbf{3}\hspace{0.5pt},\hspace{-1.5pt}{\color{flavour1}\mathbf{\overline{5}}}]\indices{{\color{flavour1}c_1}{\color{flavour2}c_2}\,\mathfrak{g}_3}{{\color{flavour2}c_4}{\color{flavour1}c_5}}\big\}$}
}
}

\defnBox{distinctFlavourAmpSquared}{\var{input}\patternTwo}{returns the \emph{colour-summed} squared amplitude involving distinctly flavoured fermions, as a product of conjugate partial amplitudes contracted against the \funL{colourTensorOverlapMatrix} of \funL{colourTensor} objects.
}

\defnBox{sameFlavourAmp}{\var{input}\patternTwo}{returns \built{Total}@\funL{sameFlavourAmpTerms}\brace{\var{input}}\,.
}

\defnBox{sameFlavourAmpTerms}{\var{input}\patternTwo}{takes the same \var{input} options as \funL{partialAmpBasis} and returns the \built{List} of contributions to the single-flavour amplitude---given as a sum over multi-flavour amplitudes, varying the flavour-pairing of fermions with anti-fermions. 
\mathematicaBox{
\mathematicaSequence{\funL[1]{sameFlavourAmp}\brace{2}}{\big\{\funL{amp}\brace{\{\fun{f}\brace{1},\fun{f}\brace{2},\fun{fb}\brace{2},\fun{fb}\brace{1}\,\},\!\{\!\{1,4\},\{2,3\}\!\}}\\
\mbox{}\hspace{5pt}\funL{colourTensor}\brace{\{\fun{f}\brace{1},\fun{f}\brace{2},\fun{fb}\brace{2},\fun{fb}\brace{1}\,\},\!\{\!\{1,4\},\{2,3\}\!\}}\big\}\\[-14pt]}
\mathematicaSequence{\funL[1]{nice}@\%}{$\big\{\mathcal{A}_{2}({\color{flavour1}\mathbf{\underline{1}}}\hspace{0.5pt},\hspace{-1.5pt}{\color{flavour2}\mathbf{\underline{2}}}\hspace{0.5pt},\hspace{-1.5pt}{\color{flavour2}\mathbf{\overline{3}}}\hspace{0.5pt},\hspace{-1.5pt}{\color{flavour1}\mathbf{\overline{4}}})\,\mathcal{C}[{\color{flavour1}\mathbf{\underline{1}}}\hspace{0.5pt},\hspace{-1.5pt}{\color{flavour2}\mathbf{\underline{2}}}\hspace{0.5pt},\hspace{-1.5pt}{\color{flavour2}\mathbf{\overline{3}}}\hspace{0.5pt},\hspace{-1.5pt}{\color{flavour1}\mathbf{\overline{4}}}]\indices{{\color{flavour1}c_1}{\color{flavour2}c_2}}{{\color{flavour2}c_3}{\color{flavour1}c_4}},\mathcal{A}_{2}({\color{flavour1}\mathbf{\underline{1}}}\hspace{0.5pt},\hspace{-1.5pt}{\color{flavour2}\mathbf{\underline{2}}}\hspace{0.5pt},\hspace{-1.5pt}{\color{flavour2}\mathbf{\overline{4}}}\hspace{0.5pt},\hspace{-1.5pt}{\color{flavour1}\mathbf{\overline{3}}})\,\mathcal{C}[{\color{flavour1}\mathbf{\underline{1}}}\hspace{0.5pt},\hspace{-1.5pt}{\color{flavour2}\mathbf{\underline{2}}}\hspace{0.5pt},\hspace{-1.5pt}{\color{flavour2}\mathbf{\overline{4}}}\hspace{0.5pt},\hspace{-1.5pt}{\color{flavour1}\mathbf{\overline{3}}}]\indices{{\color{flavour1}c_1}{\color{flavour2}c_2}}{{\color{flavour1}c_3}{\color{flavour2}c_4}}\big\}$}
\mathematicaSequence{\funL[1]{nice}@\funL[1]{sameFlavourAmp}\brace{2,1}}{$\big\{\mathcal{A}_{2}^{1}({\color{flavour1}\mathbf{\underline{1}}}\hspace{0.5pt},\hspace{-1.5pt}\mathbf{3}\hspace{0.5pt},\hspace{-1.5pt}{\color{flavour2}\mathbf{\underline{2}}}\hspace{0.5pt},\hspace{-1.5pt}{\color{flavour2}\mathbf{\overline{4}}}\hspace{0.5pt},\hspace{-1.5pt}{\color{flavour1}\mathbf{\overline{5}}})\,\mathcal{C}[{\color{flavour1}\mathbf{\underline{1}}}\hspace{0.5pt},\hspace{-1.5pt}\mathbf{3}\hspace{0.5pt},\hspace{-1.5pt}{\color{flavour2}\mathbf{\underline{2}}}\hspace{0.5pt},\hspace{-1.5pt}{\color{flavour2}\mathbf{\overline{4}}}\hspace{0.5pt},\hspace{-1.5pt}{\color{flavour1}\mathbf{\overline{5}}}]\indices{{\color{flavour1}c_1}{\color{flavour2}c_2}\,\mathfrak{g}_3}{{\color{flavour2}c_4}{\color{flavour1}c_5}},\mathcal{A}_{2}^{1}({\color{flavour1}\mathbf{\underline{1}}}\hspace{0.5pt},\hspace{-1.5pt}{\color{flavour2}\mathbf{\underline{2}}}\hspace{0.5pt},\hspace{-1.5pt}\mathbf{3}\hspace{0.5pt},\hspace{-1.5pt}{\color{flavour2}\mathbf{\overline{4}}}\hspace{0.5pt},\hspace{-1.5pt}{\color{flavour1}\mathbf{\overline{5}}})\,\mathcal{C}[{\color{flavour1}\mathbf{\underline{1}}}\hspace{0.5pt},\hspace{-1.5pt}{\color{flavour2}\mathbf{\underline{2}}}\hspace{0.5pt},\hspace{-1.5pt}\mathbf{3}\hspace{0.5pt},\hspace{-1.5pt}{\color{flavour2}\mathbf{\overline{4}}}\hspace{0.5pt},\hspace{-1.5pt}{\color{flavour1}\mathbf{\overline{5}}}]\indices{{\color{flavour1}c_1}{\color{flavour2}c_2}\,\mathfrak{g}_3}{{\color{flavour2}c_4}{\color{flavour1}c_5}},\\\mbox{}\hspace{7pt}\mathcal{A}_{2}^{1}({\color{flavour1}\mathbf{\underline{1}}}\hspace{0.5pt},\hspace{-1.5pt}{\color{flavour2}\mathbf{\underline{2}}}\hspace{0.5pt},\hspace{-1.5pt}{\color{flavour2}\mathbf{\overline{4}}}\hspace{0.5pt},\hspace{-1.5pt}\mathbf{3}\hspace{0.5pt},\hspace{-1.5pt}{\color{flavour1}\mathbf{\overline{5}}})\,\mathcal{C}[{\color{flavour1}\mathbf{\underline{1}}}\hspace{0.5pt},\hspace{-1.5pt}{\color{flavour2}\mathbf{\underline{2}}}\hspace{0.5pt},\hspace{-1.5pt}{\color{flavour2}\mathbf{\overline{4}}}\hspace{0.5pt},\hspace{-1.5pt}\mathbf{3}\hspace{0.5pt},\hspace{-1.5pt}{\color{flavour1}\mathbf{\overline{5}}}]\indices{{\color{flavour1}c_1}{\color{flavour2}c_2}\,\mathfrak{g}_3}{{\color{flavour2}c_4}{\color{flavour1}c_5}},\mathcal{A}_{2}^{1}({\color{flavour1}\mathbf{\underline{1}}}\hspace{0.5pt},\hspace{-1.5pt}\mathbf{3}\hspace{0.5pt},\hspace{-1.5pt}{\color{flavour2}\mathbf{\underline{2}}}\hspace{0.5pt},\hspace{-1.5pt}{\color{flavour2}\mathbf{\overline{5}}}\hspace{0.5pt},\hspace{-1.5pt}{\color{flavour1}\mathbf{\overline{4}}})\,\mathcal{C}[{\color{flavour1}\mathbf{\underline{1}}}\hspace{0.5pt},\hspace{-1.5pt}\mathbf{3}\hspace{0.5pt},\hspace{-1.5pt}{\color{flavour2}\mathbf{\underline{2}}}\hspace{0.5pt},\hspace{-1.5pt}{\color{flavour2}\mathbf{\overline{5}}}\hspace{0.5pt},\hspace{-1.5pt}{\color{flavour1}\mathbf{\overline{4}}}]\indices{{\color{flavour1}c_1}{\color{flavour2}c_2}\,\mathfrak{g}_3}{{\color{flavour1}c_4}{\color{flavour2}c_5}},\\\mbox{}\hspace{7pt}\mathcal{A}_{2}^{1}({\color{flavour1}\mathbf{\underline{1}}}\hspace{0.5pt},\hspace{-1.5pt}{\color{flavour2}\mathbf{\underline{2}}}\hspace{0.5pt},\hspace{-1.5pt}\mathbf{3}\hspace{0.5pt},\hspace{-1.5pt}{\color{flavour2}\mathbf{\overline{5}}}\hspace{0.5pt},\hspace{-1.5pt}{\color{flavour1}\mathbf{\overline{4}}})\,\mathcal{C}[{\color{flavour1}\mathbf{\underline{1}}}\hspace{0.5pt},\hspace{-1.5pt}{\color{flavour2}\mathbf{\underline{2}}}\hspace{0.5pt},\hspace{-1.5pt}\mathbf{3}\hspace{0.5pt},\hspace{-1.5pt}{\color{flavour2}\mathbf{\overline{5}}}\hspace{0.5pt},\hspace{-1.5pt}{\color{flavour1}\mathbf{\overline{4}}}]\indices{{\color{flavour1}c_1}{\color{flavour2}c_2}\,\mathfrak{g}_3}{{\color{flavour1}c_4}{\color{flavour2}c_5}},\mathcal{A}_{2}^{1}({\color{flavour1}\mathbf{\underline{1}}}\hspace{0.5pt},\hspace{-1.5pt}{\color{flavour2}\mathbf{\underline{2}}}\hspace{0.5pt},\hspace{-1.5pt}{\color{flavour2}\mathbf{\overline{5}}}\hspace{0.5pt},\hspace{-1.5pt}\mathbf{3}\hspace{0.5pt},\hspace{-1.5pt}{\color{flavour1}\mathbf{\overline{4}}})\,\mathcal{C}[{\color{flavour1}\mathbf{\underline{1}}}\hspace{0.5pt},\hspace{-1.5pt}{\color{flavour2}\mathbf{\underline{2}}}\hspace{0.5pt},\hspace{-1.5pt}{\color{flavour2}\mathbf{\overline{5}}}\hspace{0.5pt},\hspace{-1.5pt}\mathbf{3}\hspace{0.5pt},\hspace{-1.5pt}{\color{flavour1}\mathbf{\overline{4}}}]\indices{{\color{flavour1}c_1}{\color{flavour2}c_2}\,\mathfrak{g}_3}{{\color{flavour1}c_4}{\color{flavour2}c_5}}\big\}$}
}
}

\defnBox{sameFlavourAmpSquared}{\var{input}\patternTwo}{returns the \emph{colour-summed} squared amplitude involving indistinguishable fermions, as a product of conjugate partial amplitudes contracted against the \funL{colourTensorOverlapMatrix} of \funL{colourTensor} objects.
}

\newpage
\vspace{0pt}\sectionAppendix{Colour Tensors for Charged Matter Amplitudes}{appendix:colour_tensors}\vspace{-0pt}

\vspace{0pt}\subsectionAppendix{Specifying Charge Generators for Matter Particles}{appendix:specifying_colour_tensors}\vspace{-10pt}

\defnBox{setChargeGenerators}{\var{input}\pattern}{given any set of user-provided charge generators specified by \var{input}---\built{List} or \built{SparseArray} of `generators'---sets (or resets) a number of \built{Global} objects related to the representation, colour tensors, and the corresponding Lie algebra.\\[-10pt]

Alternatively, calling \fun{setChargeGenerators} with \var{input} chosen among the set \{\fun{a}\brace{1}\,,\fun{a}\brace{2}\,,\fun{a}\brace{3}\,,\fun{g}\brace{2}\} will make use of pre-saved generators for the fundamental representation of these Lie algebras specified in the Chevalley basis; if \var{input} is chosen from \{\fun{u}\brace{1}\,,\fun{su}\brace{2}\,,\fun{su}\brace{3}\,,\fun{su}\brace{4}\}, then the charge generators will be defined to be the familiar (Hermitian)/unitary generator matrices for these algebras.\\[-10pt]

\textbf{Note}: Upon initialization of the package \fpackage, the command \fun{setChargeGenerators}\brace{\fun{a}\brace{2}} is called. 

}

\vspace{5pt}\subsubsectionAppendix{Global Objects Defined Upon Initializing Charge Generators}{appendix:global_objects_related_to_charges}\vspace{-10pt}

\defnBox{adjointGenerators}{}{a \built{Global} object which is \built{Set} by the function \fun{setChargeGenerators} representing the generators of the adjoint representation $\mathbf{\r{ad}}(\mathbf{\b{R}})$ for some representation $\mathbf{\b{R}}$. Specifically, it is merely a \built{List} of generators---given by \built{List}@\built{Transpose}\brace{\funL[1]{adjointRepresentation},$\{$2,1,3$\}\!$}---corresponding to \built{SparseArray} matrices of size $\mathrm{dim}(\mathfrak{g})\!\times\!\mathrm{dim}(\mathfrak{g})$.
\mathematicaBox{
\mathematicaSequence[1]{\funL[1]{setChargeGenerators}\brace{\fun{a}\brace{1}\,};\\
\funL[1]{nice}\brace{\funL[1]{adjointGenerators}}
}{$\rule{0pt}{26pt}\left\{\left(\begin{array}{@{$\,\,$}c@{$\,\,$}c@{$\,\,$}c@{}}
{\color{dim}0}&{\color{dim}0}&{\color{dim}0}\\[-5pt]
\fwboxR{0pt}{\text{-}}1&{\color{dim}0}&{\color{dim}0}\\[-5pt]
{\color{dim}0}&2&{\color{dim}0}\end{array}\right),\left(\begin{array}{@{}c@{$\,\,$}c@{$\,\,\,\,$}c@{}}
1&{\color{dim}0}&{\color{dim}0}\\[-5pt]
{\color{dim}0}&{\color{dim}0}&{\color{dim}0}\\[-5pt]
{\color{dim}0}&{\color{dim}0}&\fwboxR{0pt}{\text{-}}1\end{array}\right),\left(\begin{array}{@{}c@{$\,\,\,\,$}c@{$\,\,$}c@{}}
{\color{dim}0}&\fwboxR{0pt}{\text{-}}2&{\color{dim}0}\\[-5pt]
{\color{dim}0}&{\color{dim}0}&1\\[-5pt]
{\color{dim}0}&{\color{dim}0}&{\color{dim}0}\end{array}\right)\right\}$}
\mathematicaSequence[1]{\funL[1]{setChargeGenerators}\brace{\fun{su}\brace{2}\,};\\
\funL[1]{nice}\brace{\funL[1]{adjointGenerators}}
}{$\rule{0pt}{26pt}\left\{\left(\begin{array}{@{}c@{$\,\,$}c@{$\,\,\,\,$}c@{}}
{\color{dim}0}&{\color{dim}0}&{\color{dim}0}\\[-5pt]
{\color{dim}0}&{\color{dim}0}&\fwboxR{0pt}{\text{-}}\built{I}\\[-5pt]
{\color{dim}0}&\built{I}&{\color{dim}0}\end{array}\right),\left(\begin{array}{@{$\,\,$}c@{$\,\,$}c@{$\,\,$}c@{}}
{\color{dim}0}&{\color{dim}0}&\built{I}\\[-5pt]
{\color{dim}0}&{\color{dim}0}&{\color{dim}0}\\[-5pt]
\fwboxR{0pt}{\text{-}}\built{I}&{\color{dim}0}&{\color{dim}0}\end{array}\right),\left(\begin{array}{@{}c@{$\,\,\,\,$}c@{$\,\,$}c@{}}
{\color{dim}0}&\fwboxR{0pt}{\text{-}}\built{I}&{\color{dim}0}\\[-5pt]
\built{I}&{\color{dim}0}&{\color{dim}0}\\[-5pt]
{\color{dim}0}&{\color{dim}0}&{\color{dim}0}\end{array}\right)\right\}$}
}
}

\defnBox{chargeGenerators}{}{a \built{Global} object which is \built{Set} by the function \fun{setChargeGenerators} expressing the generators of the user-specified representation/algebra $\mathbf{\b{R}}$. Specifically, it is a \built{List} of generators---given by \built{List} of \built{Length} $\mathrm{dim}(\mathfrak{g})$ consisting of \built{SparseArray} objects each of size $\mathrm{dim}(\mathbf{\b{R}})\!\times\!\mathrm{dim}(\mathbf{\b{R}})$.\\[-10pt]

\textbf{Note}: The data-format of \funL[1]{chargeGenerators} is exactly as above, regardless of how the user specifies this information when calling \fun{setChargeGenerators}.\\[-12pt]
\mathematicaBox{
\mathematicaSequence[1]{\funL[1]{setChargeGenerators}\brace{\fun{a}\brace{1}\,};\\
\funL[1]{nice}\brace{\funL[1]{chargeGenerators}}
}{$\rule{0pt}{26pt}\left\{\left(\begin{array}{@{}c@{$\,\,$}c@{}}
{\color{dim}0}&{\color{dim}0}\\[-1pt]
1&{\color{dim}0}\end{array}\right),\left(\begin{array}{@{}c@{$\,\,\,\,$}c@{}}
\frac{1}{2}&{\color{dim}0}\\[-1pt]
{\color{dim}0}&\fwboxR{0pt}{\text{-}}\frac{1}{2}\end{array}\right),\left(\begin{array}{@{}c@{$\,\,$}c@{}}
{\color{dim}0}&1\\[-1pt]
{\color{dim}0}&{\color{dim}0}\end{array}\right)\right\}$}
\mathematicaSequence[1]{\funL[1]{setChargeGenerators}\brace{\fun{su}\brace{2}\,};\\
\funL[1]{nice}\brace{\funL[1]{chargeGenerators}}
}{$\rule{0pt}{26pt}\left\{\left(\begin{array}{@{}c@{$\,\,$}c@{}}
{\color{dim}0}&\frac{1}{2}\\[-1pt]
\frac{1}{2}&{\color{dim}0}\end{array}\right),\left(\begin{array}{@{}c@{$\,\,\,\,$}c@{}}
{\color{dim}0}&\fwboxR{0pt}{\text{-}}\frac{\built{I}}{2}\\[-1pt]
\frac{\built{I}}{2}&{\color{dim}0}\end{array}\right),\left(\begin{array}{@{}c@{$\,\,\,\,$}c@{}}
\frac{1}{2}&{\color{dim}0}\\[-1pt]
{\color{dim}0}&\fwboxR{0pt}{\text{-}}\frac{1}{2}\end{array}\right)\right\}$}
}
}

\defnBox{inverseKillingMetric}{}{is a \built{Global} object which is \built{Set} by the function \fun{setChargeGenerators}\brace{} encoding the inverse-killing metric for the algebra generated by the representation defining the charges of fermions. Specifically, it is a rank-$(2,0)$ tensor encoded by a \built{SparseArray} of dimension $\mathrm{dim}(\mathfrak{g})\!\times\!\mathrm{dim}(\mathfrak{g})$.
\mathematicaBox{
\mathematicaSequence[1]{\funL[1]{setChargeGenerators}\brace{\fun{g}\brace{2}\,};\\
\funL[1]{nice}\brace{\funL[1]{inverseKillingMetric}}
}{$\rule{0pt}{86pt}\left(\begin{array}{@{}c@{$\,\,$}c@{$\,\,$}c@{$\,\,$}c@{$\,\,$}c@{$\,\,$}c@{$\,\,$}c@{$\,\,$}c@{$\,\,$}c@{$\,\,$}c@{$\,\,$}c@{$\,\,$}c@{$\,\,$}c@{$\,\,$}c@{}}
{\color{dim}0}&{\color{dim}0}&{\color{dim}0}&{\color{dim}0}&{\color{dim}0}&{\color{dim}0}&{\color{dim}0}&{\color{dim}0}&{\color{dim}0}&{\color{dim}0}&{\color{dim}0}&{\color{dim}0}&{\color{dim}0}&2\\[-5pt]
{\color{dim}0}&{\color{dim}0}&{\color{dim}0}&{\color{dim}0}&{\color{dim}0}&{\color{dim}0}&{\color{dim}0}&{\color{dim}0}&{\color{dim}0}&{\color{dim}0}&{\color{dim}0}&{\color{dim}0}&2&{\color{dim}0}\\[-5pt]
{\color{dim}0}&{\color{dim}0}&{\color{dim}0}&{\color{dim}0}&{\color{dim}0}&{\color{dim}0}&{\color{dim}0}&{\color{dim}0}&{\color{dim}0}&{\color{dim}0}&{\color{dim}0}&6&{\color{dim}0}&{\color{dim}0}\\[-5pt]
{\color{dim}0}&{\color{dim}0}&{\color{dim}0}&{\color{dim}0}&{\color{dim}0}&{\color{dim}0}&{\color{dim}0}&{\color{dim}0}&{\color{dim}0}&{\color{dim}0}&6&{\color{dim}0}&{\color{dim}0}&{\color{dim}0}\\[-5pt]
{\color{dim}0}&{\color{dim}0}&{\color{dim}0}&{\color{dim}0}&{\color{dim}0}&{\color{dim}0}&{\color{dim}0}&{\color{dim}0}&{\color{dim}0}&6&{\color{dim}0}&{\color{dim}0}&{\color{dim}0}&{\color{dim}0}\\[-5pt]
{\color{dim}0}&{\color{dim}0}&{\color{dim}0}&{\color{dim}0}&{\color{dim}0}&{\color{dim}0}&{\color{dim}0}&{\color{dim}0}&2&{\color{dim}0}&{\color{dim}0}&{\color{dim}0}&{\color{dim}0}&{\color{dim}0}\\[-5pt]
{\color{dim}0}&{\color{dim}0}&{\color{dim}0}&{\color{dim}0}&{\color{dim}0}&{\color{dim}0}&12&6&{\color{dim}0}&{\color{dim}0}&{\color{dim}0}&{\color{dim}0}&{\color{dim}0}&{\color{dim}0}\\[-5pt]
{\color{dim}0}&{\color{dim}0}&{\color{dim}0}&{\color{dim}0}&{\color{dim}0}&{\color{dim}0}&6&4&{\color{dim}0}&{\color{dim}0}&{\color{dim}0}&{\color{dim}0}&{\color{dim}0}&{\color{dim}0}\\[-5pt]
{\color{dim}0}&{\color{dim}0}&{\color{dim}0}&{\color{dim}0}&{\color{dim}0}&2&{\color{dim}0}&{\color{dim}0}&{\color{dim}0}&{\color{dim}0}&{\color{dim}0}&{\color{dim}0}&{\color{dim}0}&{\color{dim}0}\\[-5pt]
{\color{dim}0}&{\color{dim}0}&{\color{dim}0}&{\color{dim}0}&6&{\color{dim}0}&{\color{dim}0}&{\color{dim}0}&{\color{dim}0}&{\color{dim}0}&{\color{dim}0}&{\color{dim}0}&{\color{dim}0}&{\color{dim}0}\\[-5pt]
{\color{dim}0}&{\color{dim}0}&{\color{dim}0}&6&{\color{dim}0}&{\color{dim}0}&{\color{dim}0}&{\color{dim}0}&{\color{dim}0}&{\color{dim}0}&{\color{dim}0}&{\color{dim}0}&{\color{dim}0}&{\color{dim}0}\\[-5pt]
{\color{dim}0}&{\color{dim}0}&6&{\color{dim}0}&{\color{dim}0}&{\color{dim}0}&{\color{dim}0}&{\color{dim}0}&{\color{dim}0}&{\color{dim}0}&{\color{dim}0}&{\color{dim}0}&{\color{dim}0}&{\color{dim}0}\\[-5pt]
{\color{dim}0}&2&{\color{dim}0}&{\color{dim}0}&{\color{dim}0}&{\color{dim}0}&{\color{dim}0}&{\color{dim}0}&{\color{dim}0}&{\color{dim}0}&{\color{dim}0}&{\color{dim}0}&{\color{dim}0}&{\color{dim}0}\\[-5pt]
2&{\color{dim}0}&{\color{dim}0}&{\color{dim}0}&{\color{dim}0}&{\color{dim}0}&{\color{dim}0}&{\color{dim}0}&{\color{dim}0}&{\color{dim}0}&{\color{dim}0}&{\color{dim}0}&{\color{dim}0}&{\color{dim}0}\end{array}\right)$}
}
}

\defnBox{killingMetric}{}{is a \built{Global} object which is \built{Set} by the function \funL{setChargeGenerators} encoding the killing metric for the algebra generated by the representation defining the charges of fermions. Specifically, it is a rank-$(0,2)$ tensor encoded by a \built{SparseArray} of dimension $\mathrm{dim}(\mathfrak{g})\!\times\!\mathrm{dim}(\mathfrak{g})$.
\mathematicaBox{
\mathematicaSequence[1]{\funL[1]{setChargeGenerators}\brace{\fun{a}\brace{2}\,};\\
\funL[1]{nice}\brace{\funL[1]{killingMetric}}
}{$\rule{0pt}{52pt}\left(\begin{array}{@{}c@{$\,\,$}c@{$\,\,$}c@{$\,\,\,\,$}c@{$\,\,\,\,$}c@{$\,\,$}c@{$\,\,$}c@{$\,\,$}c@{}}
{\color{dim}0}&{\color{dim}0}&{\color{dim}0}&{\color{dim}0}&{\color{dim}0}&{\color{dim}0}&{\color{dim}0}&1\\[-5pt]
{\color{dim}0}&{\color{dim}0}&{\color{dim}0}&{\color{dim}0}&{\color{dim}0}&{\color{dim}0}&1&{\color{dim}0}\\[-5pt]
{\color{dim}0}&{\color{dim}0}&{\color{dim}0}&{\color{dim}0}&{\color{dim}0}&1&{\color{dim}0}&{\color{dim}0}\\[-5pt]
{\color{dim}0}&{\color{dim}0}&{\color{dim}0}&2&\fwboxR{0pt}{\text{-}}1&{\color{dim}0}&{\color{dim}0}&{\color{dim}0}\\[-5pt]
{\color{dim}0}&{\color{dim}0}&{\color{dim}0}&\fwboxR{0pt}{\text{-}}1&2&{\color{dim}0}&{\color{dim}0}&{\color{dim}0}\\[-5pt]
{\color{dim}0}&{\color{dim}0}&1&{\color{dim}0}&{\color{dim}0}&{\color{dim}0}&{\color{dim}0}&{\color{dim}0}\\[-5pt]
{\color{dim}0}&1&{\color{dim}0}&{\color{dim}0}&{\color{dim}0}&{\color{dim}0}&{\color{dim}0}&{\color{dim}0}\\[-5pt]
1&{\color{dim}0}&{\color{dim}0}&{\color{dim}0}&{\color{dim}0}&{\color{dim}0}&{\color{dim}0}&{\color{dim}0}\end{array}\right)$}
\mathematicaSequence[1]{\funL[1]{setChargeGenerators}\brace{\fun{su}\brace{3}\,};\\
\funL[1]{nice}\brace{\funL[1]{killingMetric}}
}{$\rule{0pt}{52pt}\left(\begin{array}{@{}c@{$\,\,$}c@{$\,\,$}c@{$\,\,$}c@{$\,\,$}c@{$\,\,$}c@{$\,\,$}c@{$\,\,$}c@{}}
2&{\color{dim}0}&{\color{dim}0}&{\color{dim}0}&{\color{dim}0}&{\color{dim}0}&{\color{dim}0}&{\color{dim}0}\\[-5pt]
{\color{dim}0}&2&{\color{dim}0}&{\color{dim}0}&{\color{dim}0}&{\color{dim}0}&{\color{dim}0}&{\color{dim}0}\\[-5pt]
{\color{dim}0}&{\color{dim}0}&2&{\color{dim}0}&{\color{dim}0}&{\color{dim}0}&{\color{dim}0}&{\color{dim}0}\\[-5pt]
{\color{dim}0}&{\color{dim}0}&{\color{dim}0}&2&{\color{dim}0}&{\color{dim}0}&{\color{dim}0}&{\color{dim}0}\\[-5pt]
{\color{dim}0}&{\color{dim}0}&{\color{dim}0}&{\color{dim}0}&2&{\color{dim}0}&{\color{dim}0}&{\color{dim}0}\\[-5pt]
{\color{dim}0}&{\color{dim}0}&{\color{dim}0}&{\color{dim}0}&{\color{dim}0}&2&{\color{dim}0}&{\color{dim}0}\\[-5pt]
{\color{dim}0}&{\color{dim}0}&{\color{dim}0}&{\color{dim}0}&{\color{dim}0}&{\color{dim}0}&2&{\color{dim}0}\\[-5pt]
{\color{dim}0}&{\color{dim}0}&{\color{dim}0}&{\color{dim}0}&{\color{dim}0}&{\color{dim}0}&{\color{dim}0}&2\end{array}\right)$}
}
}

\newpage
\vspace{5pt}\subsectionAppendix{Building Concrete Colour Tensors}{appendix:building_colour_tensors}\vspace{-14pt}
%
\defnBox{buildColourTensors}{\var{expression}\pattern}{takes a \funL{colourTensor} object and converts it to an explicit tensor, representing it as a \built{SparseArray} object, using the charge generators specified upon calling \funL{setChargeGenerators}.
\mathematicaBox{
\mathematicaSequence[1]{\funL[1]{setChargeGenerators}\brace{\fun{g}\brace{2}};\\
\funL[1]{buildColourTensors}\brace{\funL[1]{distinctFlavourAmpTerms}\brace{2,1}}}{\vspace{-20pt}\rule[-54pt]{0pt}{65pt}}

\mathematicaSequence{\built{First}\brace{\built{ArrayRules}\brace{\built{First}@\%}}}{$\{1,1,7,1,1\}\to{-}\frac{2}{3}\funL{amp}\brace{\{\fun{f}\brace{1},\fun{g}\brace{3},\fun{f}\brace{2},\fun{fb}\brace{2},\fun{fb}\brace{1}\,\},\!\{\!\{1,5\},\{3,4\}\!\}}$}
}
~\\[-100pt]
\fwboxL{0pt}{\hspace{15pt}\vspace{-20pt}$\big\{\scalebox{0.9}{\sparse{1024}{7,7,14,7,7}},\,\scalebox{0.9}{\sparse{1378}{7,7,14,7,7}},$}\\[2pt]\fwboxL{0pt}{\hspace{22pt}$\scalebox{0.9}{\sparse{1024}{7,7,14,7,7}}$\big\}}~\\[34pt]

We can use \fun{buildColourTensors} to evaluate concrete objects such as \funL{colourTensorOverlapMatrix}:
\mathematicaBox{%
\mathematicaSequence[1]{tensors=\funL[1]{sameFlavourAmpTerms}\brace{2,1}/.\fun{amp}\brace{\var{x}\patternTwo}$\,\mapsto$1;\\
\funL[1]{nice}@\%}{$\rule{0pt}{18pt}\mbox{}\hspace{-5pt}\big\{\mathcal{C}[{\color{flavour1}\mathbf{\underline{1}}}\hspace{1.03pt},\hspace{-1.5pt}\mathbf{3}\hspace{1.03pt},\hspace{-1.5pt}{\color{flavour2}\mathbf{\underline{2}}}\hspace{1.03pt},\hspace{-1.5pt}{\color{flavour2}\mathbf{\overline{4}}}\hspace{1.03pt},\hspace{-1.5pt}{\color{flavour1}\mathbf{\overline{5}}}]\indices{{\color{flavour1}c_{1}}\,{\color{flavour2}c_{2}}\,\mathfrak{g}_{3}}{{\color{flavour2}c_{4}}\,{\color{flavour1}c_{5}}},\mathcal{C}[{\color{flavour1}\mathbf{\underline{1}}}\hspace{1.03pt},\hspace{-1.5pt}{\color{flavour2}\mathbf{\underline{2}}}\hspace{1.03pt},\hspace{-1.5pt}\mathbf{3}\hspace{1.03pt},\hspace{-1.5pt}{\color{flavour2}\mathbf{\overline{4}}}\hspace{1.03pt},\hspace{-1.5pt}{\color{flavour1}\mathbf{\overline{5}}}]\indices{{\color{flavour1}c_{1}}\,{\color{flavour2}c_{2}}\,\mathfrak{g}_{3}}{{\color{flavour2}c_{4}}\,{\color{flavour1}c_{5}}},\mathcal{C}[{\color{flavour1}\mathbf{\underline{1}}}\hspace{1.03pt},\hspace{-1.5pt}{\color{flavour2}\mathbf{\underline{2}}}\hspace{1.03pt},\hspace{-1.5pt}{\color{flavour2}\mathbf{\overline{4}}}\hspace{1.03pt},\hspace{-1.5pt}\mathbf{3}\hspace{1.03pt},\hspace{-1.5pt}{\color{flavour1}\mathbf{\overline{5}}}]\indices{{\color{flavour1}c_{1}}\,{\color{flavour2}c_{2}}\,\mathfrak{g}_{3}}{{\color{flavour2}c_{4}}\,{\color{flavour1}c_{5}}},$\\$\mathcal{C}[{\color{flavour1}\mathbf{\underline{1}}}\hspace{1.03pt},\hspace{-1.5pt}\mathbf{3}\hspace{1.03pt},\hspace{-1.5pt}{\color{flavour2}\mathbf{\underline{2}}}\hspace{1.03pt},\hspace{-1.5pt}{\color{flavour2}\mathbf{\overline{5}}}\hspace{1.03pt},\hspace{-1.5pt}{\color{flavour1}\mathbf{\overline{4}}}]\indices{{\color{flavour1}c_{1}}\,{\color{flavour2}c_{2}}\,\mathfrak{g}_{3}}{{\color{flavour1}c_{4}}\,{\color{flavour2}c_{5}}},\mathcal{C}[{\color{flavour1}\mathbf{\underline{1}}}\hspace{1.03pt},\hspace{-1.5pt}{\color{flavour2}\mathbf{\underline{2}}}\hspace{1.03pt},\hspace{-1.5pt}\mathbf{3}\hspace{1.03pt},\hspace{-1.5pt}{\color{flavour2}\mathbf{\overline{5}}}\hspace{1.03pt},\hspace{-1.5pt}{\color{flavour1}\mathbf{\overline{4}}}]\indices{{\color{flavour1}c_{1}}\,{\color{flavour2}c_{2}}\,\mathfrak{g}_{3}}{{\color{flavour1}c_{4}}\,{\color{flavour2}c_{5}}},\mathcal{C}[{\color{flavour1}\mathbf{\underline{1}}}\hspace{1.03pt},\hspace{-1.5pt}{\color{flavour2}\mathbf{\underline{2}}}\hspace{1.03pt},\hspace{-1.5pt}{\color{flavour2}\mathbf{\overline{5}}}\hspace{1.03pt},\hspace{-1.5pt}\mathbf{3}\hspace{1.03pt},\hspace{-1.5pt}{\color{flavour1}\mathbf{\overline{4}}}]\indices{{\color{flavour1}c_{1}}\,{\color{flavour2}c_{2}}\,\mathfrak{g}_{3}}{{\color{flavour1}c_{4}}\,{\color{flavour2}c_{5}}}\big\}\vspace{-5pt}$}
\mathematicaSequence[1]{\funL[1]{setChargeGenerators}\brace{\fun{g}\brace{2}};\\
\funL[1]{nice}@\funL[1]{buildColourTensors}\brace{\fun{colourTensorOverlapMatrix}@tensors}}{\vspace{-10pt}$\left(\begin{array}{@{}c@{$\,\,$}c@{$\,\,$}c@{$\,\,\,\,$}c@{$\,\,$}c@{$\,\,\,\,$}c@{}}
{\color{dim}0}&14&28&{\color{dim}0}&{\color{dim}0}&{\color{dim}0}\\[-5pt]
14&28&14&{\color{dim}0}&{\color{dim}0}&{\color{dim}0}\\[-5pt]
28&14&{\color{dim}0}&{\color{dim}0}&{\color{dim}0}&{\color{dim}0}\\[-5pt]
{\color{dim}0}&{\color{dim}0}&{\color{dim}0}&14&14&\fwboxR{0pt}{\text{-}}14\\[-5pt]
{\color{dim}0}&{\color{dim}0}&{\color{dim}0}&14&28&14\\[-5pt]
{\color{dim}0}&{\color{dim}0}&{\color{dim}0}&\fwboxR{0pt}{\text{-}}14&14&14\end{array}\right)$}
}
\vspace{-5pt}
Or we can use it to verify the equivalence between different choices of basis:\\[-24pt]

\mathematicaBox{%
\mathematicaSequence[2]{default=\funL[1]{distinctFlavourAmp}\brace{2,1};\\
alt=(default/.\{\fun{f}\brace{1}$\to$\fun{f}\brace{2},\fun{f}\brace{2}$\to$\fun{f}\brace{1},\fun{fb}\brace{1}$\to$\fun{fb}\brace{2},\fun{fb}\brace{2}$\to$\fun{fb}\brace{1}\,\};\\
\built{Column}@\funL[1]{nice}/@\{default,alt\}}{\vspace{0pt}$\begin{array}{@{}l@{}}\\[-24pt]\phantom{{+}}\mathcal{A}_{2}^{1}({\color{flavour1}\mathbf{\underline{1}}}\hspace{1.02pt},\hspace{-1.5pt}{\color{flavour2}\mathbf{\underline{2}}}\hspace{1.02pt},\hspace{-1.5pt}{\color{flavour2}\mathbf{\overline{4}}}\hspace{1.02pt},\hspace{-1.5pt}\mathbf{3}\hspace{1.02pt},\hspace{-1.5pt}{\color{flavour1}\mathbf{\overline{5}}})\mathcal{C}[{\color{flavour1}\mathbf{\underline{1}}}\hspace{1.03pt},\hspace{-1.5pt}{\color{flavour2}\mathbf{\underline{2}}}\hspace{1.03pt},\hspace{-1.5pt}{\color{flavour2}\mathbf{\overline{4}}}\hspace{1.03pt},\hspace{-1.5pt}\mathbf{3}\hspace{1.03pt},\hspace{-1.5pt}{\color{flavour1}\mathbf{\overline{5}}}]\indices{{\color{flavour1}c_{1}}\,{\color{flavour2}c_{2}}\,\mathfrak{g}_{3}}{{\color{flavour2}c_{4}}\,{\color{flavour1}c_{5}}}{+}\mathcal{A}_{2}^{1}({\color{flavour1}\mathbf{\underline{1}}}\hspace{1.02pt},\hspace{-1.5pt}{\color{flavour2}\mathbf{\underline{2}}}\hspace{1.02pt},\hspace{-1.5pt}\mathbf{3}\hspace{1.02pt},\hspace{-1.5pt}{\color{flavour2}\mathbf{\overline{4}}}\hspace{1.02pt},\hspace{-1.5pt}{\color{flavour1}\mathbf{\overline{5}}})\mathcal{C}[{\color{flavour1}\mathbf{\underline{1}}}\hspace{1.03pt},\hspace{-1.5pt}{\color{flavour2}\mathbf{\underline{2}}}\hspace{1.03pt},\hspace{-1.5pt}\mathbf{3}\hspace{1.03pt},\hspace{-1.5pt}{\color{flavour2}\mathbf{\overline{4}}}\hspace{1.03pt},\hspace{-1.5pt}{\color{flavour1}\mathbf{\overline{5}}}]\indices{{\color{flavour1}c_{1}}\,{\color{flavour2}c_{2}}\,\mathfrak{g}_{3}}{{\color{flavour2}c_{4}}\,{\color{flavour1}c_{5}}}\\{+}\mathcal{A}_{2}^{1}({\color{flavour1}\mathbf{\underline{1}}}\hspace{1.02pt},\hspace{-1.5pt}\mathbf{3}\hspace{1.02pt},\hspace{-1.5pt}{\color{flavour2}\mathbf{\underline{2}}}\hspace{1.02pt},\hspace{-1.5pt}{\color{flavour2}\mathbf{\overline{4}}}\hspace{1.02pt},\hspace{-1.5pt}{\color{flavour1}\mathbf{\overline{5}}})\mathcal{C}[{\color{flavour1}\mathbf{\underline{1}}}\hspace{1.03pt},\hspace{-1.5pt}\mathbf{3}\hspace{1.03pt},\hspace{-1.5pt}{\color{flavour2}\mathbf{\underline{2}}}\hspace{1.03pt},\hspace{-1.5pt}{\color{flavour2}\mathbf{\overline{4}}}\hspace{1.03pt},\hspace{-1.5pt}{\color{flavour1}\mathbf{\overline{5}}}]\indices{{\color{flavour1}c_{1}}\,{\color{flavour2}c_{2}}\,\mathfrak{g}_{3}}{{\color{flavour2}c_{4}}\,{\color{flavour1}c_{5}}}\\[5pt]
\phantom{{+}}\mathcal{A}_{2}^{1}({\color{flavour2}\mathbf{\underline{2}}}\hspace{1.02pt},\hspace{-1.5pt}{\color{flavour1}\mathbf{\underline{1}}}\hspace{1.02pt},\hspace{-1.5pt}{\color{flavour1}\mathbf{\overline{5}}}\hspace{1.02pt},\hspace{-1.5pt}\mathbf{3}\hspace{1.02pt},\hspace{-1.5pt}{\color{flavour2}\mathbf{\overline{4}}})\mathcal{C}[{\color{flavour2}\mathbf{\underline{2}}}\hspace{1.03pt},\hspace{-1.5pt}{\color{flavour1}\mathbf{\underline{1}}}\hspace{1.03pt},\hspace{-1.5pt}{\color{flavour1}\mathbf{\overline{5}}}\hspace{1.03pt},\hspace{-1.5pt}\mathbf{3}\hspace{1.03pt},\hspace{-1.5pt}{\color{flavour2}\mathbf{\overline{4}}}]\indices{{\color{flavour1}c_{1}}\,{\color{flavour2}c_{2}}\,\mathfrak{g}_{3}}{{\color{flavour2}c_{4}}\,{\color{flavour1}c_{5}}}{+}\mathcal{A}_{2}^{1}({\color{flavour2}\mathbf{\underline{2}}}\hspace{1.02pt},\hspace{-1.5pt}{\color{flavour1}\mathbf{\underline{1}}}\hspace{1.02pt},\hspace{-1.5pt}\mathbf{3}\hspace{1.02pt},\hspace{-1.5pt}{\color{flavour1}\mathbf{\overline{5}}}\hspace{1.02pt},\hspace{-1.5pt}{\color{flavour2}\mathbf{\overline{4}}})\mathcal{C}[{\color{flavour2}\mathbf{\underline{2}}}\hspace{1.03pt},\hspace{-1.5pt}{\color{flavour1}\mathbf{\underline{1}}}\hspace{1.03pt},\hspace{-1.5pt}\mathbf{3}\hspace{1.03pt},\hspace{-1.5pt}{\color{flavour1}\mathbf{\overline{5}}}\hspace{1.03pt},\hspace{-1.5pt}{\color{flavour2}\mathbf{\overline{4}}}]\indices{{\color{flavour1}c_{1}}\,{\color{flavour2}c_{2}}\,\mathfrak{g}_{3}}{{\color{flavour2}c_{4}}\,{\color{flavour1}c_{5}}}\\
{+}\mathcal{A}_{2}^{1}({\color{flavour2}\mathbf{\underline{2}}}\hspace{1.02pt},\hspace{-1.5pt}\mathbf{3}\hspace{1.02pt},\hspace{-1.5pt}{\color{flavour1}\mathbf{\underline{1}}}\hspace{1.02pt},\hspace{-1.5pt}{\color{flavour1}\mathbf{\overline{5}}}\hspace{1.02pt},\hspace{-1.5pt}{\color{flavour2}\mathbf{\overline{4}}})\mathcal{C}[{\color{flavour2}\mathbf{\underline{2}}}\hspace{1.03pt},\hspace{-1.5pt}\mathbf{3}\hspace{1.03pt},\hspace{-1.5pt}{\color{flavour1}\mathbf{\underline{1}}}\hspace{1.03pt},\hspace{-1.5pt}{\color{flavour1}\mathbf{\overline{5}}}\hspace{1.03pt},\hspace{-1.5pt}{\color{flavour2}\mathbf{\overline{4}}}]\indices{{\color{flavour1}c_{1}}\,{\color{flavour2}c_{2}}\,\mathfrak{g}_{3}}{{\color{flavour2}c_{4}}\,{\color{flavour1}c_{5}}}\end{array}$}
\mathematicaSequence[1]{\funL[1]{setChargeGenerators}\brace{\fun{g}\brace{2}};\\
\funL[1]{buildColourTensors}\brace{default-alt}\vspace{10pt}}{\vspace{-5pt}$\begin{array}{@{}c@{}}\\[-31pt]\includegraphics[scale=1]{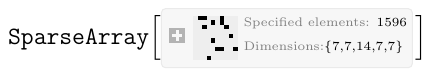}\end{array}\vspace{-12pt}$}
\mathematicaSequence{\funL[1]{toSingleFlavour}@\%\vspace{10pt}}{\vspace{-5pt}$\begin{array}{@{}c@{}}\\[-31pt]\includegraphics[scale=1]{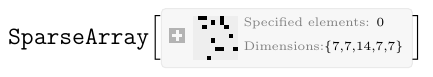}\end{array}\vspace{-5pt}$}
}
~\\[-50pt]
}

\newpage
\vspace{5pt}\subsectionAppendix{Linear Relations Between Concrete Colour Tensors}{appendix:identities_among_colourTensors}\vspace{-10pt}
%
~\\[-40pt]\vspace{20pt}
\defnBox{colourTensorRelations}{\var{expression}\pattern}{for all the \fun{colourTensor} (or \fun{colorFactor}) objects appearing in \var{expression}, returns an elimination \built{Rule} expressing any linear relations satisfied among them \emph{for the specified set of} \built{chargeGenerators}.
\mathematicaBox{%
\mathematicaSequence[2]{\funL[1]{setChargeGenerators}\brace{\fun{a}\brace{2}\,};\\
relations=\funL[1]{colourTensorRelations}@\funL[1]{sameFlavourAmp}\brace{2,1};\\
\built{Column}@\funL[1]{nice}@relations\\[-7pt]}{\vspace{-5pt}$\begin{array}{@{}l@{}l@{}l@{}}~\\[-20pt]
\mathcal{C}[{\color{flavour1}\mathbf{\underline{1}}}\hspace{1.03pt},\hspace{-1.5pt}{\color{flavour2}\mathbf{\underline{2}}}\hspace{1.03pt},\hspace{-1.5pt}\mathbf{3}\hspace{1.03pt},\hspace{-1.5pt}{\color{flavour2}\mathbf{\overline{5}}}\hspace{1.03pt},\hspace{-1.5pt}{\color{flavour1}\mathbf{\overline{4}}}]\indices{{\color{flavour1}c_{1}}\,{\color{flavour2}c_{2}}\,\mathfrak{g}_{3}}{{\color{flavour1}c_{4}}\,{\color{flavour2}c_{5}}}&\to&\phantom{{+}}\frac{10}{3}\,\mathcal{C}[{\color{flavour1}\mathbf{\underline{1}}}\hspace{1.03pt},\hspace{-1.5pt}\mathbf{3}\hspace{1.03pt},\hspace{-1.5pt}{\color{flavour2}\mathbf{\underline{2}}}\hspace{1.03pt},\hspace{-1.5pt}{\color{flavour2}\mathbf{\overline{4}}}\hspace{1.03pt},\hspace{-1.5pt}{\color{flavour1}\mathbf{\overline{5}}}]\indices{{\color{flavour1}c_{1}}\,{\color{flavour2}c_{2}}\,\mathfrak{g}_{3}}{{\color{flavour2}c_{4}}\,{\color{flavour1}c_{5}}}\!{+}\frac{8}{3}\,\mathcal{C}[{\color{flavour1}\mathbf{\underline{1}}}\hspace{1.03pt},\hspace{-1.5pt}{\color{flavour2}\mathbf{\underline{2}}}\hspace{1.03pt},\hspace{-1.5pt}{\color{flavour2}\mathbf{\overline{4}}}\hspace{1.03pt},\hspace{-1.5pt}\mathbf{3}\hspace{1.03pt},\hspace{-1.5pt}{\color{flavour1}\mathbf{\overline{5}}}]\indices{{\color{flavour1}c_{1}}\,{\color{flavour2}c_{2}}\,\mathfrak{g}_{3}}{{\color{flavour2}c_{4}}\,{\color{flavour1}c_{5}}}\!\\
&&{+}\,2\,\,\,\mathcal{C}[{\color{flavour1}\mathbf{\underline{1}}}\hspace{1.03pt},\hspace{-1.5pt}\mathbf{3}\hspace{1.03pt},\hspace{-1.5pt}{\color{flavour2}\mathbf{\underline{2}}}\hspace{1.03pt},\hspace{-1.5pt}{\color{flavour2}\mathbf{\overline{5}}}\hspace{1.03pt},\hspace{-1.5pt}{\color{flavour1}\mathbf{\overline{4}}}]\indices{{\color{flavour1}c_{1}}\,{\color{flavour2}c_{2}}\,\mathfrak{g}_{3}}{{\color{flavour1}c_{4}}\,{\color{flavour2}c_{5}}}\!{-}\,3\,\,\mathcal{C}[{\color{flavour1}\mathbf{\underline{1}}}\hspace{1.03pt},\hspace{-1.5pt}{\color{flavour2}\mathbf{\underline{2}}}\hspace{1.03pt},\hspace{-1.5pt}\mathbf{3}\hspace{1.03pt},\hspace{-1.5pt}{\color{flavour2}\mathbf{\overline{4}}}\hspace{1.03pt},\hspace{-1.5pt}{\color{flavour1}\mathbf{\overline{5}}}]\indices{{\color{flavour1}c_{1}}\,{\color{flavour2}c_{2}}\,\mathfrak{g}_{3}}{{\color{flavour2}c_{4}}\,{\color{flavour1}c_{5}}}\!\\[5pt]
\mathcal{C}[{\color{flavour1}\mathbf{\underline{1}}}\hspace{1.03pt},\hspace{-1.5pt}{\color{flavour2}\mathbf{\underline{2}}}\hspace{1.03pt},\hspace{-1.5pt}{\color{flavour2}\mathbf{\overline{5}}}\hspace{1.03pt},\hspace{-1.5pt}\mathbf{3}\hspace{1.03pt},\hspace{-1.5pt}{\color{flavour1}\mathbf{\overline{4}}}]\indices{{\color{flavour1}c_{1}}\,{\color{flavour2}c_{2}}\,\mathfrak{g}_{3}}{{\color{flavour1}c_{4}}\,{\color{flavour2}c_{5}}}&\to&\phantom{{+}}3\,\mathcal{C}[{\color{flavour1}\mathbf{\underline{1}}}\hspace{1.03pt},\hspace{-1.5pt}{\color{flavour2}\mathbf{\underline{2}}}\hspace{1.03pt},\hspace{-1.5pt}{\color{flavour2}\mathbf{\overline{4}}}\hspace{1.03pt},\hspace{-1.5pt}\mathbf{3}\hspace{1.03pt},\hspace{-1.5pt}{\color{flavour1}\mathbf{\overline{5}}}]\indices{{\color{flavour1}c_{1}}\,{\color{flavour2}c_{2}}\,\mathfrak{g}_{3}}{{\color{flavour2}c_{4}}\,{\color{flavour1}c_{5}}}\!{+}3\,\mathcal{C}[{\color{flavour1}\mathbf{\underline{1}}}\hspace{1.03pt},\hspace{-1.5pt}\mathbf{3}\hspace{1.03pt},\hspace{-1.5pt}{\color{flavour2}\mathbf{\underline{2}}}\hspace{1.03pt},\hspace{-1.5pt}{\color{flavour2}\mathbf{\overline{4}}}\hspace{1.03pt},\hspace{-1.5pt}{\color{flavour1}\mathbf{\overline{5}}}]\indices{{\color{flavour1}c_{1}}\,{\color{flavour2}c_{2}}\,\mathfrak{g}_{3}}{{\color{flavour2}c_{4}}\,{\color{flavour1}c_{5}}}\!\\
&&{+}\phantom{1}\,\mathcal{C}[{\color{flavour1}\mathbf{\underline{1}}}\hspace{1.03pt},\hspace{-1.5pt}\mathbf{3}\hspace{1.03pt},\hspace{-1.5pt}{\color{flavour2}\mathbf{\underline{2}}}\hspace{1.03pt},\hspace{-1.5pt}{\color{flavour2}\mathbf{\overline{5}}}\hspace{1.03pt},\hspace{-1.5pt}{\color{flavour1}\mathbf{\overline{4}}}]\indices{{\color{flavour1}c_{1}}\,{\color{flavour2}c_{2}}\,\mathfrak{g}_{3}}{{\color{flavour1}c_{4}}\,{\color{flavour2}c_{5}}}\!{-}3\,\mathcal{C}[{\color{flavour1}\mathbf{\underline{1}}}\hspace{1.03pt},\hspace{-1.5pt}{\color{flavour2}\mathbf{\underline{2}}}\hspace{1.03pt},\hspace{-1.5pt}\mathbf{3}\hspace{1.03pt},\hspace{-1.5pt}{\color{flavour2}\mathbf{\overline{4}}}\hspace{1.03pt},\hspace{-1.5pt}{\color{flavour1}\mathbf{\overline{5}}}]\indices{{\color{flavour1}c_{1}}\,{\color{flavour2}c_{2}}\,\mathfrak{g}_{3}}{{\color{flavour2}c_{4}}\,{\color{flavour1}c_{5}}}\!
\end{array}$}
\mathematicaSequence[1]{\funL[1]{setChargeGenerators}\brace{\fun{g}\brace{2}\,};\\
relations=\funL[1]{colourTensorRelations}@\funL[1]{sameFlavourAmp}\brace{2,1};}{$\{\}$
}
}
An interesting application is the comparison against DDM for the case of $\mathbf{\r{ad}}(\mathbf{\b{R}})$-charged matter:
\mathematicaBox{%
\mathematicaSequence[5]{\funL[1]{setChargeGenerators}\brace{\fun{a}\brace{2}\,};\\
\funL[1]{setChargeGenerators}\brace{\funL[1]{adjointGenerators}\,};\\
newForm=\funL[1]{sameFlavourAmpTerms}\brace{2,1};\\
ddmColorFactors=\fun{colorFactor}\brace{1,\#\#,5}\&@@@\built{Permutations}\brace{\built{Range}\brace{2,4}\,};\\
relationsRule=\funL[1]{colourTensorRelations}\brace{\{ddmColorFactors,newForm\}};\\
\built{Column}@\funL[1]{nice}@\%\\[-5pt]}{\vspace{5pt}$\begin{array}{@{}lll@{}}~\\[-42pt]
\mathcal{C}[{\color{flavour1}\mathbf{\underline{1}}}\hspace{0.5pt},\hspace{-1.5pt}\mathbf{3}\hspace{0.5pt},\hspace{-1.5pt}{\color{flavour2}\mathbf{\underline{2}}}\hspace{0.5pt},\hspace{-1.5pt}{\color{flavour2}\mathbf{\overline{4}}}\hspace{0.5pt},\hspace{-1.5pt}{\color{flavour1}\mathbf{\overline{5}}}]&\to&\phantom{{+}}\egCFc{-}\egCFd\\[-10.5pt]
\mathcal{C}[{\color{flavour1}\mathbf{\underline{1}}}\hspace{0.5pt},\hspace{-1.5pt}{\color{flavour2}\mathbf{\underline{2}}}\hspace{0.5pt},\hspace{-1.5pt}\mathbf{3}\hspace{0.5pt},\hspace{-1.5pt}{\color{flavour2}\mathbf{\overline{4}}}\hspace{0.5pt},\hspace{-1.5pt}{\color{flavour1}\mathbf{\overline{5}}}]&\to&\phantom{{+}}\egCFa{+}\egCFf{-}\egCFd{-}\egCFe\\[-10.5pt]
\mathcal{C}[{\color{flavour1}\mathbf{\underline{1}}}\hspace{0.5pt},\hspace{-1.5pt}{\color{flavour2}\mathbf{\underline{2}}}\hspace{0.5pt},\hspace{-1.5pt}{\color{flavour2}\mathbf{\overline{4}}}\hspace{0.5pt},\hspace{-1.5pt}\mathbf{3}\hspace{0.5pt},\hspace{-1.5pt}{\color{flavour1}\mathbf{\overline{5}}}]&\to&\phantom{{+}}\egCFb{-}\egCFe\\[-10.5pt]
\mathcal{C}[{\color{flavour1}\mathbf{\underline{1}}}\hspace{0.5pt},\hspace{-1.5pt}\mathbf{3}\hspace{0.5pt},\hspace{-1.5pt}{\color{flavour2}\mathbf{\underline{2}}}\hspace{0.5pt},\hspace{-1.5pt}{\color{flavour2}\mathbf{\overline{5}}}\hspace{0.5pt},\hspace{-1.5pt}{\color{flavour1}\mathbf{\overline{4}}}]&\to&{-}\egCFd\\[-10.5pt]
\mathcal{C}[{\color{flavour1}\mathbf{\underline{1}}}\hspace{0.5pt},\hspace{-1.5pt}{\color{flavour2}\mathbf{\underline{2}}}\hspace{0.5pt},\hspace{-1.5pt}\mathbf{3}\hspace{0.5pt},\hspace{-1.5pt}{\color{flavour2}\mathbf{\overline{5}}}\hspace{0.5pt},\hspace{-1.5pt}{\color{flavour1}\mathbf{\overline{4}}}]&\to&\phantom{{+}}\egCFf{-}\egCFd{-}\egCFe\\[-10.5pt]
\mathcal{C}[{\color{flavour1}\mathbf{\underline{1}}}\hspace{0.5pt},\hspace{-1.5pt}{\color{flavour2}\mathbf{\underline{2}}}\hspace{0.5pt},\hspace{-1.5pt}{\color{flavour2}\mathbf{\overline{5}}}\hspace{0.5pt},\hspace{-1.5pt}\mathbf{3}\hspace{0.5pt},\hspace{-1.5pt}{\color{flavour1}\mathbf{\overline{4}}}]&\to&\phantom{{+}}\egCFf{-}\egCFd\\[-12pt]
\end{array}$\vspace{2pt}}
\mathematicaSequence{\smash{\funL[1]{nice}@\built{Expand}@(\funL[1]{toSingleFlavour}\brace{\built{Total}@newForm}/.relationsRule)}}{
$\phantom{{+}}{A}_{2}^{1}({\color{flavour0}\mathbf{\underline{1}}}\hspace{1.02pt},\hspace{-1.5pt}{\color{flavour0}\mathbf{\underline{2}}}\hspace{1.02pt},\hspace{-1.5pt}\mathbf{3}\hspace{1.02pt},\hspace{-1.5pt}{\color{flavour0}\mathbf{\overline{4}}}\hspace{1.02pt},\hspace{-1.5pt}{\color{flavour0}\mathbf{\overline{5}}})\egCFa{+}{A}_{2}^{1}({\color{flavour0}\mathbf{\underline{1}}}\hspace{1.02pt},\hspace{-1.5pt}{\color{flavour0}\mathbf{\underline{2}}}\hspace{1.02pt},\hspace{-1.5pt}{\color{flavour0}\mathbf{\overline{4}}}\hspace{1.02pt},\hspace{-1.5pt}\mathbf{3}\hspace{1.02pt},\hspace{-1.5pt}{\color{flavour0}\mathbf{\overline{5}}})\egCFb$\\[-10pt]${+}{A}_{2}^{1}({\color{flavour0}\mathbf{\underline{1}}}\hspace{1.02pt},\hspace{-1.5pt}\mathbf{3}\hspace{1.02pt},\hspace{-1.5pt}{\color{flavour0}\mathbf{\underline{2}}}\hspace{1.02pt},\hspace{-1.5pt}{\color{flavour0}\mathbf{\overline{4}}}\hspace{1.02pt},\hspace{-1.5pt}{\color{flavour0}\mathbf{\overline{5}}})\egCFc{+}{A}_{2}^{1}({\color{flavour0}\mathbf{\underline{1}}}\hspace{1.02pt},\hspace{-1.5pt}\mathbf{3}\hspace{1.02pt},\hspace{-1.5pt}{\color{flavour0}\mathbf{\overline{4}}}\hspace{1.02pt},\hspace{-1.5pt}{\color{flavour0}\mathbf{\underline{2}}}\hspace{1.02pt},\hspace{-1.5pt}{\color{flavour0}\mathbf{\overline{5}}})\egCFd$\\[-10pt]
${+}{A}_{2}^{1}({\color{flavour0}\mathbf{\underline{1}}}\hspace{1.02pt},\hspace{-1.5pt}{\color{flavour0}\mathbf{\overline{4}}}\hspace{1.02pt},\hspace{-1.5pt}{\color{flavour0}\mathbf{\underline{2}}}\hspace{1.02pt},\hspace{-1.5pt}\mathbf{3}\hspace{1.02pt},\hspace{-1.5pt}{\color{flavour0}\mathbf{\overline{5}}})\egCFe{+}{A}_{2}^{1}({\color{flavour0}\mathbf{\underline{1}}}\hspace{1.02pt},\hspace{-1.5pt}{\color{flavour0}\mathbf{\overline{4}}}\hspace{1.02pt},\hspace{-1.5pt}\mathbf{3}\hspace{1.02pt},\hspace{-1.5pt}{\color{flavour0}\mathbf{\underline{2}}}\hspace{1.02pt},\hspace{-1.5pt}{\color{flavour0}\mathbf{\overline{5}}})\egCFf$\\[-8pt]}
}~\\[-60pt]
}

\vspace{5pt}\sectionAppendix{Interfacing with \package~Package}{appendix:interfacing_with_tree_amplitudes}\vspace{-10pt}

\defnBox{toAnalytic}{\var{expression}\pattern}{uses the \package\ package to express partial amplitudes in terms of spinors.
\mathematicaBox{
\mathematicaSequence{egPartial=\built{First}\brace{\funL[1]{partialAmpBasis}\brace{2,\{\fun{p},\fun{m}\}}\,}}{\funL{amp}\brace{\{\fun{f}\brace{1},\fun{p}\brace{3},\fun{m}\brace{4},\fun{f}\brace{2},\fun{fb}\brace{2},\fun{fb}\brace{1}\,\},\!\{\!\{1,6\},\{4,5\}\!\}}}
\mathematicaSequence{\funL[1]{toAnalytic}\brace{egPartial}}{$\displaystyle\frac{\fun{ab}\brace{1,4}\hspace{2pt}\fun{asb}\brace{4,\fun{p}\brace{5,6},2}^{\hspace{2pt}3}}{\fun{ab}\brace{1,3}\hspace{2pt}\fun{ab}\brace{3,4}\hspace{2pt}\fun{asb}\brace{1,\fun{p}\brace{3,4},2}\hspace{2pt}\fun{asb}\brace{4,\fun{p}\brace{1,3},6}\hspace{2pt}\fun{sb}\brace{2,5}\hspace{2pt}\fun{s}\brace{1,3,4}}$\\
$\rule{0pt}{20pt}\mbox{}\hspace{-8pt}\displaystyle{-}\frac{\fun{ab}\brace{5,6}^{\hspace{2pt}2}\hspace{2pt}\fun{asb}\brace{1,\fun{p}\brace{2,4},3}\hspace{2pt}\fun{sb}\brace{2,3}\,^{\hspace{2pt}3}}{\fun{ab}\brace{6,1}\hspace{2pt}\fun{asb}\brace{1,\fun{p}\brace{3,4},2}\hspace{2pt}\fun{asb}\brace{5,\fun{p}\brace{1,6},3}\hspace{2pt}\hspace{2pt}\fun{sb}\brace{3,4}\,\hspace{2pt}\fun{sb}\brace{4,2}\hspace{2pt}\fun{s}\brace{5,6,1}}$\\$\rule{0pt}{20pt}\mbox{}\hspace{-8pt}\displaystyle{-}\frac{\fun{ab}\brace{4,5}^{\hspace{2pt}3}\hspace{2pt}\fun{sb}\brace{1,3}\,^{\hspace{2pt}2}\hspace{2pt}\fun{sb}\brace{3,6}\,}{\fun{ab}\brace{2,5}\hspace{2pt}\fun{asb}\brace{4,\fun{p}\brace{1,3},6}\hspace{2pt}\fun{asb}\brace{5,\fun{p}\brace{1,6},3}\hspace{2pt}\fun{sb}\brace{6,1}\,\,\fun{s}\brace{6,1,3}}$}
\mathematicaSequence{\funL[1]{nice}@\%}{$\displaystyle\frac{\ab{14}\langle 4|(56)|2]^{3}}{\ab{13}\ab{34}\langle 1|(34)|2]\langle 4|(13)|6][25]s_{134}}{-}\frac{\ab{56}^{2}\langle 1|(24)|3][23]^{3}}{\ab{61}\langle 1|(34)|2]\langle 5|(16)|3][34][42]s_{561}}$\\
$\displaystyle\rule{0pt}{20pt}\mbox{}\hspace{-8pt}{-}\frac{\ab{45}^{3}[13]^{2}[36]}{\ab{25}\langle 4|(13)|6]\langle 5|(16)|3][61]s_{613}}$}
}
}

\defnBox{toNumeric}{\var{expression}\pattern}{ takes an amplitude, and yields the numerical evaluation of the amplitude using a set of reference kinematics. If no kinematics have been specified by the user, then \fun{useReferenceKinematics} (for the appropriate multiplicity) of the \package\ package will be called. 
\mathematicaBox{
\mathematicaSequence{\funL[1]{toNumeric}\brace{\funL[1]{partialAmpBasis}\brace{3}\,}~\\[16pt]}{$\displaystyle\left\{{-}\frac{425240}{43465653},{-}\frac{708733052744}{40891080952953},\frac{2638400}{1341306351},\frac{9861420160}{77688614619}\right\}$\\[24pt]}~\\[-104pt]
&\scalebox{0.6}{\hspace{-5pt}Explicit kinematics appropriate for 6-particle scattering must be assigned}\\[-5pt]
&\scalebox{0.6}{\hspace{5pt}Initializing kinematics using \funL[1]{useReferenceKinematics}\brace{6}}\\[30pt]
}
}

\addtocontents{toc}{\protect~\\[-20pt]\mbox{\vspace{0pt}}\protect\hrulefill\par}
\renewcommand{\indexname}{Alphabetic Glossary of Functions Defined by the Package}

\newpage
\hypertarget{index}{}\printindex

\newpage
\addcontentsline{toc}{section}{References}

\providecommand{\href}[2]{#2}\begingroup\raggedright\endgroup

\end{document}

%% file: header_data.tex
\let\olditemize\itemize\renewcommand{\itemize}{\vspace{-4.pt}\olditemize\setlength{\itemsep}{0.75pt}\setlength{\parskip}{0pt}\setlength{\parsep}{0pt}}
\let\oldenumerate\enumerate\renewcommand{\enumerate}{\vspace{-4pt}\oldenumerate\setlength{\itemsep}{1pt}\setlength{\parskip}{0pt}\setlength{\parsep}{0pt}}
\makeatletter\renewcommand\section{\addtocontents{toc}{\protect\addvspace{-15.5\p@}}\@startsection {section}{1}{\z@}{-0.0ex \@plus .2ex \@minus 0.2ex}{1ex \@plus.1ex\@minus 0.5ex}{\normalfont\large\bfseries}}
\renewcommand\subsection{\addtocontents{toc}{\protect\addvspace{-2.25\p@}}\@startsection {subsection}{2}{\z@}{0.5ex \@plus .2ex \@minus 0.2ex}{0.75ex \@plus.1ex\@minus 0.5ex}{\normalfont\bfseries}}
\renewcommand\subsubsection{\addtocontents{toc}{\protect\addvspace{-2.25\p@}}\@startsection {subsubsection}{3}{\z@}{0.5ex \@plus .2ex \@minus 0.2ex}{0.75ex \@plus.1ex\@minus 0.5ex}{\normalfont\bfseries}}
\renewcommand\paragraph{\addtocontents{toc}{\protect\addvspace{-2.25\p@}}\@startsection {paragraph}{4}{\z@}{0.5ex \@plus .2ex \@minus 0.2ex}{0.75ex \@plus.1ex\@minus 0.5ex}{\normalfont\bfseries}}
\newcommand{\eq}[1]{\vspace{-4.5pt}\begin{equation}#1\vspace{-4.5pt}\end{equation}}

\newcommand{\fwbox}[2]{\text{\makebox[#1][c]{$\hspace{-150pt}\displaystyle#2\hspace{-150pt}$}}}
\newcommand{\fwboxL}[2]{\text{\makebox[#1][l]{$#2$}}}
\newcommand{\fwboxR}[2]{\text{\makebox[#1][r]{$#2$}}}
\newcommand{\equivR}{\fwbox{14.5pt}{\hspace{-0pt}\fwboxR{0pt}{\raisebox{0.47pt}{\hspace{1.25pt}:\hspace{-4pt}}}=\fwboxL{0pt}{}}}
\newcommand{\equivL}{\fwbox{14.5pt}{\fwboxR{0pt}{}=\fwboxL{0pt}{\raisebox{0.47pt}{\hspace{-4pt}:\hspace{1.25pt}}}}}

\newcommand{\bigger}[1]{\raisebox{-0.95pt}{\scalebox{1.25}{$#1$}}}

\renewcommand{\phi}{\varphi}
\renewcommand{\bar}{\overline}

\newcommand{\ab}[1]{\langle #1\rangle}

\newcommand{\package}{\textbf{\tt{\textbf{\textbf{tree}}}\rule[-1.05pt]{7.5pt}{.75pt}\tt{\textbf{amplitudes}}}}
\newcommand{\fpackage}{\textbf{\tt{\textbf{fermionic}}\rule[-1.05pt]{7.5pt}{.75pt}\tt{\textbf{amplitudes}}}}

\newcommand{\sectionAppendix}[2]{\section{\texorpdfstring{#1}{\Alph{section}.\, #1}\label{#2}}}
\newcommand{\subsectionAppendix}[2]{\subsection{\texorpdfstring{#1}{\Alph{section}.\arabic{subsection}\, #1}\label{#2}}}
\newcommand{\subsubsectionAppendix}[2]{\subsubsection{\texorpdfstring{#1}{\Alph{section}.\arabic{subsection}.\arabic{subsubsection}\, #1}\label{#2}}}

\definecolor{hred}{rgb}{0.575,0.0,0.225}
\definecolor{mhvBlue}{rgb}{0.3,0.2,0.75}
\definecolor{varcolor}{rgb}{0.1,0.55,0.25}
\definecolor{functioncolor}{rgb}{0.1,0.35,0.75}
\definecolor{paper_blue}{rgb}{0.3,0.2,0.75}
\definecolor{paper_red}{rgb}{0.65,0.1,0.15}
\definecolor{paper_green}{rgb}{0.05,0.35,0.125}
\definecolor{paper_grey}{gray}{0.375}
\definecolor{hblue}{rgb}{0,0,0.725}
\definecolor{hred}{rgb}{0.575,0.0,0.225}
\definecolor{dim}{rgb}{0.55,0.55,0.55}
\definecolor{deemph}{rgb}{0.25,0.25,0.25}
\definecolor{hteal}{rgb}{0.0,0.545,0.7451}
\definecolor{rindou1}{rgb}{0.4431,0.2862,0.7960}
\definecolor{rindou2}{rgb}{0.0078,0.1215,0.4392}
\definecolor{lapis}{rgb}{0.0.0470,0.2941,0.5568}
\definecolor{emerald}{rgb}{0.31, 0.78, 0.47}
\definecolor{pinegreen}{rgb}{0.0, 0.47, 0.44}
\definecolor{jade}{rgb}{0.0, 0.66, 0.42}
\definecolor{teal}{rgb}{0.0, 0.5, 0.5}
\definecolor{hblue}{rgb}{0,0,0.725}
\definecolor{hred}{rgb}{0.575,0.0,0.225}
\definecolor{hgreen}{rgb}{0.0,0.4,0.2}
\definecolor{hteal}{rgb}{0.0,0.545,0.7451}

\definecolor{flavour0}{rgb}{0,0.4,0.9}
\definecolor{flavour1}{rgb}{0.3,0.3,0.9}
\definecolor{flavour2}{rgb}{0.9,0.3,0.3}
\definecolor{flavour3}{rgb}{0.0,0.5,0.4}
\definecolor{flavour4}{rgb}{0.4,0.2,0.7}
\definecolor{flavour5}{rgb}{0,0.4,0.9}
\definecolor{flavour6}{rgb}{0.9,0,0.5}

\renewcommand{\r}[1]{{\color{hred}#1}}
\renewcommand{\b}[1]{{\color{hblue}#1}}

\newcommand{\g}[1]{{\color{paper_green}#1}}

\renewcommand{\t}[1]{{\color{hteal}#1}}

\newcounter{lineNo}
\def\inItemQ{0}
\renewcommand{\brace}[1]{\hspace{-2pt}{\texttt{[}}#1{\texttt{]}}\hspace{-3pt}}

\newcommand{\var}[1]{{{\color{varcolor}{\sl#1}}}}
\newcommand{\fun}[1]{\text{\texttt{\textbf{\texttt{\color{functioncolor}#1}}}}}

\newcommand{\built}[1]{{\color{black}\textbf{\texttt{#1}}}}

\newcommand{\indices}[2]{{\hspace{-0.0pt}}^{\smash{#1}}_{\phantom{\smash{#1}}\smash{#2}}}

\newcommand{\pattern}{{\color{varcolor}\rule[-1.05pt]{7.5pt}{.65pt}}}
\newcommand{\patternTwo}{{\color{varcolor}\rule[-1.05pt]{12pt}{.65pt}}}

\newcommand{\defnBox}[4][2]{\def\inItemQ{1}~\\[-12pt]\noindent\fwbox{24pt}{}\fwboxL{425pt}{\begin{minipage}[t]{0.932\textwidth}%
\ifthenelse{\equal{#1}{5}}{\hypertarget{gaugeTheory:#2}{\hspace{-22pt}{$\bullet$\,\,\ifthenelse{\equal{#3}{}}{\built{#2}}{\fun{#2}}}}}{%
\ifthenelse{\equal{#1}{2}}{\hypertarget{#2}{\hspace{-22pt}{$\bullet$\,\,\ifthenelse{\equal{#3}{}}{\built{#2}}{\fun{#2}}}}}{\hspace{-22pt}{$\bullet$\,\,\ifthenelse{\equal{#3}{}}{\built{#2}}{\fun{#2}}}}}%
\ifthenelse{\equal{#1}{2}}{\ifthenelse{\equal{#3}{}}{\index{#2@\built{#2}|textbf}}{\index{#2@\fun{#2}|textbf}}}{\ifthenelse{\equal{#1}{5}}{\index{$\a$a(gaugeTheory)#2@\fun{#2} (state)|textbf}}{\ifthenelse{\equal{#3}{}}{\index{#2@\built{#2}|textnormal}}{\index{#2@\fun{#2}|textnormal}}}}%
\ifthenelse{\equal{#3}{}}{\hspace{-1.75pt}}{\brace{#3}}%
\hspace{0pt}\textnormal{\texttt{:\!}}\,\,#4~\\[-12pt]\end{minipage}\hspace{\fill}~\def\inItemQ{0}\vspace{5pt}\hspace{\fill}}\\[-12pt]}

\newcommand{\defnBoxTwo}[5][2]{\def\inItemQ{1}~\\[-12pt]\noindent\fwbox{24pt}{}\fwboxL{425pt}{\begin{minipage}[t]{0.932\textwidth}\ifthenelse{\equal{#1}{2}}{\hypertarget{#2}{\hspace{-22pt}{$\bullet$\,\,\ifthenelse{\equal{#3}{}}{\built{#2}}{\fun{#2}}}}}{\hspace{-22pt}{$\bullet$\,\,\ifthenelse{\equal{#3}{}}{\built{#2}}{\fun{#2}}}}\ifthenelse{\equal{#1}{2}}{\ifthenelse{\equal{#3}{}}{\index{#2@\built{#2}|textbf}}{\index{#2@\fun{#2}|textbf}}}{\ifthenelse{\equal{#3}{}}{\index{#2@\built{#2}}}{\index{#2@\fun{#2}}}}\ifthenelse{\equal{#3}{}}{\hspace{-1.75pt}}{\brace{#3}}\hspace{1.5pt}\ifthenelse{\equal{#4}{}}{\hspace{-1.75pt}}{\brace{#4}}\hspace{0pt}\textnormal{\texttt{:\!}}\,\,#5~\\[-12pt]\end{minipage}\hspace{\fill}~\def\inItemQ{0}\vspace{5pt}\hspace{\fill}}\\[-12pt]}

\newcommand{\mathematicaBox}[1]{~\\[0pt]~\\[-12pt]\noindent\fwboxL{450pt}{\setcounter{lineNo}{0}\ifthenelse{\inItemQ=1}{\phantom{}\hspace{-24.55pt}}{\phantom{}\hspace{-0.5pt}}\boxed{\ifthenelse{\inItemQ=1}{\begin{minipage}[t]{1.0585\textwidth}}{\begin{minipage}{0.9865\textwidth}}\begin{tabular}{@{}l@{$\;\;\;$}p{11cm}@{}}#1\end{tabular}\end{minipage}\hspace{\fill}}}\hspace{\fill}\\[2pt]}

\newcommand{\mathematicaSequence}[3][0]{\stepcounter{lineNo}\ifthenelse{\value{lineNo}=1}{}{~\\[-12pt]}
{\color{paper_blue}{\scriptsize{\tt In[}\arabic{lineNo}{\tt]}}\raisebox{-0.65pt}{{\scriptsize{\tt\raisebox{0.75pt}{:}=}}}}&\begin{minipage}[t]{14cm}{\tt #2}\end{minipage}\\[2pt]
\addtocounter{lineNo}{#1}\ifthenelse{\equal{#3}{}}{}{{\color{paper_blue}{\scriptsize{\tt Out[\arabic{lineNo}]}}\raisebox{-0.65pt}{{\scriptsize{\tt=}}}}&\begin{minipage}[t]{14cm}{\tt #3}\end{minipage}\\[1pt]}}

\newcommand{\fourBr}[1][0]{\hyperlink{fourBr:ab}{\fun{ab}}\ifthenelse{\equal{#1}{0}}{\index{abFour@\fun{ab}\,(four-bracket)|textnormal}}{[#1]\index{abFour@\fun{ab}\,(four-bracket)|textnormal}}}

\newcommand{\funL}[2][2]{\ifthenelse{\equal{#1}{2}}{\hyperlink{#2}{\fun{#2}\hspace{0pt}}\index{#2@\fun{#2}}}{\ifthenelse{\equal{#1}{1}}{\hyperlink{#2}{\textbf{\texttt{{\color{black}#2}}}}}{\ifthenelse{\equal{#1}{3}}{\hyperlink{#2}{\textbf{\texttt{{\color{black}#2}}}}\index{#2@\built{#2}}}{\hyperlink{#2}{\textbf{\texttt{{\color{black}#2}}}}\index{#2@\fun{#2}}}}}}

\newcommand{\builtL}[2][2]{\ifthenelse{\equal{#1}{2}}{\hyperlink{#2}{\built{#2}}\index{#2@\built{#2}}}{\ifthenelse{\equal{#1}{1}}{\hyperlink{#2}{\textbf{\texttt{{\color{black}#2}}}}}{\ifthenelse{\equal{#1}{3}{{\hyperlink{#2}{\textbf{\texttt{{\color{black}#2}}}\index{#2@\fun{#2}}}}
}{\hyperlink{#2}{\textbf{\texttt{{\color{black}#2}}}\index{#2@\built{#2}}}}}}}}

\newcommand{\stateYM}[2][0]{\ifthenelse{\equal{#1}{0}}{\hyperlink{sYM:#2}{\fun{#2}}\index{$\a$a(sYM)#2@\fun{#2} (sYM state)}}{\hypertarget{sYM:#2}{\fun{#2}}\index{$\a$a(sYM)#2@\fun{#2} (sYM state)|textbf}}}

\newcommand{\stateGR}[2][0]{\ifthenelse{\equal{#1}{0}}{\hyperlink{sGR:#2}{\fun{#2}}\index{$\a$a(superGR)#2@\fun{#2} (sGR state)}}{\hypertarget{sGR:#2}{\fun{#2}}\index{$\a$a(superGR)#2@\fun{#2} (sGR state)|textbf}}}

\newcommand{\stateF}[2][0]{\ifthenelse{\equal{#1}{0}}{\hyperlink{gaugeTheory:#2}{\fun{#2}}\index{$\a$a(gaugeTheory)#2@\fun{#2} (state)}}{\hypertarget{gaugeTheory:#2}{\fun{#2}}\index{$\a$a(gaugeTheory)#2@\fun{#2} (state)|textbf}}}

\newcommand{\stateB}[2][0]{\ifthenelse{\equal{#1}{0}}{\hyperlink{gaugeTheory:#2}{\fun{#2}}\index{$\a$a(gaugeTheory)#2@\fun{#2} (state)}}{\hypertarget{gaugeTheory:#2}{\fun{#2}}\index{$\a$a(gaugeTheory)#2@\fun{#2} (state)|textbf}}}

\newcommand{\sparse}[2]{\text{\built{SparseArray}}\Big[\begin{tikzpicture}[baseline=-3pt]\draw[draw=black!8.23,line width=0.5pt,fill=black!3.13,rounded corners=2pt](0,-.5)rectangle(4.25,.5);\draw[draw=none,fill=black!29.8,rounded corners=0pt](0.135,-0.09)rectangle(0.405,0.18);\draw[line width=1.2pt,black!3.13](0.135+0.05,0.045)--(0.405-0.05,0.045);
\draw[line width=1.2pt,black!3.13](0.27,-0.09+0.05)--(.27,0.18-0.05);\node[anchor=180]at(1.25,.25){\text{{\tiny{\color[rgb]{0.501,0.501,0.501}{Specified elements:} {\color{black}#1}}}}};
\draw[draw=none,fill=black!6.66](0.54,-0.75/2)rectangle(0.54+0.75,0.75/2);
\draw[draw=none,fill=black!100](0.54+1*.075,-0.75/2+9*.075)rectangle(0.54+3*.075,0.75/2);
\draw[draw=none,fill=black!100](0.54+4*.075,-0.75/2+8*.075)rectangle(0.54+5*.075,-0.75/2+9*0.075);
\draw[draw=none,fill=black!100](0.54+6*.075,-0.75/2+8*.075)rectangle(0.54+7*.075,-0.75/2+10*0.075);
\draw[draw=none,fill=black!100](0.54+2*.075,-0.75/2+7*.075)rectangle(0.54+3*.075,-0.75/2+8*0.075);
\draw[draw=none,fill=black!100](0.54+7*.075,-0.75/2+7*.075)rectangle(0.54+8*.075,-0.75/2+8*0.075);
\draw[draw=none,fill=black!100](0.54+8*.075,-0.75/2+5*.075)rectangle(0.54+9*.075,-0.75/2+6*0.075);
\draw[draw=none,fill=black!100](0.54+4*.075,-0.75/2+4*.075)rectangle(0.54+5*.075,-0.75/2+5*0.075);
\draw[draw=none,fill=black!100](0.54+5*.075,-0.75/2+2*.075)rectangle(0.54+7*.075,-0.75/2+3*0.075);
\draw[draw=none,fill=black!100](0.54+9*.075,-0.75/2+2*.075)rectangle(0.54+10*.075,-0.75/2+3*0.075);
\draw[draw=none,fill=black!100](0.54+3*.075,-0.75/2+0*.075)rectangle(0.54+4*.075,-0.75/2+1*0.075);
\node[anchor=180]at(1.25,-0.15){\text{{\tiny{\color[rgb]{0.501,0.501,0.501}{Dimensions:}{\color{black}\{#2\}}}}}};\end{tikzpicture}\Big]}

\tikzset{prop/.style={line width=1,line cap=round,rounded corners=0.05pt}}
\tikzset{bosonProp/.style={decorate, decoration={snake,amplitude=.4mm,segment length=2.49mm,post length=0mm},black,line width=1,line cap=round,rounded corners=0.05pt}}
\tikzset{fermionProp/.style={prop,decoration={markings,mark connection node=connode,mark=at position 0.5 with {\node[transform shape,scale=0.24,shape={dart},dart tip angle=40,fill] (connode) {};}},postaction={decorate}}}
\tikzset{ddot/.style={fill=black,circle,minimum size=2.pt,inner sep=0}}\tikzset{dot/.style={fill=black,circle,minimum size=1.05pt,inner sep=0}}
\newcommand{\fdiagramMaa}{\begin{tikzpicture}[baseline=-1.5]\useasboundingbox(-0.35,-.95)rectangle(1.85,.95);
\coordinate(g2)at(-0.125,0);\coordinate(g1)at(0.75,-0.66);\coordinate(g3)at(0.75,0.66);\coordinate(g4)at(1.625,0);
\coordinate(f1)at(0,-0.5);\coordinate(fb1)at(1.5,-0.5);\coordinate(f2)at(0,0.5);\coordinate(fb2)at(1.5,0.5);\coordinate(v0)at(0.75,0);
\coordinate(v1)at(0.75,-0.315);\coordinate(v11)at(0.5,-0.3375);\coordinate(v12)at(1,-0.3375);\coordinate(v2)at(0.75,0.315);\coordinate(v21)at(0.5,0.3375);\coordinate(v22)at(1,0.3375);
\draw[black,bosonProp](v1)--(v0);\draw[black,bosonProp](v0)--(v2);\node[ddot]at(v0){};
\draw[bosonProp](v0)--(g2);\node[dot]at(g2){};
\draw[flavour2,fermionProp](f2)to[bend right=25/2](v2);\draw[flavour2,fermionProp](v2)to[bend right=25/2](fb2);\node[ddot]at(v2){};
\draw[flavour1,fermionProp](f1)to[bend left=25/2](v1);\draw[flavour1,fermionProp](v1)to[bend left=25/2](fb1);\node[ddot]at(v1){};
\node[anchor=east,inner sep=1pt]at(g2){{\footnotesize${3}$}};
\node[anchor=east,inner sep=1pt]at(f1){{\footnotesize${\color{flavour1}\underline{1}}$}};
\node[anchor=east,inner sep=1pt]at(f2){{\footnotesize${\color{flavour2}\underline{2}}$}};
\node[anchor=west,inner sep=1pt]at(fb2){{\footnotesize${\color{flavour2}\overline{4}}$}};
\node[anchor=west,inner sep=1pt]at(fb1){{\footnotesize${\color{flavour1}\overline{5}}$}};
\end{tikzpicture}}

\newcommand{\fdiagramMab}{\begin{tikzpicture}[baseline=-1.5]\useasboundingbox(-0.35,-.95)rectangle(1.85,.95);
\coordinate(g2)at(-0.125,0);\coordinate(g1)at(0.75,-0.66);\coordinate(g3)at(0.75,0.66);\coordinate(g4)at(1.625,0);
\coordinate(f1)at(0,-0.5);\coordinate(fb1)at(1.5,-0.5);\coordinate(f2)at(0,0.5);\coordinate(fb2)at(1.5,0.5);\coordinate(v0)at(0.75,0);
\coordinate(v1)at(0.75,-0.315);\coordinate(v11)at(0.5,-0.3375);\coordinate(v12)at(1,-0.3375);\coordinate(v2)at(0.75,0.315);\coordinate(v21)at(0.5,0.3375);\coordinate(v22)at(1,0.3375);
\draw[black,bosonProp](v1)--(v22);
\draw[bosonProp](v21)--(g2);\node[dot]at(g2){};
\draw[flavour2,fermionProp](f2)to[bend right=25/3](v21);\draw[flavour2,fermionProp](v21)to[bend right=25/3](v22);\draw[flavour2,fermionProp](v22)to[bend right=25/3](fb2);\node[ddot]at(v21){};\node[ddot]at(v22){};
\draw[flavour1,fermionProp](f1)to[bend left=25/2](v1);\draw[flavour1,fermionProp](v1)to[bend left=25/2](fb1);\node[ddot]at(v1){};
\node[anchor=east,inner sep=1pt]at(g2){{\footnotesize${3}$}};
\node[anchor=east,inner sep=1pt]at(f1){{\footnotesize${\color{flavour1}\underline{1}}$}};
\node[anchor=east,inner sep=1pt]at(f2){{\footnotesize${\color{flavour2}\underline{2}}$}};
\node[anchor=west,inner sep=1pt]at(fb2){{\footnotesize${\color{flavour2}\overline{4}}$}};
\node[anchor=west,inner sep=1pt]at(fb1){{\footnotesize${\color{flavour1}\overline{5}}$}};
\end{tikzpicture}}

\newcommand{\fdiagramMac}{\begin{tikzpicture}[baseline=-1.5]\useasboundingbox(-0.35,-.95)rectangle(1.85,.95);
\coordinate(g2)at(-0.125,0);\coordinate(g1)at(0.75,-0.66);\coordinate(g3)at(0.75,0.66);\coordinate(g4)at(1.625,0);
\coordinate(f1)at(0,-0.5);\coordinate(fb1)at(1.5,-0.5);\coordinate(f2)at(0,0.5);\coordinate(fb2)at(1.5,0.5);\coordinate(v0)at(0.75,0);
\coordinate(v1)at(0.75,-0.315);\coordinate(v11)at(0.5,-0.3375);\coordinate(v12)at(1,-0.3375);\coordinate(v2)at(0.75,0.315);\coordinate(v21)at(0.5,0.3375);\coordinate(v22)at(1,0.3375);
\draw[black,bosonProp](v2)--(v12);
\draw[bosonProp](v11)--(g2);\node[dot]at(g2){};
\draw[flavour2,fermionProp](f2)to[bend right=25/2](v2);\draw[flavour2,fermionProp](v2)to[bend right=25/2](fb2);\node[ddot]at(v2){};
\draw[flavour1,fermionProp](f1)to[bend left=25/3](v11);\draw[flavour1,fermionProp](v11)to[bend left=25/3](v12);\draw[flavour1,fermionProp](v12)to[bend left=25/3](fb1);\node[ddot]at(v11){};\node[ddot]at(v12){};
\node[anchor=east,inner sep=1pt]at(g2){{\footnotesize${3}$}};
\node[anchor=east,inner sep=1pt]at(f1){{\footnotesize${\color{flavour1}\underline{1}}$}};
\node[anchor=east,inner sep=1pt]at(f2){{\footnotesize${\color{flavour2}\underline{2}}$}};
\node[anchor=west,inner sep=1pt]at(fb2){{\footnotesize${\color{flavour2}\overline{4}}$}};
\node[anchor=west,inner sep=1pt]at(fb1){{\footnotesize${\color{flavour1}\overline{5}}$}};
\end{tikzpicture}}

\newcommand{\fdiagramMcc}{\begin{tikzpicture}[baseline=-1.5]\useasboundingbox(-0.35,-.95)rectangle(1.85,.95);
\coordinate(g2)at(-0.125,0);\coordinate(g1)at(0.75,-0.66);\coordinate(g3)at(0.75,0.66);\coordinate(g4)at(1.625,0);
\coordinate(f1)at(0,-0.5);\coordinate(fb1)at(1.5,-0.5);\coordinate(f2)at(0,0.5);\coordinate(fb2)at(1.5,0.5);\coordinate(v0)at(0.75,0);
\coordinate(v1)at(0.75,-0.315);\coordinate(v11)at(0.5,-0.3375);\coordinate(v12)at(1,-0.3375);\coordinate(v2)at(0.75,0.315);\coordinate(v21)at(0.5,0.3375);\coordinate(v22)at(1,0.3375);
\draw[black,bosonProp](v2)--(v11);
\draw[bosonProp](v12)--(g4);\node[dot]at(g4){};
\draw[flavour2,fermionProp](f2)to[bend right=25/2](v2);\draw[flavour2,fermionProp](v2)to[bend right=25/2](fb2);\node[ddot]at(v2){};
\draw[flavour1,fermionProp](f1)to[bend left=25/3](v11);\draw[flavour1,fermionProp](v11)to[bend left=25/3](v12);\draw[flavour1,fermionProp](v12)to[bend left=25/3](fb1);\node[ddot]at(v11){};\node[ddot]at(v12){};
\node[anchor=west,inner sep=1pt]at(g4){{\footnotesize${3}$}};
\node[anchor=east,inner sep=1pt]at(f1){{\footnotesize${\color{flavour1}\underline{1}}$}};
\node[anchor=east,inner sep=1pt]at(f2){{\footnotesize${\color{flavour2}\underline{2}}$}};
\node[anchor=west,inner sep=1pt]at(fb2){{\footnotesize${\color{flavour2}\overline{4}}$}};
\node[anchor=west,inner sep=1pt]at(fb1){{\footnotesize${\color{flavour1}\overline{5}}$}};
\end{tikzpicture}}

\newcommand{\fdiagramMcb}{\begin{tikzpicture}[baseline=-1.5]\useasboundingbox(-0.35,-.95)rectangle(1.85,.95);
\coordinate(g2)at(-0.125,0);\coordinate(g1)at(0.75,-0.66);\coordinate(g3)at(0.75,0.66);\coordinate(g4)at(1.625,0);
\coordinate(f1)at(0,-0.5);\coordinate(fb1)at(1.5,-0.5);\coordinate(f2)at(0,0.5);\coordinate(fb2)at(1.5,0.5);\coordinate(v0)at(0.75,0);
\coordinate(v1)at(0.75,-0.315);\coordinate(v11)at(0.5,-0.3375);\coordinate(v12)at(1,-0.3375);\coordinate(v2)at(0.75,0.315);\coordinate(v21)at(0.5,0.3375);\coordinate(v22)at(1,0.3375);
\draw[black,bosonProp](v1)--(v21);
\draw[bosonProp](v22)--(g4);\node[dot]at(g4){};
\draw[flavour2,fermionProp](f2)to[bend right=25/3](v21);\draw[flavour2,fermionProp](v21)to[bend right=25/3](v22);\draw[flavour2,fermionProp](v22)to[bend right=25/3](fb2);\node[ddot]at(v21){};\node[ddot]at(v22){};
\draw[flavour1,fermionProp](f1)to[bend left=25/2](v1);\draw[flavour1,fermionProp](v1)to[bend left=25/2](fb1);\node[ddot]at(v1){};
\node[anchor=west,inner sep=1pt]at(g4){{\footnotesize${3}$}};
\node[anchor=east,inner sep=1pt]at(f1){{\footnotesize${\color{flavour1}\underline{1}}$}};
\node[anchor=east,inner sep=1pt]at(f2){{\footnotesize${\color{flavour2}\underline{2}}$}};
\node[anchor=west,inner sep=1pt]at(fb2){{\footnotesize${\color{flavour2}\overline{4}}$}};
\node[anchor=west,inner sep=1pt]at(fb1){{\footnotesize${\color{flavour1}\overline{5}}$}};
\end{tikzpicture}}

\newcommand{\fdiagramMca}{\begin{tikzpicture}[baseline=-1.5]\useasboundingbox(-0.35,-.95)rectangle(1.85,.95);
\coordinate(g2)at(-0.125,0);\coordinate(g1)at(0.75,-0.66);\coordinate(g3)at(0.75,0.66);\coordinate(g4)at(1.625,0);
\coordinate(f1)at(0,-0.5);\coordinate(fb1)at(1.5,-0.5);\coordinate(f2)at(0,0.5);\coordinate(fb2)at(1.5,0.5);\coordinate(v0)at(0.75,0);
\coordinate(v1)at(0.75,-0.315);\coordinate(v11)at(0.5,-0.3375);\coordinate(v12)at(1,-0.3375);\coordinate(v2)at(0.75,0.315);\coordinate(v21)at(0.5,0.3375);\coordinate(v22)at(1,0.3375);
\draw[black,bosonProp](v1)--(v0);\draw[black,bosonProp](v0)--(v2);\node[ddot]at(v0){};
\draw[bosonProp](v0)--(g4);\node[dot]at(g4){};
\draw[flavour2,fermionProp](f2)to[bend right=25/2](v2);\draw[flavour2,fermionProp](v2)to[bend right=25/2](fb2);\node[ddot]at(v2){};
\draw[flavour1,fermionProp](f1)to[bend left=25/2](v1);\draw[flavour1,fermionProp](v1)to[bend left=25/2](fb1);\node[ddot]at(v1){};
\node[anchor=west,inner sep=1pt]at(g4){{\footnotesize${3}$}};
\node[anchor=east,inner sep=1pt]at(f1){{\footnotesize${\color{flavour1}\underline{1}}$}};
\node[anchor=east,inner sep=1pt]at(f2){{\footnotesize${\color{flavour2}\underline{2}}$}};
\node[anchor=west,inner sep=1pt]at(fb2){{\footnotesize${\color{flavour2}\overline{4}}$}};
\node[anchor=west,inner sep=1pt]at(fb1){{\footnotesize${\color{flavour1}\overline{5}}$}};
\end{tikzpicture}}

\newcommand{\fdiagramMba}{\begin{tikzpicture}[baseline=-1.5]\useasboundingbox(-0.35,-.95)rectangle(1.85,.95);
\coordinate(g2)at(-0.125,0);\coordinate(g1)at(0.75,-0.66);\coordinate(g3)at(0.75,0.66);\coordinate(g4)at(1.625,0);
\coordinate(f1)at(0,-0.5);\coordinate(fb1)at(1.5,-0.5);\coordinate(f2)at(0,0.5);\coordinate(fb2)at(1.5,0.5);\coordinate(v0)at(0.75,0);
\coordinate(v1)at(0.75,-0.315);\coordinate(v11)at(0.5,-0.3375);\coordinate(v12)at(1,-0.3375);\coordinate(v2)at(0.75,0.315);\coordinate(v21)at(0.5,0.3375);\coordinate(v22)at(1,0.3375);
\draw[black,bosonProp](v1)--(v21);
\draw[bosonProp](v22)--(g3);\node[dot]at(g3){};
\draw[flavour2,fermionProp](f2)to[bend right=25/3](v21);\draw[flavour2,fermionProp](v21)to[bend right=25/3](v22);\draw[flavour2,fermionProp](v22)to[bend right=25/3](fb2);\node[ddot]at(v21){};\node[ddot]at(v22){};
\draw[flavour1,fermionProp](f1)to[bend left=25/2](v1);\draw[flavour1,fermionProp](v1)to[bend left=25/2](fb1);\node[ddot]at(v1){};
\node[anchor=south,inner sep=1pt]at(g3){{\footnotesize${3}$}};
\node[anchor=east,inner sep=1pt]at(f1){{\footnotesize${\color{flavour1}\underline{1}}$}};
\node[anchor=east,inner sep=1pt]at(f2){{\footnotesize${\color{flavour2}\underline{2}}$}};
\node[anchor=west,inner sep=1pt]at(fb2){{\footnotesize${\color{flavour2}\overline{4}}$}};
\node[anchor=west,inner sep=1pt]at(fb1){{\footnotesize${\color{flavour1}\overline{5}}$}};
\end{tikzpicture}}

\newcommand{\fdiagramMbb}{\begin{tikzpicture}[baseline=-1.5]\useasboundingbox(-0.35,-.95)rectangle(1.85,.95);
\coordinate(g2)at(-0.125,0);\coordinate(g1)at(0.75,-0.66);\coordinate(g3)at(0.75,0.66);\coordinate(g4)at(1.625,0);
\coordinate(f1)at(0,-0.5);\coordinate(fb1)at(1.5,-0.5);\coordinate(f2)at(0,0.5);\coordinate(fb2)at(1.5,0.5);\coordinate(v0)at(0.75,0);
\coordinate(v1)at(0.75,-0.315);\coordinate(v11)at(0.5,-0.3375);\coordinate(v12)at(1,-0.3375);\coordinate(v2)at(0.75,0.315);\coordinate(v21)at(0.5,0.3375);\coordinate(v22)at(1,0.3375);
\draw[black,bosonProp](v1)--(v22);
\draw[bosonProp](v21)--(g3);\node[dot]at(g3){};
\draw[flavour2,fermionProp](f2)to[bend right=25/3](v21);\draw[flavour2,fermionProp](v21)to[bend right=25/3](v22);\draw[flavour2,fermionProp](v22)to[bend right=25/3](fb2);\node[ddot]at(v21){};\node[ddot]at(v22){};
\draw[flavour1,fermionProp](f1)to[bend left=25/2](v1);\draw[flavour1,fermionProp](v1)to[bend left=25/2](fb1);\node[ddot]at(v1){};
\node[anchor=south,inner sep=1pt]at(g3){{\footnotesize${3}$}};
\node[anchor=east,inner sep=1pt]at(f1){{\footnotesize${\color{flavour1}\underline{1}}$}};
\node[anchor=east,inner sep=1pt]at(f2){{\footnotesize${\color{flavour2}\underline{2}}$}};
\node[anchor=west,inner sep=1pt]at(fb2){{\footnotesize${\color{flavour2}\overline{4}}$}};
\node[anchor=west,inner sep=1pt]at(fb1){{\footnotesize${\color{flavour1}\overline{5}}$}};
\end{tikzpicture}}

\newcommand{\fdiagramMacc}{\begin{tikzpicture}[baseline=-1.5]\useasboundingbox(-0.35,-.95)rectangle(1.85,.95);
\coordinate(g2)at(-0.125,0);\coordinate(g1)at(0.75,-0.66);\coordinate(g3)at(0.75,0.66);\coordinate(g4)at(1.625,0);
\coordinate(f1)at(0,-0.5);\coordinate(fb1)at(1.5,-0.5);\coordinate(f2)at(0,0.5);\coordinate(fb2)at(1.5,0.5);\coordinate(v0)at(0.75,0);
\coordinate(v1)at(0.75,-0.315);\coordinate(v11)at(0.5,-0.3375);\coordinate(v12)at(1,-0.3375);\coordinate(v2)at(0.75,0.315);\coordinate(v21)at(0.5,0.3375);\coordinate(v22)at(1,0.3375);
\draw[black,bosonProp](v2)--(v12);
\draw[bosonProp](v11)--(g1);\node[dot]at(g1){};
\draw[flavour2,fermionProp](f2)to[bend right=25/2](v2);\draw[flavour2,fermionProp](v2)to[bend right=25/2](fb2);\node[ddot]at(v2){};
\draw[flavour1,fermionProp](f1)to[bend left=25/3](v11);\draw[flavour1,fermionProp](v11)to[bend left=25/3](v12);\draw[flavour1,fermionProp](v12)to[bend left=25/3](fb1);\node[ddot]at(v11){};\node[ddot]at(v12){};
\node[anchor=north,inner sep=1pt]at(g1){{\footnotesize${3}$}};
\node[anchor=east,inner sep=1pt]at(f1){{\footnotesize${\color{flavour1}\underline{1}}$}};
\node[anchor=east,inner sep=1pt]at(f2){{\footnotesize${\color{flavour2}\underline{2}}$}};
\node[anchor=west,inner sep=1pt]at(fb2){{\footnotesize${\color{flavour2}\overline{4}}$}};
\node[anchor=west,inner sep=1pt]at(fb1){{\footnotesize${\color{flavour1}\overline{5}}$}};
\end{tikzpicture}}

\newcommand{\fdiagramMccc}{\begin{tikzpicture}[baseline=-1.5]\useasboundingbox(-0.35,-.95)rectangle(1.85,.95);
\coordinate(g2)at(-0.125,0);\coordinate(g1)at(0.75,-0.66);\coordinate(g3)at(0.75,0.66);\coordinate(g4)at(1.625,0);
\coordinate(f1)at(0,-0.5);\coordinate(fb1)at(1.5,-0.5);\coordinate(f2)at(0,0.5);\coordinate(fb2)at(1.5,0.5);\coordinate(v0)at(0.75,0);
\coordinate(v1)at(0.75,-0.315);\coordinate(v11)at(0.5,-0.3375);\coordinate(v12)at(1,-0.3375);\coordinate(v2)at(0.75,0.315);\coordinate(v21)at(0.5,0.3375);\coordinate(v22)at(1,0.3375);
\draw[black,bosonProp](v2)--(v11);
\draw[bosonProp](v12)--(g1);\node[dot]at(g1){};
\draw[flavour2,fermionProp](f2)to[bend right=25/2](v2);\draw[flavour2,fermionProp](v2)to[bend right=25/2](fb2);\node[ddot]at(v2){};
\draw[flavour1,fermionProp](f1)to[bend left=25/3](v11);\draw[flavour1,fermionProp](v11)to[bend left=25/3](v12);\draw[flavour1,fermionProp](v12)to[bend left=25/3](fb1);\node[ddot]at(v11){};\node[ddot]at(v12){};
\node[anchor=north,inner sep=1pt]at(g1){{\footnotesize${3}$}};
\node[anchor=east,inner sep=1pt]at(f1){{\footnotesize${\color{flavour1}\underline{1}}$}};
\node[anchor=east,inner sep=1pt]at(f2){{\footnotesize${\color{flavour2}\underline{2}}$}};
\node[anchor=west,inner sep=1pt]at(fb2){{\footnotesize${\color{flavour2}\overline{4}}$}};
\node[anchor=west,inner sep=1pt]at(fb1){{\footnotesize${\color{flavour1}\overline{5}}$}};
\end{tikzpicture}}

\newcommand{\fdiagramSFba}{\begin{tikzpicture}[baseline=-1.5]\useasboundingbox(-0.35,-.95)rectangle(1.85,.95);
\coordinate(g2)at(-0.125,0);\coordinate(g1)at(0.75,-0.66);\coordinate(g3)at(0.75,0.66);\coordinate(g4)at(1.625,0);
\coordinate(f1)at(0,-0.5);\coordinate(fb1)at(1.5,-0.5);\coordinate(f2)at(0,0.5);\coordinate(fb2)at(1.5,0.5);\coordinate(v0)at(0.75,0);
\coordinate(v1)at(0.75,-0.315);\coordinate(v11)at(0.5,-0.3375);\coordinate(v12)at(1,-0.3375);\coordinate(v2)at(0.75,0.315);\coordinate(v21)at(0.5,0.3375);\coordinate(v22)at(1,0.3375);
\draw[black,bosonProp](v1)--(v0);\draw[black,bosonProp](v0)--(v2);\node[ddot]at(v0){};
\draw[bosonProp](v0)--(g2);\node[dot]at(g2){};
\draw[flavour0,fermionProp](fb2)to[bend left=25/2](v2);\draw[flavour0,fermionProp](v2)to[bend left=25/2](f2);\node[ddot]at(v2){};
\draw[flavour0,fermionProp](f1)to[bend left=25/2](v1);\draw[flavour0,fermionProp](v1)to[bend left=25/2](fb1);\node[ddot]at(v1){};
\node[anchor=east,inner sep=1pt]at(g2){{\footnotesize${3}$}};
\node[anchor=east,inner sep=1pt]at(f1){{\footnotesize${\color{flavour0}\underline{1}}$}};
\node[anchor=east,inner sep=1pt]at(f2){{\footnotesize${\color{flavour0}\overline{4}}$}};
\node[anchor=west,inner sep=1pt]at(fb2){{\footnotesize${\color{flavour0}\underline{2}}$}};
\node[anchor=west,inner sep=1pt]at(fb1){{\footnotesize${\color{flavour0}\overline{5}}$}};
\end{tikzpicture}}

\newcommand{\fdiagramSFbb}{\begin{tikzpicture}[baseline=-1.5]\useasboundingbox(-0.35,-.95)rectangle(1.85,.95);
\coordinate(g2)at(-0.125,0);\coordinate(g1)at(0.75,-0.66);\coordinate(g3)at(0.75,0.66);\coordinate(g4)at(1.625,0);
\coordinate(f1)at(0,-0.5);\coordinate(fb1)at(1.5,-0.5);\coordinate(f2)at(0,0.5);\coordinate(fb2)at(1.5,0.5);\coordinate(v0)at(0.75,0);
\coordinate(v1)at(0.75,-0.315);\coordinate(v11)at(0.5,-0.3375);\coordinate(v12)at(1,-0.3375);\coordinate(v2)at(0.75,0.315);\coordinate(v21)at(0.5,0.3375);\coordinate(v22)at(1,0.3375);
\draw[black,bosonProp](v1)--(v22);
\draw[bosonProp](v21)--(g2);\node[dot]at(g2){};
\draw[flavour0,fermionProp](fb2)to[bend left=25/3](v22);\draw[flavour0,fermionProp](v22)to[bend left=25/3](v21);\draw[flavour0,fermionProp](v21)to[bend left=25/3](f2);\node[ddot]at(v21){};\node[ddot]at(v22){};
\draw[flavour0,fermionProp](f1)to[bend left=25/2](v1);\draw[flavour0,fermionProp](v1)to[bend left=25/2](fb1);\node[ddot]at(v1){};
\node[anchor=east,inner sep=1pt]at(g2){{\footnotesize${3}$}};
\node[anchor=east,inner sep=1pt]at(f1){{\footnotesize${\color{flavour0}\underline{1}}$}};
\node[anchor=east,inner sep=1pt]at(f2){{\footnotesize${\color{flavour0}\overline{4}}$}};
\node[anchor=west,inner sep=1pt]at(fb2){{\footnotesize${\color{flavour0}\underline{2}}$}};
\node[anchor=west,inner sep=1pt]at(fb1){{\footnotesize${\color{flavour0}\overline{5}}$}};
\end{tikzpicture}}

\newcommand{\fdiagramSFbc}{\begin{tikzpicture}[baseline=-1.5]\useasboundingbox(-0.35,-.95)rectangle(1.85,.95);
\coordinate(g2)at(-0.125,0);\coordinate(g1)at(0.75,-0.66);\coordinate(g3)at(0.75,0.66);\coordinate(g4)at(1.625,0);
\coordinate(f1)at(0,-0.5);\coordinate(fb1)at(1.5,-0.5);\coordinate(f2)at(0,0.5);\coordinate(fb2)at(1.5,0.5);\coordinate(v0)at(0.75,0);
\coordinate(v1)at(0.75,-0.315);\coordinate(v11)at(0.5,-0.3375);\coordinate(v12)at(1,-0.3375);\coordinate(v2)at(0.75,0.315);\coordinate(v21)at(0.5,0.3375);\coordinate(v22)at(1,0.3375);
\draw[black,bosonProp](v2)--(v12);
\draw[bosonProp](v11)--(g2);\node[dot]at(g2){};
\draw[flavour0,fermionProp](fb2)to[bend left=25/2](v2);\draw[flavour0,fermionProp](v2)to[bend left=25/2](f2);\node[ddot]at(v2){};
\draw[flavour0,fermionProp](f1)to[bend left=25/3](v11);\draw[flavour0,fermionProp](v11)to[bend left=25/3](v12);\draw[flavour0,fermionProp](v12)to[bend left=25/3](fb1);\node[ddot]at(v11){};\node[ddot]at(v12){};
\node[anchor=east,inner sep=1pt]at(g2){{\footnotesize${3}$}};
\node[anchor=east,inner sep=1pt]at(f1){{\footnotesize${\color{flavour0}\underline{1}}$}};
\node[anchor=east,inner sep=1pt]at(f2){{\footnotesize${\color{flavour0}\overline{4}}$}};
\node[anchor=west,inner sep=1pt]at(fb2){{\footnotesize${\color{flavour0}\underline{2}}$}};
\node[anchor=west,inner sep=1pt]at(fb1){{\footnotesize${\color{flavour0}\overline{5}}$}};
\end{tikzpicture}}

\newcommand{\fdiagramSFca}{\begin{tikzpicture}[baseline=-1.5]\useasboundingbox(-0.35,-.95)rectangle(1.85,.95);
\coordinate(g2)at(-0.125,0);\coordinate(g1)at(0.75,-0.66);\coordinate(g3)at(0.75,0.66);\coordinate(g4)at(1.625,0);
\coordinate(f1)at(0,-0.5);\coordinate(fb1)at(1.5,-0.5);\coordinate(f2)at(0,0.5);\coordinate(fb2)at(1.5,0.5);\coordinate(v0)at(0.75,0);
\coordinate(v1)at(0.75,-0.315);\coordinate(v11)at(0.5,-0.3375);\coordinate(v12)at(1,-0.3375);\coordinate(v2)at(0.75,0.315);\coordinate(v21)at(0.5,0.3375);\coordinate(v22)at(1,0.3375);
\draw[black,bosonProp](v2)--(v0);\draw[black,bosonProp](v0)--(v1);\node[ddot]at(v0){};
\draw[bosonProp](v0)--(g4);\node[dot]at(g4){};
\draw[flavour0,fermionProp](fb2)to[bend left=25/2](v2);\draw[flavour0,fermionProp](v2)to[bend left=25/2](f2);\node[ddot]at(v2){};
\draw[flavour0,fermionProp](f1)to[bend left=25/2](v1);\draw[flavour0,fermionProp](v1)to[bend left=25/2](fb1);\node[ddot]at(v1){};
\node[anchor=west,inner sep=1pt]at(g4){{\footnotesize${3}$}};
\node[anchor=east,inner sep=1pt]at(f1){{\footnotesize${\color{flavour0}\underline{1}}$}};
\node[anchor=east,inner sep=1pt]at(f2){{\footnotesize${\color{flavour0}\overline{4}}$}};
\node[anchor=west,inner sep=1pt]at(fb2){{\footnotesize${\color{flavour0}\underline{2}}$}};
\node[anchor=west,inner sep=1pt]at(fb1){{\footnotesize${\color{flavour0}\overline{5}}$}};
\end{tikzpicture}}

\newcommand{\fdiagramSFcb}{\begin{tikzpicture}[baseline=-1.5]\useasboundingbox(-0.35,-.95)rectangle(1.85,.95);
\coordinate(g2)at(-0.125,0);\coordinate(g1)at(0.75,-0.66);\coordinate(g3)at(0.75,0.66);\coordinate(g4)at(1.625,0);
\coordinate(f1)at(0,-0.5);\coordinate(fb1)at(1.5,-0.5);\coordinate(f2)at(0,0.5);\coordinate(fb2)at(1.5,0.5);\coordinate(v0)at(0.75,0);
\coordinate(v1)at(0.75,-0.315);\coordinate(v11)at(0.5,-0.3375);\coordinate(v12)at(1,-0.3375);\coordinate(v2)at(0.75,0.315);\coordinate(v21)at(0.5,0.3375);\coordinate(v22)at(1,0.3375);
\draw[black,bosonProp](v2)--(v11);
\draw[bosonProp](v12)--(g4);\node[dot]at(g4){};
\draw[flavour0,fermionProp](fb2)to[bend left=25/2](v2);\draw[flavour0,fermionProp](v2)to[bend left=25/2](f2);\node[ddot]at(v2){};
\draw[flavour0,fermionProp](f1)to[bend left=25/3](v11);\draw[flavour0,fermionProp](v11)to[bend left=25/3](v12);\draw[flavour0,fermionProp](v12)to[bend left=25/3](fb1);\node[ddot]at(v11){};\node[ddot]at(v12){};
\node[anchor=west,inner sep=1pt]at(g4){{\footnotesize${3}$}};
\node[anchor=east,inner sep=1pt]at(f1){{\footnotesize${\color{flavour0}\underline{1}}$}};
\node[anchor=east,inner sep=1pt]at(f2){{\footnotesize${\color{flavour0}\overline{4}}$}};
\node[anchor=west,inner sep=1pt]at(fb2){{\footnotesize${\color{flavour0}\underline{2}}$}};
\node[anchor=west,inner sep=1pt]at(fb1){{\footnotesize${\color{flavour0}\overline{5}}$}};
\end{tikzpicture}}

\newcommand{\fdiagramSFcc}{\begin{tikzpicture}[baseline=-1.5]\useasboundingbox(-0.35,-.95)rectangle(1.85,.95);
\coordinate(g2)at(-0.125,0);\coordinate(g1)at(0.75,-0.66);\coordinate(g3)at(0.75,0.66);\coordinate(g4)at(1.625,0);
\coordinate(f1)at(0,-0.5);\coordinate(fb1)at(1.5,-0.5);\coordinate(f2)at(0,0.5);\coordinate(fb2)at(1.5,0.5);\coordinate(v0)at(0.75,0);
\coordinate(v1)at(0.75,-0.315);\coordinate(v11)at(0.5,-0.3375);\coordinate(v12)at(1,-0.3375);\coordinate(v2)at(0.75,0.315);\coordinate(v21)at(0.5,0.3375);\coordinate(v22)at(1,0.3375);
\draw[black,bosonProp](v1)--(v21);
\draw[bosonProp](v22)--(g4);\node[dot]at(g4){};
\draw[flavour0,fermionProp](fb2)to[bend left=25/3](v22);\draw[flavour0,fermionProp](v22)to[bend left=25/3](v21);\draw[flavour0,fermionProp](v21)to[bend left=25/3](f2);\node[ddot]at(v21){};\node[ddot]at(v22){};
\draw[flavour0,fermionProp](f1)to[bend left=25/2](v1);\draw[flavour0,fermionProp](v1)to[bend left=25/2](fb1);\node[ddot]at(v1){};
\node[anchor=west,inner sep=1pt]at(g4){{\footnotesize${3}$}};
\node[anchor=east,inner sep=1pt]at(f1){{\footnotesize${\color{flavour0}\underline{1}}$}};
\node[anchor=east,inner sep=1pt]at(f2){{\footnotesize${\color{flavour0}\overline{4}}$}};
\node[anchor=west,inner sep=1pt]at(fb2){{\footnotesize${\color{flavour0}\underline{2}}$}};
\node[anchor=west,inner sep=1pt]at(fb1){{\footnotesize${\color{flavour0}\overline{5}}$}};
\end{tikzpicture}}

\newcommand{\fdiagramSFda}{\begin{tikzpicture}[baseline=-1.5]\useasboundingbox(-0.35,-.95)rectangle(1.85,.95);
\coordinate(g2)at(-0.125,0);\coordinate(g1)at(0.75,-0.66);\coordinate(g3)at(0.75,0.66);\coordinate(g4)at(1.16,0);
\coordinate(f1)at(0,-0.75);\coordinate(fb1)at(1.,-0.75);\coordinate(f2)at(0,0.75);\coordinate(fb2)at(1.,0.75);\coordinate(v0)at(0.75,0);
\coordinate(v1)at(0.185,0);\coordinate(v11)at(0.1625,-0.25);\coordinate(v12)at(0.1625,0.25);
\coordinate(v2)at(0.815,0);\coordinate(v21)at(0.8375,-0.25);\coordinate(v22)at(0.8375,0.25);
\draw[black,bosonProp](v1)--(v21);
\draw[bosonProp](v22)--(g4);\node[dot]at(g4){};
\draw[flavour0,fermionProp](fb2)to[bend right=25/3](v22);\draw[flavour0,fermionProp](v22)to[bend right=25/3](v21);\draw[flavour0,fermionProp](v21)to[bend right=25/3](fb1);\node[ddot]at(v21){};\node[ddot]at(v22){};
\draw[flavour0,fermionProp](f1)to[bend right=25/2](v1);\draw[flavour0,fermionProp](v1)to[bend right=25/2](f2);\node[ddot]at(v1){};
\node[anchor=west,inner sep=1pt]at(g4){{\footnotesize${3}$}};
\node[anchor=east,inner sep=1pt]at(f1){{\footnotesize${\color{flavour0}\underline{1}}$}};
\node[anchor=east,inner sep=1pt]at(f2){{\footnotesize${\color{flavour0}\overline{4}}$}};
\node[anchor=west,inner sep=1pt]at(fb2){{\footnotesize${\color{flavour0}\underline{2}}$}};
\node[anchor=west,inner sep=1pt]at(fb1){{\footnotesize${\color{flavour0}\overline{5}}$}};
\end{tikzpicture}}

\newcommand{\fdiagramSFdb}{\begin{tikzpicture}[baseline=-1.5]\useasboundingbox(-0.35,-.95)rectangle(1.85,.95);
\coordinate(g2)at(-0.16,0);\coordinate(g1)at(0.75,-0.66);\coordinate(g3)at(0.75,0.66);\coordinate(g4)at(1.16,0);
\coordinate(f1)at(0,-0.75);\coordinate(fb1)at(1.,-0.75);\coordinate(f2)at(0,0.75);\coordinate(fb2)at(1.,0.75);\coordinate(v0)at(0.75,0);
\coordinate(v1)at(0.185,0);\coordinate(v11)at(0.1625,-0.25);\coordinate(v12)at(0.1625,0.25);
\coordinate(v2)at(0.815,0);\coordinate(v21)at(0.8375,-0.25);\coordinate(v22)at(0.8375,0.25);
\draw[black,bosonProp](v1)--(v22);
\draw[bosonProp](v21)--(g4);\node[dot]at(g4){};
\draw[flavour0,fermionProp](fb2)to[bend right=25/3](v22);\draw[flavour0,fermionProp](v22)to[bend right=25/3](v21);\draw[flavour0,fermionProp](v21)to[bend right=25/3](fb1);\node[ddot]at(v21){};\node[ddot]at(v22){};
\draw[flavour0,fermionProp](f1)to[bend right=25/2](v1);\draw[flavour0,fermionProp](v1)to[bend right=25/2](f2);\node[ddot]at(v1){};
\node[anchor=west,inner sep=1pt]at(g4){{\footnotesize${3}$}};
\node[anchor=east,inner sep=1pt]at(f1){{\footnotesize${\color{flavour0}\underline{1}}$}};
\node[anchor=east,inner sep=1pt]at(f2){{\footnotesize${\color{flavour0}\overline{4}}$}};
\node[anchor=west,inner sep=1pt]at(fb2){{\footnotesize${\color{flavour0}\underline{2}}$}};
\node[anchor=west,inner sep=1pt]at(fb1){{\footnotesize${\color{flavour0}\overline{5}}$}};
\end{tikzpicture}}

\newcommand{\fdiagramSFaa}{\begin{tikzpicture}[baseline=-1.5]\useasboundingbox(-0.35,-.95)rectangle(1.85,.95);
\coordinate(g2)at(-0.16,0);\coordinate(g1)at(0.75,-0.66);\coordinate(g3)at(0.75,0.66);\coordinate(g4)at(1.16,0);
\coordinate(f1)at(0,-0.75);\coordinate(fb1)at(1.,-0.75);\coordinate(f2)at(0,0.75);\coordinate(fb2)at(1.,0.75);\coordinate(v0)at(0.75,0);
\coordinate(v1)at(0.185,0);\coordinate(v11)at(0.1625,-0.25);\coordinate(v12)at(0.1625,0.25);
\coordinate(v2)at(0.815,0);\coordinate(v21)at(0.8375,-0.25);\coordinate(v22)at(0.8375,0.25);
\draw[black,bosonProp](v2)--(v12);
\draw[bosonProp](v11)--(g2);\node[dot]at(g2){};
\draw[flavour0,fermionProp](fb2)to[bend right=25/2](v2);\draw[flavour0,fermionProp](v2)to[bend right=25/2](fb1);\node[ddot]at(v2){};
\draw[flavour0,fermionProp](f1)to[bend right=25/3](v11);\draw[flavour0,fermionProp](v11)to[bend right=25/3](v12);\draw[flavour0,fermionProp](v12)to[bend right=25/3](f2);\node[ddot]at(v11){};\node[ddot]at(v12){};
\node[anchor=east,inner sep=1pt]at(g2){{\footnotesize${3}$}};
\node[anchor=east,inner sep=1pt]at(f1){{\footnotesize${\color{flavour0}\underline{1}}$}};
\node[anchor=east,inner sep=1pt]at(f2){{\footnotesize${\color{flavour0}\overline{4}}$}};
\node[anchor=west,inner sep=1pt]at(fb2){{\footnotesize${\color{flavour0}\underline{2}}$}};
\node[anchor=west,inner sep=1pt]at(fb1){{\footnotesize${\color{flavour0}\overline{5}}$}};
\end{tikzpicture}}

\newcommand{\fdiagramSFab}{\begin{tikzpicture}[baseline=-1.5]\useasboundingbox(-0.35,-.95)rectangle(1.85,.95);
\coordinate(g2)at(-0.16,0);\coordinate(g1)at(0.75,-0.66);\coordinate(g3)at(0.75,0.66);\coordinate(g4)at(1.16,0);
\coordinate(f1)at(0,-0.75);\coordinate(fb1)at(1.,-0.75);\coordinate(f2)at(0,0.75);\coordinate(fb2)at(1.,0.75);\coordinate(v0)at(0.75,0);
\coordinate(v1)at(0.185,0);\coordinate(v11)at(0.1625,-0.25);\coordinate(v12)at(0.1625,0.25);
\coordinate(v2)at(0.815,0);\coordinate(v21)at(0.8375,-0.25);\coordinate(v22)at(0.8375,0.25);
\draw[black,bosonProp](v2)--(v11);
\draw[bosonProp](v12)--(g2);\node[dot]at(g2){};
\draw[flavour0,fermionProp](fb2)to[bend right=25/2](v2);\draw[flavour0,fermionProp](v2)to[bend right=25/2](fb1);\node[ddot]at(v2){};
\draw[flavour0,fermionProp](f1)to[bend right=25/3](v11);\draw[flavour0,fermionProp](v11)to[bend right=25/3](v12);\draw[flavour0,fermionProp](v12)to[bend right=25/3](f2);\node[ddot]at(v11){};\node[ddot]at(v12){};
\node[anchor=east,inner sep=1pt]at(g2){{\footnotesize${3}$}};
\node[anchor=east,inner sep=1pt]at(f1){{\footnotesize${\color{flavour0}\underline{1}}$}};
\node[anchor=east,inner sep=1pt]at(f2){{\footnotesize${\color{flavour0}\overline{4}}$}};
\node[anchor=west,inner sep=1pt]at(fb2){{\footnotesize${\color{flavour0}\underline{2}}$}};
\node[anchor=west,inner sep=1pt]at(fb1){{\footnotesize${\color{flavour0}\overline{5}}$}};
\end{tikzpicture}}

\newcommand{\sfFinalD}{\begin{tikzpicture}[baseline=-1.5]\useasboundingbox(-0.35,-.95)rectangle(1.85,.95);
\coordinate(g2)at(-0.125,0);\coordinate(g1)at(0.75,-0.66);\coordinate(g3)at(0.75,0.66);\coordinate(g4)at(1.625,0);
\coordinate(f1)at(0,-0.5);\coordinate(fb1)at(1.5,-0.5);\coordinate(f2)at(0,0.5);\coordinate(fb2)at(1.5,0.5);\coordinate(v0)at(0.75,0);
\coordinate(v1)at(0.75,-0.315);\coordinate(v11)at(0.5,-0.3375);\coordinate(v12)at(1,-0.3375);\coordinate(v2)at(0.75,0.315);\coordinate(v21)at(0.5,0.3375);\coordinate(v22)at(1,0.3375);
\draw[black,bosonProp](v1)--(v22);
\draw[bosonProp](v21)--(g2);\node[dot]at(g2){};
\draw[flavour0,fermionProp](f2)to[bend right=25/3](v21);\draw[flavour0,fermionProp](v21)to[bend right=25/3](v22);\draw[flavour0,fermionProp](v22)to[bend right=25/3](fb2);\node[ddot]at(v21){};\node[ddot]at(v22){};
\draw[flavour0,fermionProp](f1)to[bend left=25/2](v1);\draw[flavour0,fermionProp](v1)to[bend left=25/2](fb1);\node[ddot]at(v1){};
\node[anchor=east,inner sep=1pt]at(g2){{\footnotesize${3}$}};
\node[anchor=east,inner sep=1pt]at(f1){{\footnotesize${\color{flavour0}\underline{1}}$}};
\node[anchor=east,inner sep=1pt]at(f2){{\footnotesize${\color{flavour0}\underline{2}}$}};
\node[anchor=west,inner sep=1pt]at(fb2){{\footnotesize${\color{flavour0}\overline{4}}$}};
\node[anchor=west,inner sep=1pt]at(fb1){{\footnotesize${\color{flavour0}\overline{5}}$}};
\end{tikzpicture}}

\newcommand{\sfFinalDb}{\begin{tikzpicture}[baseline=-1.5]\useasboundingbox(-0.35,-.95)rectangle(1.85,.95);
\coordinate(g2)at(-0.125,0);\coordinate(g1)at(0.75,-0.66);\coordinate(g3)at(0.75,0.66);\coordinate(g4)at(1.625,0);
\coordinate(f1)at(0,-0.5);\coordinate(fb1)at(1.5,-0.5);\coordinate(f2)at(0,0.5);\coordinate(fb2)at(1.5,0.5);\coordinate(v0)at(0.75,0);
\coordinate(v1)at(0.75,-0.315);\coordinate(v11)at(0.5,-0.3375);\coordinate(v12)at(1,-0.3375);\coordinate(v2)at(0.75,0.315);\coordinate(v21)at(0.5,0.3375);\coordinate(v22)at(1,0.3375);
\draw[black,bosonProp](v1)--(v22);
\draw[bosonProp](v21)--(g2);\node[dot]at(g2){};
\draw[flavour0,fermionProp](f2)to[bend right=25/3](v21);\draw[flavour0,fermionProp](v21)to[bend right=25/3](v22);\draw[flavour0,fermionProp](v22)to[bend right=25/3](fb2);\node[ddot]at(v21){};\node[ddot]at(v22){};
\draw[flavour0,fermionProp](f1)to[bend left=25/2](v1);\draw[flavour0,fermionProp](v1)to[bend left=25/2](fb1);\node[ddot]at(v1){};
\node[anchor=east,inner sep=1pt]at(g2){{\footnotesize${3}$}};
\node[anchor=east,inner sep=1pt]at(f1){{\footnotesize${\color{flavour0}\underline{1}}$}};
\node[anchor=east,inner sep=1pt]at(f2){{\footnotesize${\color{flavour0}\underline{2}}$}};
\node[anchor=west,inner sep=1pt]at(fb2){{\footnotesize${\color{flavour0}\overline{5}}$}};
\node[anchor=west,inner sep=1pt]at(fb1){{\footnotesize${\color{flavour0}\overline{4}}$}};
\end{tikzpicture}}

\newcommand{\sfFinalB}{\begin{tikzpicture}[baseline=-1.5]\useasboundingbox(-0.35,-.95)rectangle(1.85,.95);
\coordinate(g2)at(-0.125,0);\coordinate(g1)at(0.75,-0.66);\coordinate(g3)at(0.75,0.66);\coordinate(g4)at(1.625,0);
\coordinate(f1)at(0,-0.5);\coordinate(fb1)at(1.5,-0.5);\coordinate(f2)at(0,0.5);\coordinate(fb2)at(1.5,0.5);\coordinate(v0)at(0.75,0);
\coordinate(v1)at(0.75,-0.315);\coordinate(v11)at(0.5,-0.3375);\coordinate(v12)at(1,-0.3375);\coordinate(v2)at(0.75,0.315);\coordinate(v21)at(0.5,0.3375);\coordinate(v22)at(1,0.3375);
\draw[black,bosonProp](v2)--(v12);
\draw[bosonProp](v11)--(g1);\node[dot]at(g1){};
\draw[flavour0,fermionProp](f2)to[bend right=25/2](v2);\draw[flavour0,fermionProp](v2)to[bend right=25/2](fb2);\node[ddot]at(v2){};
\draw[flavour0,fermionProp](f1)to[bend left=25/3](v11);\draw[flavour0,fermionProp](v11)to[bend left=25/3](v12);\draw[flavour0,fermionProp](v12)to[bend left=25/3](fb1);\node[ddot]at(v11){};\node[ddot]at(v12){};
\node[anchor=north,inner sep=1pt]at(g1){{\footnotesize${3}$}};
\node[anchor=east,inner sep=1pt]at(f1){{\footnotesize${\color{flavour0}\underline{1}}$}};
\node[anchor=east,inner sep=1pt]at(f2){{\footnotesize${\color{flavour0}\underline{2}}$}};
\node[anchor=west,inner sep=1pt]at(fb2){{\footnotesize${\color{flavour0}\overline{4}}$}};
\node[anchor=west,inner sep=1pt]at(fb1){{\footnotesize${\color{flavour0}\overline{5}}$}};
\end{tikzpicture}}

\newcommand{\sfFinalBb}{\begin{tikzpicture}[baseline=-1.5]\useasboundingbox(-0.35,-.95)rectangle(1.85,.95);
\coordinate(g2)at(-0.125,0);\coordinate(g1)at(0.75,-0.66);\coordinate(g3)at(0.75,0.66);\coordinate(g4)at(1.625,0);
\coordinate(f1)at(0,-0.5);\coordinate(fb1)at(1.5,-0.5);\coordinate(f2)at(0,0.5);\coordinate(fb2)at(1.5,0.5);\coordinate(v0)at(0.75,0);
\coordinate(v1)at(0.75,-0.315);\coordinate(v11)at(0.5,-0.3375);\coordinate(v12)at(1,-0.3375);\coordinate(v2)at(0.75,0.315);\coordinate(v21)at(0.5,0.3375);\coordinate(v22)at(1,0.3375);
\draw[black,bosonProp](v2)--(v12);
\draw[bosonProp](v11)--(g1);\node[dot]at(g1){};
\draw[flavour0,fermionProp](f2)to[bend right=25/2](v2);\draw[flavour0,fermionProp](v2)to[bend right=25/2](fb2);\node[ddot]at(v2){};
\draw[flavour0,fermionProp](f1)to[bend left=25/3](v11);\draw[flavour0,fermionProp](v11)to[bend left=25/3](v12);\draw[flavour0,fermionProp](v12)to[bend left=25/3](fb1);\node[ddot]at(v11){};\node[ddot]at(v12){};
\node[anchor=north,inner sep=1pt]at(g1){{\footnotesize${3}$}};
\node[anchor=east,inner sep=1pt]at(f1){{\footnotesize${\color{flavour0}\underline{1}}$}};
\node[anchor=east,inner sep=1pt]at(f2){{\footnotesize${\color{flavour0}\underline{2}}$}};
\node[anchor=west,inner sep=1pt]at(fb2){{\footnotesize${\color{flavour0}\overline{5}}$}};
\node[anchor=west,inner sep=1pt]at(fb1){{\footnotesize${\color{flavour0}\overline{4}}$}};
\end{tikzpicture}}

\newcommand{\sfFinalC}{\begin{tikzpicture}[baseline=-1.5]\useasboundingbox(-0.35,-.95)rectangle(1.85,.95);
\coordinate(g2)at(-0.125,0);\coordinate(g1)at(0.75,-0.66);\coordinate(g3)at(0.75,0.66);\coordinate(g4)at(1.625,0);
\coordinate(f1)at(0,-0.5);\coordinate(fb1)at(1.5,-0.5);\coordinate(f2)at(0,0.5);\coordinate(fb2)at(1.5,0.5);\coordinate(v0)at(0.75,0);
\coordinate(v1)at(0.75,-0.315);\coordinate(v11)at(0.5,-0.3375);\coordinate(v12)at(1,-0.3375);\coordinate(v2)at(0.75,0.315);\coordinate(v21)at(0.5,0.3375);\coordinate(v22)at(1,0.3375);
\draw[black,bosonProp](v2)--(v11);
\draw[bosonProp](v12)--(g1);\node[dot]at(g1){};
\draw[flavour0,fermionProp](f2)to[bend right=25/2](v2);\draw[flavour0,fermionProp](v2)to[bend right=25/2](fb2);\node[ddot]at(v2){};
\draw[flavour0,fermionProp](f1)to[bend left=25/3](v11);\draw[flavour0,fermionProp](v11)to[bend left=25/3](v12);\draw[flavour0,fermionProp](v12)to[bend left=25/3](fb1);\node[ddot]at(v11){};\node[ddot]at(v12){};
\node[anchor=north,inner sep=1pt]at(g1){{\footnotesize${3}$}};
\node[anchor=east,inner sep=1pt]at(f1){{\footnotesize${\color{flavour0}\underline{1}}$}};
\node[anchor=east,inner sep=1pt]at(f2){{\footnotesize${\color{flavour0}\underline{2}}$}};
\node[anchor=west,inner sep=1pt]at(fb2){{\footnotesize${\color{flavour0}\overline{4}}$}};
\node[anchor=west,inner sep=1pt]at(fb1){{\footnotesize${\color{flavour0}\overline{5}}$}};
\end{tikzpicture}}

\newcommand{\sfFinalCb}{\begin{tikzpicture}[baseline=-1.5]\useasboundingbox(-0.35,-.95)rectangle(1.85,.95);
\coordinate(g2)at(-0.125,0);\coordinate(g1)at(0.75,-0.66);\coordinate(g3)at(0.75,0.66);\coordinate(g4)at(1.625,0);
\coordinate(f1)at(0,-0.5);\coordinate(fb1)at(1.5,-0.5);\coordinate(f2)at(0,0.5);\coordinate(fb2)at(1.5,0.5);\coordinate(v0)at(0.75,0);
\coordinate(v1)at(0.75,-0.315);\coordinate(v11)at(0.5,-0.3375);\coordinate(v12)at(1,-0.3375);\coordinate(v2)at(0.75,0.315);\coordinate(v21)at(0.5,0.3375);\coordinate(v22)at(1,0.3375);
\draw[black,bosonProp](v2)--(v11);
\draw[bosonProp](v12)--(g1);\node[dot]at(g1){};
\draw[flavour0,fermionProp](f2)to[bend right=25/2](v2);\draw[flavour0,fermionProp](v2)to[bend right=25/2](fb2);\node[ddot]at(v2){};
\draw[flavour0,fermionProp](f1)to[bend left=25/3](v11);\draw[flavour0,fermionProp](v11)to[bend left=25/3](v12);\draw[flavour0,fermionProp](v12)to[bend left=25/3](fb1);\node[ddot]at(v11){};\node[ddot]at(v12){};
\node[anchor=north,inner sep=1pt]at(g1){{\footnotesize${3}$}};
\node[anchor=east,inner sep=1pt]at(f1){{\footnotesize${\color{flavour0}\underline{1}}$}};
\node[anchor=east,inner sep=1pt]at(f2){{\footnotesize${\color{flavour0}\underline{2}}$}};
\node[anchor=west,inner sep=1pt]at(fb2){{\footnotesize${\color{flavour0}\overline{5}}$}};
\node[anchor=west,inner sep=1pt]at(fb1){{\footnotesize${\color{flavour0}\overline{4}}$}};
\end{tikzpicture}}

\newcommand{\sfFinalA}{\begin{tikzpicture}[baseline=-1.5]\useasboundingbox(-0.35,-.95)rectangle(1.85,.95);
\coordinate(g2)at(-0.125,0);\coordinate(g1)at(0.75,-0.66);\coordinate(g3)at(0.75,0.66);\coordinate(g4)at(1.625,0);
\coordinate(f1)at(0,-0.5);\coordinate(fb1)at(1.5,-0.5);\coordinate(f2)at(0,0.5);\coordinate(fb2)at(1.5,0.5);\coordinate(v0)at(0.75,0);
\coordinate(v1)at(0.75,-0.315);\coordinate(v11)at(0.5,-0.3375);\coordinate(v12)at(1,-0.3375);\coordinate(v2)at(0.75,0.315);\coordinate(v21)at(0.5,0.3375);\coordinate(v22)at(1,0.3375);
\draw[black,bosonProp](v1)--(v21);
\draw[bosonProp](v22)--(g4);\node[dot]at(g4){};
\draw[flavour0,fermionProp](f2)to[bend right=25/3](v21);\draw[flavour0,fermionProp](v21)to[bend right=25/3](v22);\draw[flavour0,fermionProp](v22)to[bend right=25/3](fb2);\node[ddot]at(v21){};\node[ddot]at(v22){};
\draw[flavour0,fermionProp](f1)to[bend left=25/2](v1);\draw[flavour0,fermionProp](v1)to[bend left=25/2](fb1);\node[ddot]at(v1){};
\node[anchor=west,inner sep=1pt]at(g4){{\footnotesize${3}$}};
\node[anchor=east,inner sep=1pt]at(f1){{\footnotesize${\color{flavour0}\underline{1}}$}};
\node[anchor=east,inner sep=1pt]at(f2){{\footnotesize${\color{flavour0}\underline{2}}$}};
\node[anchor=west,inner sep=1pt]at(fb2){{\footnotesize${\color{flavour0}\overline{4}}$}};
\node[anchor=west,inner sep=1pt]at(fb1){{\footnotesize${\color{flavour0}\overline{5}}$}};
\end{tikzpicture}}

\newcommand{\sfFinalAb}{\begin{tikzpicture}[baseline=-1.5]\useasboundingbox(-0.35,-.95)rectangle(1.85,.95);
\coordinate(g2)at(-0.125,0);\coordinate(g1)at(0.75,-0.66);\coordinate(g3)at(0.75,0.66);\coordinate(g4)at(1.625,0);
\coordinate(f1)at(0,-0.5);\coordinate(fb1)at(1.5,-0.5);\coordinate(f2)at(0,0.5);\coordinate(fb2)at(1.5,0.5);\coordinate(v0)at(0.75,0);
\coordinate(v1)at(0.75,-0.315);\coordinate(v11)at(0.5,-0.3375);\coordinate(v12)at(1,-0.3375);\coordinate(v2)at(0.75,0.315);\coordinate(v21)at(0.5,0.3375);\coordinate(v22)at(1,0.3375);
\draw[black,bosonProp](v1)--(v21);
\draw[bosonProp](v22)--(g4);\node[dot]at(g4){};
\draw[flavour0,fermionProp](f2)to[bend right=25/3](v21);\draw[flavour0,fermionProp](v21)to[bend right=25/3](v22);\draw[flavour0,fermionProp](v22)to[bend right=25/3](fb2);\node[ddot]at(v21){};\node[ddot]at(v22){};
\draw[flavour0,fermionProp](f1)to[bend left=25/2](v1);\draw[flavour0,fermionProp](v1)to[bend left=25/2](fb1);\node[ddot]at(v1){};
\node[anchor=west,inner sep=1pt]at(g4){{\footnotesize${3}$}};
\node[anchor=east,inner sep=1pt]at(f1){{\footnotesize${\color{flavour0}\underline{1}}$}};
\node[anchor=east,inner sep=1pt]at(f2){{\footnotesize${\color{flavour0}\underline{2}}$}};
\node[anchor=west,inner sep=1pt]at(fb2){{\footnotesize${\color{flavour0}\overline{5}}$}};
\node[anchor=west,inner sep=1pt]at(fb1){{\footnotesize${\color{flavour0}\overline{4}}$}};
\end{tikzpicture}}

\newcommand{\egCFa}{\begin{array}{@{}c@{}}\includegraphics[scale=1]{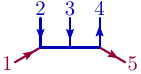}\end{array}}
\newcommand{\egCFb}{\begin{array}{@{}c@{}}\includegraphics[scale=1]{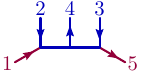}\end{array}}
\newcommand{\egCFc}{\begin{array}{@{}c@{}}\includegraphics[scale=1]{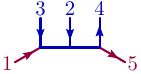}\end{array}}
\newcommand{\egCFd}{\begin{array}{@{}c@{}}\includegraphics[scale=1]{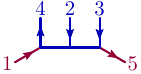}\end{array}}
\newcommand{\egCFe}{\begin{array}{@{}c@{}}\includegraphics[scale=1]{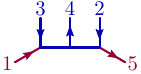}\end{array}}
\newcommand{\egCFf}{\begin{array}{@{}c@{}}\includegraphics[scale=1]{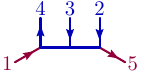}\end{array}}